\documentclass[reprint, amsmath, amssymb, aps, showkeys, showpacs, superscriptaddress
]{revtex4-2}
\usepackage{comment}
\usepackage{graphicx}
\graphicspath{{Figures/}}
\usepackage{dcolumn}
\usepackage{bm}
\usepackage{amsmath}
\usepackage{bbm}
\usepackage{subfigure}
\usepackage[tight]{units} 
\usepackage{textcomp} 
\usepackage{gensymb} 
\usepackage{bbold}
\usepackage{dsfont}
\usepackage[svgnames]{xcolor}
\usepackage{tabularx}
\usepackage{subfiles}
\usepackage{natbib}
\usepackage{appendix}

\usepackage{hyperref}
\hypersetup{
        colorlinks=true,
        citecolor=SteelBlue,
        filecolor=LimeGreen,
        linkcolor=SlateBlue,
        urlcolor=DarkOrchid
}
\usepackage{cleveref}

\newcommand{\new}[1]{{}{#1}}

\newcommand{\vplus}{\bm{ \mathcal{V} }_{+} }
\newcommand{\vplusph}{\bm{ \mathcal{V} }_{+,\text{ph}}}

\newcommand{\vtimesph}{\bm{ \mathcal{V} }_{\times,\text{ph}}}

\begin{document}

\title{Quantum precision limits of displacement noise free interferometers}
\author{Tuvia Gefen}
\affiliation{Institute for Quantum Information and Matter, California Institute of Technology, Pasadena, CA 91125, USA}
\author{Rajashik Tarafder}
\affiliation{Theoretical Astrophysics, Walter Burke Institute for Theoretical Physics, California Institute of Technology, Pasadena, CA 91125, USA}
\affiliation{LIGO Laboratory, California Institute of Technology, Pasadena, CA 91125, USA}
\author{Rana X. Adhikari}
\affiliation{LIGO Laboratory, California Institute of Technology, Pasadena, CA 91125, USA}
\author{Yanbei Chen}
\affiliation{Theoretical Astrophysics, Walter Burke Institute for Theoretical Physics, California Institute of Technology, Pasadena, CA 91125, USA}
\date{\today}

\begin{abstract}
Current laser-interferometric gravitational wave detectors suffer from a fundamental limit to their precision due to the displacement noise of optical elements contributed by various sources. Several schemes for Displacement-Noise Free Interferometers (DFI) have been proposed to mitigate their effects. The idea behind these schemes is similar to decoherence-free subspaces in quantum sensing i.e. certain modes contain information about the gravitational waves but are insensitive to the mirror motion (displacement noise). In this paper, we derive quantum precision limits for general DFI schemes, including optimal measurement basis and optimal squeezing schemes.
We introduce a triangular cavity DFI scheme and apply our general bounds to it.
Precision analysis of this scheme with different noise models shows that the DFI property leads to interesting sensitivity profiles and improved precision due to noise mitigation and larger gain from squeezing. 
\end{abstract}

\maketitle

\textit{\textbf{Introduction---}} 
Quantum metrology studies fundamental precision limits in physical measurements imposed by quantum physics. 
Recent progresses in this field have led to formulation of precision limits for a variety of sensing devices: gravitational wave (GW) detectors \cite{Caves81, Kimble01, LSC11,Tsang11, Demkowicz13, Lang13, Miao17,Branford18}, magnetometers \cite{Brask15, Baumgratz16}, atomic clocks \cite{Macieszczak14, Katori11, Chabuda16, Kaubruegger21}, nano-NMR \cite{Schmitt17,Boss17,Aharon19,Cohen20,Schmitt21}, etc. 

We focus here on optomechanical sensors and laser interferometers. These platforms have emerged as the primary instruments for the detection of GWs, with successful observations
conducted by several of these detectors \cite{abbott2016observation,abbott2017gw170817,abbott2019gwtc,abbott2021gwtc,abbott2021gwtc3}.
They are, however, severely limited by noise sources that displace the mirror positions in the interferometer:  thermal noise, Radiation Pressure Noise (RPN), seismic noise, and Newtonian gravity noise \cite{Saulson84,Hughes98, PhysRevD.86.102001, harms2015terrestrial,Buikema20}. These noises are in particular dominant in the low-frequency regime ($< 10$ Hz), thus limiting the sensitivity at this range and preventing detection of various signals such as intermediate-mass black holes,
young Neutron Stars, extreme mass ratio
in-spirals, etc. Circumventing displacement noise is thus an outstanding challenge for GW detection and optomechanical sensors in general. 
 
Interestingly, the coupling of light fields to GW signals is different from their coupling to mirror displacement, i.e. GW information is accumulated along the optical path, unlike displacement noise which is only introduced at the mirrors. This observation has led to proposals of interferometers wherein displacement noise can be canceled while not losing the GW signal~\cite{Kawamura04}. This approach is referred to as Displacement-noise Free Interferometry (DFI).

DFI for laser interferometry was originally proposed using a simplified system and later expanded to more complex systems such as speed meters and 3D-interferometers~\cite{chen06a, Chen06b, Nishizawa08, Kokeyama09, Wang16, Nishizawa22}. 
A similar approach for laser phase noise cancellation has also been proposed for LISA using Time Delay Interferometry~\cite{de2010experimental,Tinto20,Shaddock04}. However, DFI systems with requisite sensitivities remain elusive. Furthermore, a rigorous study of the quantum precision limits of these interferometers has not been conducted.  

In this paper, we use quantum metrology techniques to derive general precision limits, optimal measurements, and optimal squeezing quadratures for DFI schemes. We develop a triangular cavity DFI scheme, which combines resonance power amplification and DFI, and apply our results to analyze it. 
In addition to the improved sensitivity at low frequencies, we observe interesting effects that motivate the use of DFI and multichannel interferometers. We identify pseudo displacement-free subspaces, i.e. subspaces that are displacement-free for a limited range of frequencies. These subspaces lead to unexpected sensitivity profiles and further noise suppression. Lastly, we study the effect of squeezing and show that DFI increases the sensitivity gain from squeezing in the high displacement noise regime.

\begin{figure}[t]
\centering
\includegraphics[width=.31\textwidth]{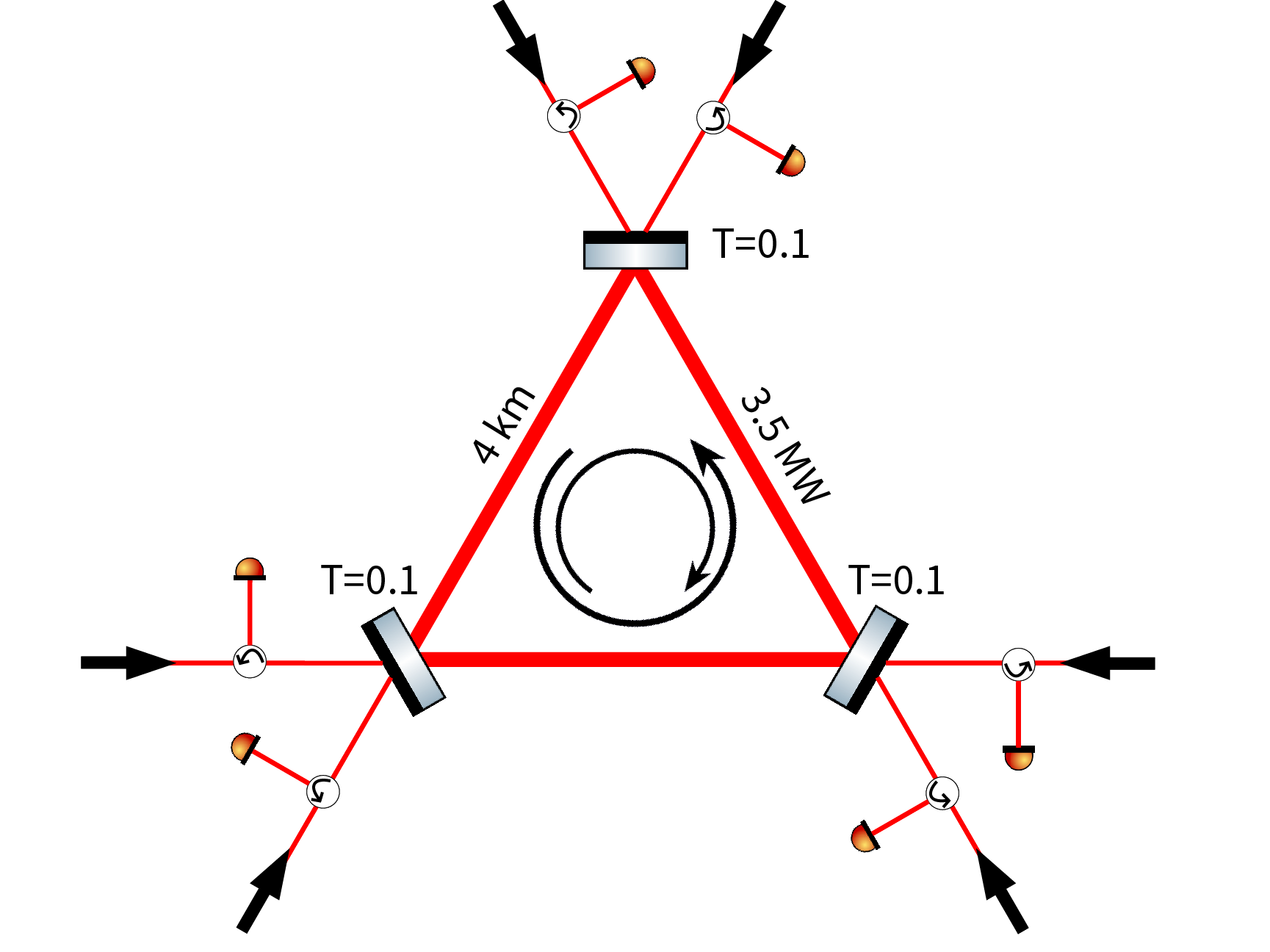}
\caption{Sketch of the DFI scheme: a symmetric triangular cavity is formed by three mirrors and six input laser fields. Six detectors are placed in the opposite direction of the input fields. The configuration leads to both a clockwise and an anti-clockwise circulating field within the cavity. We used the following parameters: Arm length: $L=4$\,km, Laser wavelength: $1064$ nm (same as advanced LIGO \cite{aasi2015advanced}). Mirrors mass: $5$ kg, Intracavity Power: $3.5\,$MW (to enhance RPN for illustrations). Power transmissivity of the mirrors: $T=0.1$.
}
\label{fig:scheme}
\end{figure}

\textit{\textbf{Formalism and Model---}} Previous DFI schemes
used several Mach-Zender interferometers \cite{chen06a, Chen06b}. However, these interferometers did not incorporate cavity resonance to amplify the power and sensitivity. 
Here, we propose a scheme that combines
DFI with cavity resonance gain: an equilateral triangular cavity with three mirrors, six input local-oscillator fields and six outputs. The six fields circulate inside the cavity - split between the clockwise
and anti-clockwise directions. The scheme and suggested parameters are described in \cref{fig:scheme}.
This triangular cavity yields power amplification: given identical mirrors transmissivities ($T$), the ratio between the intracavity power and the total input power is
$T/[3\left(1-\sqrt{1-T}\right)^{2}].$

We will show that this scheme is indeed a DFI. The intuition for this is simple: the displacement noise is generated by the three mirrors and induced on the six output fields. Since the number of mirrors is smaller than the number of output fields, we have modes that are decoupled from this noise and enable the DFI. This approach is formulated below.

We use a general formulation that holds for any system with $n$ mirrors and $k$ 
fields, such that $k>n$. The system is described using the input-output formalism \cite{Caves85, Corbitt05} and we denote the quadrature operators of the input and output fields as ${\mathbf{\hat{Q}}}_{\text{in}}=\left(\begin{array}{c}
\mathbf{\hat{a}_{1}}\\
\mathbf{\hat{a}_{2}}
\end{array}\right), \mathbf{\hat{Q}}_{\text{out}}=\left(\begin{array}{c}
\mathbf{\hat{b}_{1}}\\
\mathbf{\hat{b}_{2}}
\end{array}\right)$ respectively.
$\mathbf{\hat{a}_{1}}$, $\mathbf{\hat{b}_{1}}$
are the $k$-dimensional vectors of amplitude quadratures, and $\mathbf{\hat{a}_{2}},$ $\mathbf{\hat{b}_{2}}$ are the $k$-dimensional vectors of phase quadratures.
These quadratures satisfy the standard commutation relations:
$\left[\left(\mathbf{\hat{Q}}_{\text{out}}\right)_{l},\left(\mathbf{\hat{Q}}_{\text{out}}\right)_{k}\right]
=J_{l,k}$ with $J=i\left(\begin{array}{cc}
0 & \mathbbm{1}_{k}\\
-\mathbbm{1}_{k} & 0
\end{array}\right)$ (same for $\mathbf{\hat{Q}}_{\text{in}}$). 
 The noisy displacement of the mirrors is denoted as $\left\{ \Delta x_{i}\right\} _{i=1}^{n},$ and the amplitude of the GW polarization vector is given by $\mathbf{h}=\left( h_{+}, h_{\times}\right)^{T}.$ The input-output relations in the frequency domain are then: 
\begin{equation}
\mathbf{\hat{Q}}_{\text{out}}\left( \Omega \right)=M \left( \Omega \right)\mathbf{\hat{Q}}_{\text{in}}\left( \Omega \right)+\mathcal{V}\left( \Omega \right)\mathbf{h}\left( \Omega \right)+A\left( \Omega \right)\mathbf{\Delta x}\left( \Omega \right).
\label{eq:input-output}
\end{equation}
$\Omega=2 \pi f$ is the angular frequency, hereafter this notation will be suppressed, 
$M,$ $A,$ $\mathcal{V}$ are the transfer matrices of the input modes, displacement noise, and the GW vector respectively. Accordingly, these are $2k \times 2k$, $2k \times n$, and $2k \times 2$ dimensional matrices, that take the following general form (assuming carrier frequency is resonant with the arm length):
\begin{align}
\begin{split}
&M=\left(\begin{array}{cc}
M_{\text{int}} & 0\\
M_{21} & M_{\text{int}}
\end{array}\right), \; A=\left(\begin{array}{c}
0\\
A_{\text{ph}}
\end{array}\right),\\
&\mathcal{V}=\left(\begin{array}{c}
0\\
\mathcal{V}_{\text{ph}}
\end{array}\right)=\left(\begin{array}{cc}
0 & 0\\
\vplusph & \vtimesph
\end{array}\right).
\end{split}
\label{eq:transfer_matrices}
\end{align}
$M_{\text{int}}$ is a $k\times k$ unitary interferometer transfer matrix and $M_{21}$ is a $k \times k$ coupling matrix between the amplitude and phase quadratures due to Radiation Pressure Noise (RPN).
 $A, \mathcal{V}$ act only on the phase quadratures, with their support being $A_{\text{ph}}$($k\times n$ dimensional), $\mathcal{V}_{\text{ph}}$ ($k\times2$ dimensional).
 $\mathcal{V}_{\text{ph}}$ consists of two column vectors: $\vplusph,\vtimesph,$ these are $k$-dimensional transfer vectors of $h_{+},h_{\times}$ respectively.
A detailed description of how to calculate these transfer matrices can be found in refs. \cite{Corbitt05, supp}. 


We are now poised to define the Displacement Free Subspace (DFS): this is the space of phase quadratures of the form: $\mathbf{u}\cdot \mathbf{\hat{b}_{2}}$ with $\mathbf{u}\in\text{ker}\left(A_{\text{ph}}^{\dagger}\right).$  Since $\mathbf{u}^{\dagger} A_{\text{ph}} \mathbf{\Delta x}=0$, these quadratures are decoupled from the displacement noise term in \cref{eq:input-output} and thus resilient to this noise. Thinking of the phase quadratures as $k$-dimensional column vectors, the DFS is then the kernel of $A_{\text{ph}}^{\dagger}.$  We denote this subspace and its projection operator as $M_{\text{DFS}}$ and $\Pi_{\text{DFS}}$ respectively.
The orthogonal complement of the DFS is the coupled subspace, it is the linear span of the column vectors of $A_{\text{ph}}.$  This subspace and its projection operator are denoted as  $M_{C}$ and $\Pi_{C}$ respectively.
Since 
$A_{\text{ph}}^{\dagger}$ is an 
$n \times k$ dimensional matrix,  a sufficient condition for the existence of DFS is $k>n$, i.e. more fields than mirrors.


\textit{\textbf{Quantum precision limits---}}
Our figure of merit is the minimal detectable GW amplitude in any given polarization.
With our interferometer, the dominant polarization is approximately $h_{+}$, hence the figure of merit is the standard deviation in estimating $h_{+}$, we denote it as $\sigma$ and refer to it as the Standard Deviation (SD) or the sensitivity.
This reduces the problem to a single parameter estimation of $h_{+}$, where the sensitivity is calculated below using the Cram{\'e}r-Rao Bound.

According to the Cram{\'e}r-Rao Bound, given a readout scheme with outcomes distribution $\left\{ p\left(x\right)\right\} _{x},$ the variance, $\sigma^{2},$ of any unbiased estimator of \new{$h_{+}$} 
satisfy: $\sigma^{2}\geq F^{-1},$ with \new{$F=\langle\left(\partial_{h_{+}}\ln\left(p\right)\right)^{2}\rangle$} being the Fisher Information (FI). This lower bound is asymptotically tight \cite{cover1999elements}.

In the quantum context, further optimization over the detection schemes yields the Quantum Fisher Information (QFI), denoted as $I$, \cite{Liu19}
such that for any readout scheme $\sigma^{2}\geq I^{-1}$. 


In our case 
the QFI
has a particularly simple form \cite{stoica05,Nichols18, Branford18}:
\begin{align} I=2\left(\partial_{h_{+}}\mathbf{d}_{q}\right)^{\dagger}\Sigma_{q}^{-1}\left(\partial_{h_{+}}\mathbf{d}_{q}\right),
 \label{eq:QFIM_form}
\end{align}
where $\mathbf{d_{q}}$ and $\Sigma_{q}$ are the mean vector and covariance matrix of $\mathbf{\hat{Q}}_{\text{out}}$ respectively: 
\begin{align}
\mathbf{d_{q}}=\langle\mathbf{\hat{Q}}_{\text{out}}\rangle,\;\left(\Sigma_{q}\right)_{i,j}=\frac{1}{2}\left\langle \left\{ \hat{Q}_{\text{out},i},\hat{Q}_{\text{out},j}^{\dagger}\right\} \right\rangle -\langle\hat{Q}_{\text{out},i}\rangle\langle\hat{Q}_{\text{out},j}^{\dagger}\rangle,
\end{align}
with $\left\{ \bullet,\bullet\right\}$  being the anti-commutator of the operators.
This simple form is because the output modes are in Gaussian state,
and information about $\mathbf{h}$ is encoded only in the $\mathbf{d_{q}}$.

From \cref{eq:input-output,eq:transfer_matrices} we observe that $\partial_{h_{+}}\mathbf{d_{q}}=\vplus,$ with  $\vplus=\left(\begin{array}{cc}
0 & \vplusph\end{array}\right)^{T},$ and that $\Sigma_{q}=M\Sigma_{i}M^{\dagger}+A\Sigma_{\mathbf{\Delta x}}A^{\dagger},$ where $\Sigma_{i},$ $\Sigma_{\mathbf{\Delta x}}$ are the covariance matrices of the input quadratures and the displacement noise $\mathbf{\Delta x}$ respectively. 
Assuming the input state is vacuum and 
the displacement noise is Gaussian i.i.d.: $\mathbf{\Delta x} \sim N \left(0, \frac{1}{2}\delta ^{2} \mathbb{1}  \right),$ 
the covariance matrix is then
$\Sigma_{q}=\frac{1}{2} \left(M M^{\dagger}+\delta^{2}A A^{\dagger} \right),$ and the QFI reads:
\begin{align}
I= 4\vplus^{\dagger}\left(M M^{\dagger}+\delta^{2}A A^{\dagger}\right)^{-1}\vplus\label{eq:QFIM_full_expression}.    
\end{align}
In \cref{eq:QFIM_full_expression}, the RPN is included in the $M M^{\dagger}$ term, and the rest of the displacement noise is encoded by the additional $A A^\dagger$ term. The shot noise limit is obtained by nullifying the RPN and the displacement noise i.e. $M$ is unitary and $\delta=0$, which yields: $I=4\vplus^{\dagger}\vplus.$ This limit serves as an upper bound to any noisy QFI scenario.



The QFI (\cref{eq:QFIM_form,eq:QFIM_full_expression}) is attainable with a homodyne measurement of the quadrature
$\left(\Sigma_{q}^{-1}\vplus\right)\cdot\mathbf{\hat{Q}}_{\text{out}}$
\cite{supp,Gessner20}. Our sensitivity curves will therefore correspond to either the QFI, i.e. the SD with an optimal measurement: $\sigma=1/\sqrt{I},$ or to the FI with a specific homodyne measurement: $\sigma=1/\sqrt{F}$.    

\textit{\textbf{Precision limits of the simplified model---}} We begin with a simplified model to develop an understanding of the DFI method. The simplified model is devoid of RPN,  i.e. $M$ is unitary, and the displacement noise is taken to be a white noise, i.e. $\delta \left( \Omega \right)$ is constant. 
The QFI is therefore:
\begin{align}
I=4\vplus^{\dagger}\left(\mathbb{1}+\delta^{2} A A^{\dagger}\right)^{-1}\vplus.
\label{eq::simplified_QFI}
\end{align}

\begin{figure}[t]
\begin{center}
\subfigure[]{\includegraphics[width=.34\textwidth]{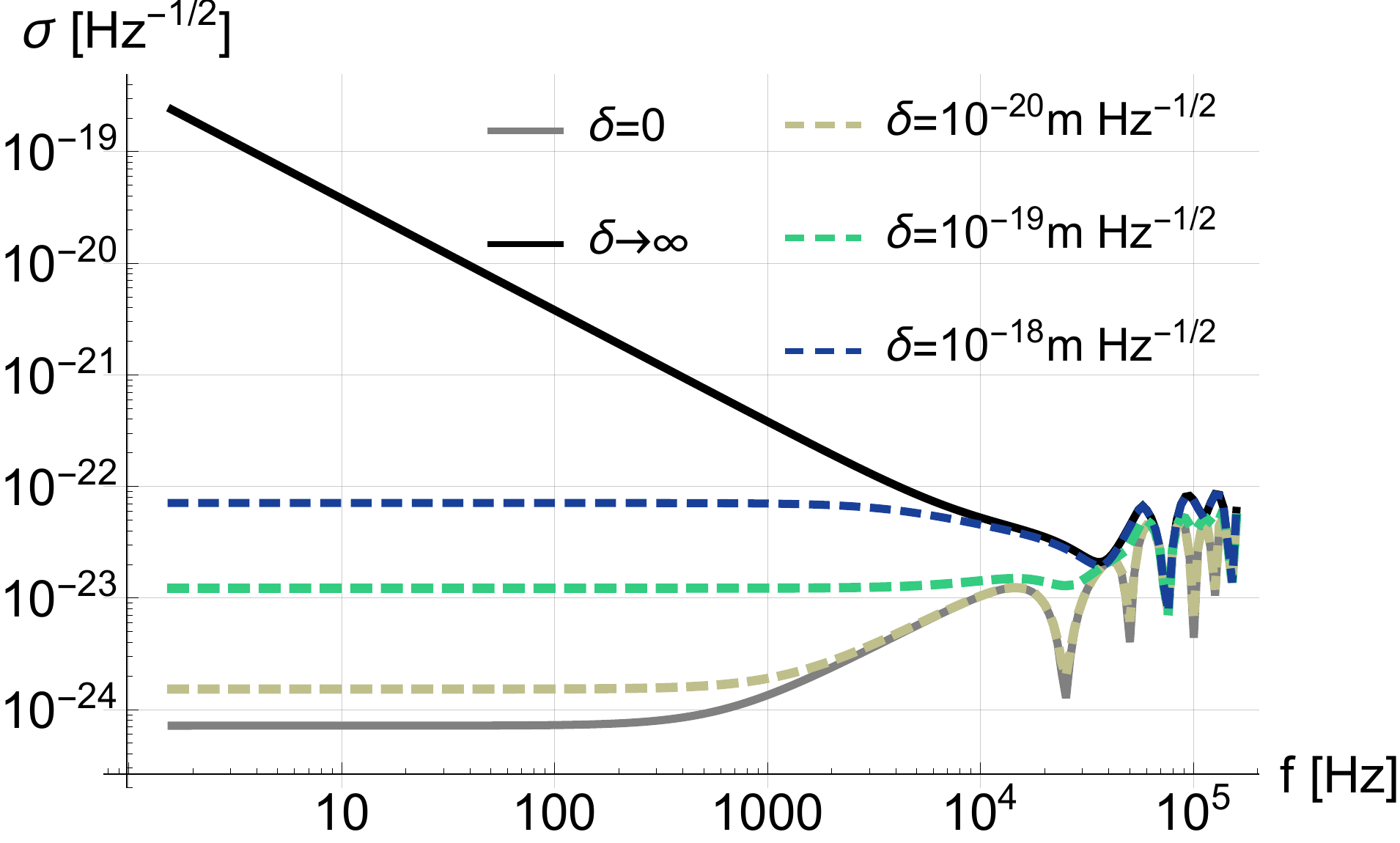}}
\subfigure[]{\includegraphics[width=0.34\textwidth]{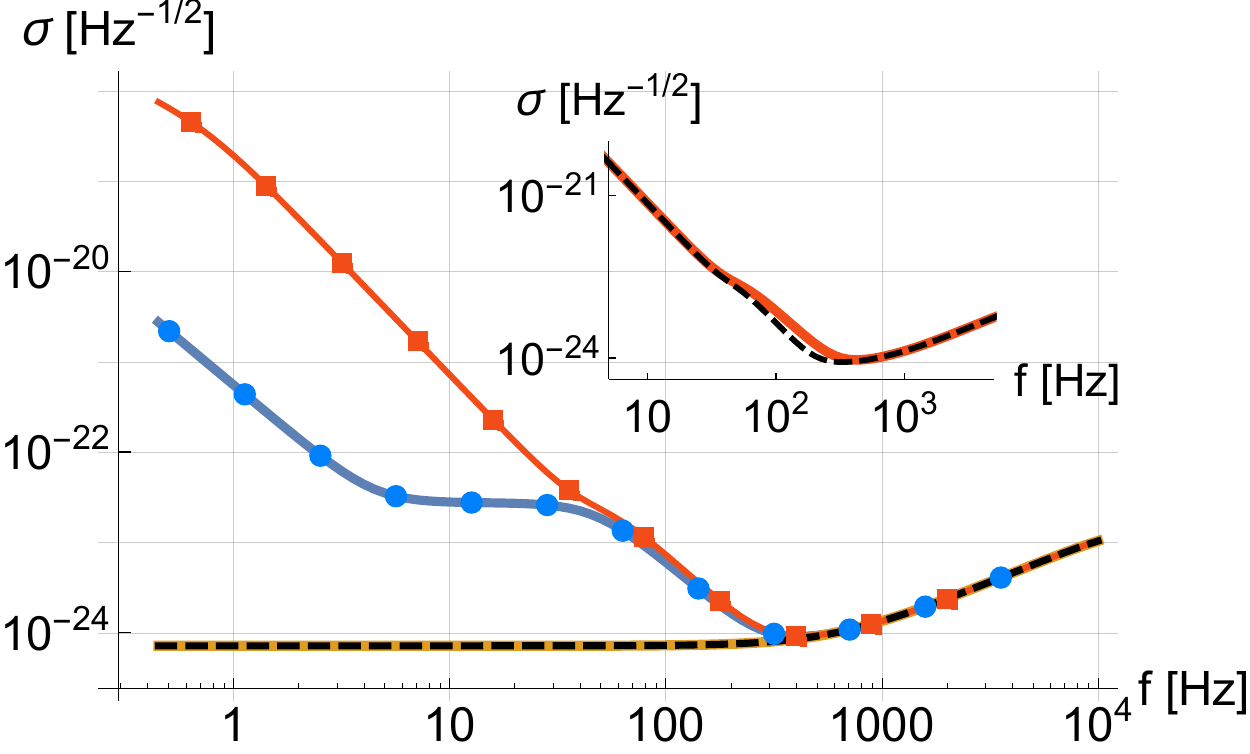}}
\end{center}
\caption{
(a) precision limits with the simplified model. The \new{SD}, $\sigma$, as a function of frequency for different levels of displacement noise ($\delta$) (\cref{eq::simplified_QFI}).
The DFI property is manifested in the fact that  $\sigma$ is finite in the limit of infinite displacement noise (solid black curve).
(b) Precision limits with realistic noise profiles: Given RPN alone, by measuring the optimal quadratures (\cref{optimal_quad_rad_pres}) the QFI (black dashed line) coincides with the shot noise limit (solid yellow line). 
\new{On the other hand,} measuring the (non-optimal) phase quadratures, yields the solid blue (circles) line (\cref{eq:rad_pressure_phase_quad}). 
Similarly, given both RPN and thermal noise, measuring the phase quadratures yields the solid red (rectangles) line. Inset: Comparison between the phase quadratures FI (solid red line) and the QFI (black dashed line) in the presence of thermal noise and RPN. }
\label{fig:sensitivity2}
\end{figure}

The sensitivity for different levels of $\delta,$ ranging from the shot noise limit ($\delta=0$) to infinite displacement noise ($\delta \rightarrow \infty$), is presented in \cref{fig:sensitivity2} (a). 
The DFI property is manifested in the fact
that as $\delta \rightarrow \infty$ the standard deviation remains finite, denoted by the black line in \cref{fig:sensitivity2} (a). We thus have finite noise in 
GW detection 
even in the presence of infinite displacement noise.

To understand the behavior of the sensitivity,
we note that the QFI can be decomposed as: 
\begin{align}
\begin{split}
&I=F_{\text{C}}+F_{\text{DFS}}\\
&=4\vplus^{\dagger}\Pi_{\text{C}}\left(\mathbb{1}+\delta^{2}A A ^{\dagger}\right)^{-1}\Pi_{\text{C}}\vplus+4\vplus^{\dagger}\Pi_{\text{DFS}}\vplus.
\label{eq:QFIM_decomposition}
\end{split}
\end{align}

The first term ($F_{\text{C}}$) is the information from the coupled subspace and the second term ($F_{\text{DFS}}$) is the information from the DFS.

In the infinite displacement noise limit ($\delta \rightarrow \infty$), the first term, $F_{C}$, vanishes and thus the QFI in this limit is:
\begin{align*}
I_{\delta \rightarrow \infty}=4\vplus^{\dagger} \Pi_{\text{DFS}}\vplus;    
\end{align*} 
i.e. we get information only from the DFS. 
As $f \rightarrow 0$ this standard deviation diverges, indicating that in this regime $\Pi_{\text{DFS}} \vplus\rightarrow 0.$
For finite $\delta$ (dashed lines in \cref{fig:sensitivity2} (a)), 
the QFI converges to $I \approx (4/\delta^{2})\vplus^{\dagger}\left(A A^{\dagger}\right)^{-1}\vplus$ at low frequencies, and thus $\sigma$ grows as $\delta$ in this limit. 

Furthermore, using \cref{eq:QFIM_decomposition} we can quantify the effectiveness of the DFI with the following coefficient: $\eta=\frac{F_{\text{DFS}}}{F_{\text{DFS}}+F_{C}},$ i.e. the fraction of the information that comes from the DFS. It will be shown that $\eta$ has an operational meaning as the gain from squeezing in the limit of large displacement noise.

\begin{figure*}[t]
\begin{center}
\subfigure[]{\includegraphics[width=0.33\textwidth]{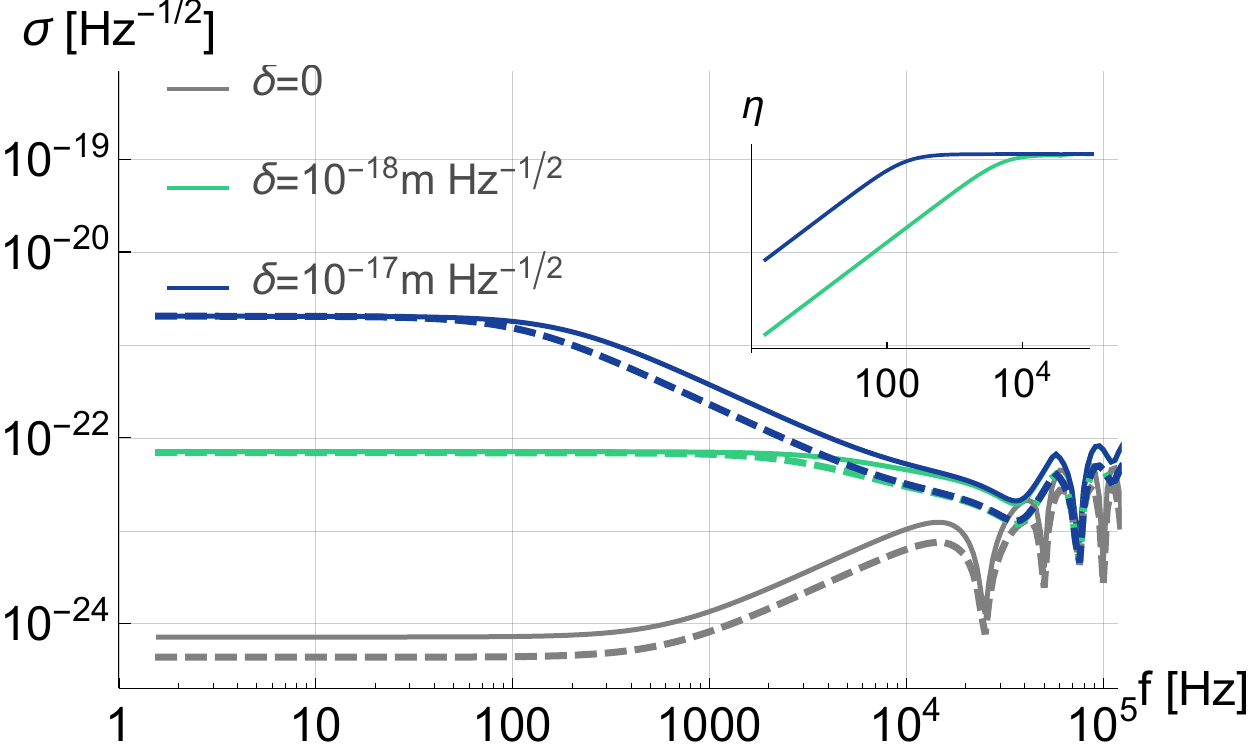}}
\subfigure[]{\includegraphics[width=0.33\textwidth]{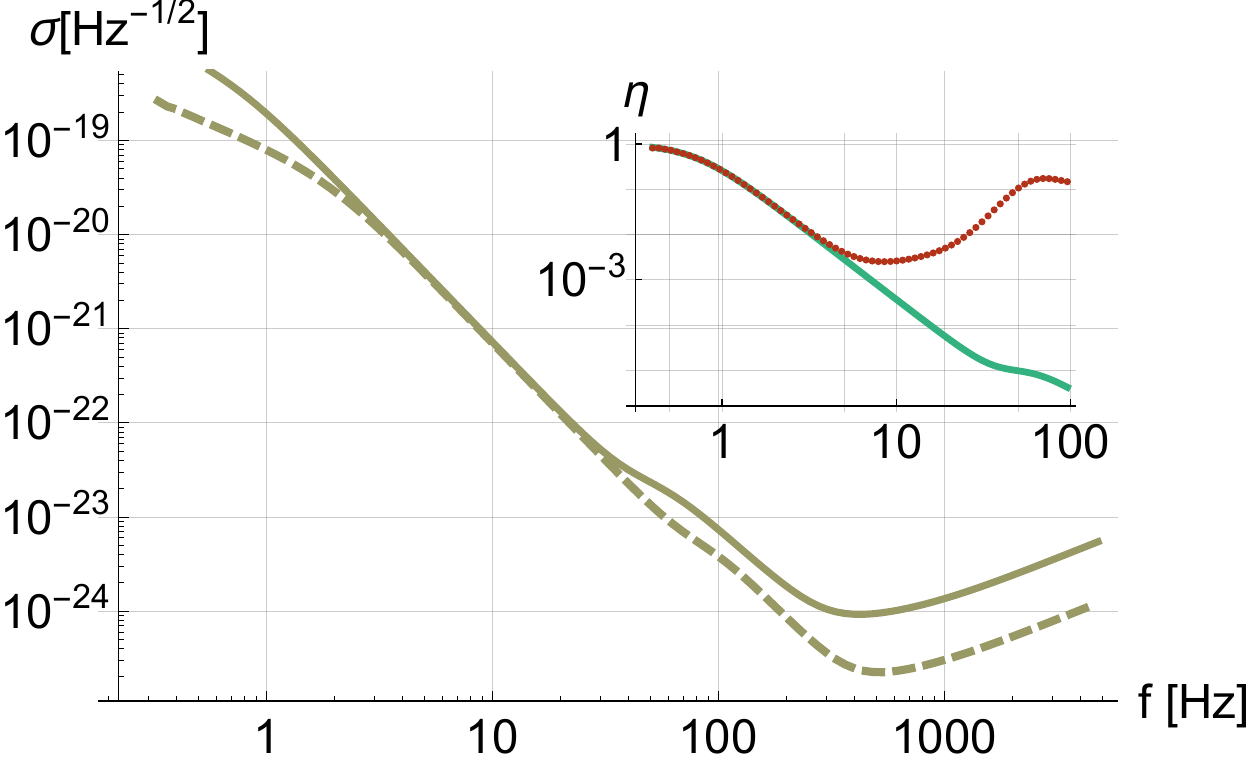}}
\subfigure[]{\includegraphics[width=0.32\textwidth]{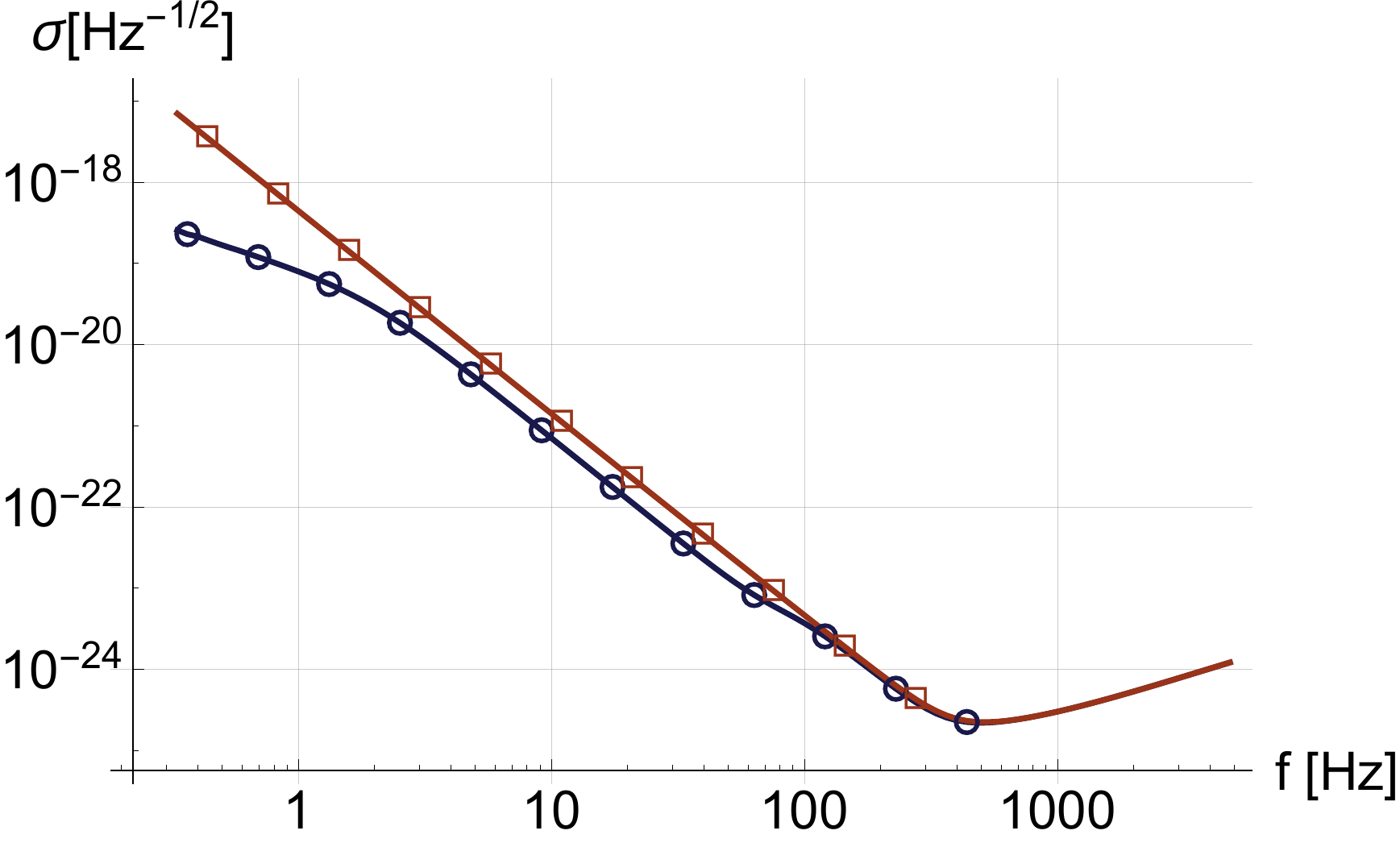}}

\end{center}
\caption{Effect of squeezing: (a) Performance with squeezing in the simplified model. Dashed lines correspond to SD ($\sigma$) with squeezing and solid lines to unsqueezed 
. For white displacement noise, squeezing becomes not effective at lower frequencies as can be also observed from the plot of $\eta$ in the inset. 
(b) Performance with squeezing given thermal noise and RPN. The solid (dashed) line corresponds to unsqueezed (optimally squeezed) SD with phase quadratures measurement. Inset: $\eta_{\text{gain}}$ (red dots) and $\eta$ (green line) as a function of frequency. 
(c) The SD with optimal squeezed input and optimal measurement (blue curve, circles) compared to the SD with the same squeezed input but a readout combination that maximizes the signal (red curve, squares).
}
\label{fig:squeezing}
\end{figure*}

\textit{\textbf{Precision limits with realistic noise profiles---}} Let us now consider the sensitivity with realistic thermal noise and RPN.
We begin by analyzing the effect of RPN alone and then study the combination of the two noises.

The effect of RPN 
is given by a non-unitary $M,$ \new{i.e. non-zero $M_{21}$ matrix (eq. \cref{eq:transfer_matrices}).}  
We assume that the mirrors are free masses, hence $M_{21} \propto \frac{1}{m \Omega^{2}},$ where $m$ is the mass of the mirrors. This typically leads to a sensitivity profile that scales as $\Omega^{-2}$ \cite{Branford18, Miao17}.

The QFI, in this case, saturates the shot noise limit (black dashed line in \cref{fig:sensitivity2} (b)), i.e. RPN is completely removed by measuring an appropriate choice of quadratures. This is a generalization of the optimal frequency-dependent readout introduced in ref. \cite{Vyatchanin96, Kimble01}. \new{Specifically,} the \new{$k$} quadratures \new{given by the column vectors of the matrix}:
\begin{align}
\new{T_{\text{dec}}=}\left(\begin{array}{c}
-M_{\text{int}}M_{21}^{\dagger}\\
\mathbbm{1}
\end{array}\right)\label{optimal_quad_rad_pres}
\end{align}
are decoupled from RPN and homodyne measurement of \new{the corresponding $k$ operators, $T_{\text{dec}}^{\dagger}\mathbf{\hat{Q}}_{\text{out}}$,} saturates the QFI and the shot noise limit.

\new{Measuring these optimal quadratures is experimentally challenging, the standard and simple readout quadratures are the phase quadratures}. Phase quadratures however are not decoupled from RPN and measuring them yields the following FI \cite{supp}: 
\begin{align}
F=4\vplusph^{\dagger}\left(\mathbb{1}+M_{21}M_{21}^{\dagger}\right)^{-1}\vplusph,
\label{eq:rad_pressure_phase_quad}
\end{align}

This expression is analogous to the QFI of the simplified model (\cref{eq::simplified_QFI}), where the term $M_{21} M_{21}^{\dagger}$ is the displacement noise caused by RPN. \new{It can be shown that $M_{21}=A_{\text{ph}} D_{x},$ where $D_{x}$ is the transfer matrix of the amplitude quadratures to the displacement of the mirrors \cite{supp}.} 
\new{The DFS} is \new{therefore} decoupled from this noise.
The corresponding sensitivity is presented in the solid blue line (circles) of \cref{fig:sensitivity2} \new{(b)}, where we observe an interesting behavior: unlike the \new{conventional sensitivity curves}, it does not diverge uniformly as $1/\Omega^{2}$ \cite{danilishin2012quantum}, instead there is a range of frequencies where the divergence \new{stops}. This plateau is due to a pseudo-DFS, a subspace that is \new{impervious} to displacement noise in this range of frequencies. Let us further elaborate on this.

In our triangular cavity scheme the phase quadratures can be decomposed to three orthogonal eigenspaces of the covariance matrix: $M_{\text{min}}\oplus M_{\text{max}}\oplus M_{\text{DFS}},$ where
$M_{\text{min}}\oplus M_{\text{max}}$ is a decomposition of $M_{\text{C}}$ to eigenspaces 
with minimal and maximal eigenvalues respectively. Since these are eigenspaces of the covariance matrix, the FI is a sum of the FI's achieved with each one of them separately, \new{i.e.:} $F=F_{\text{min}}+F_{\text{max}}+F_{\text{DFS}}.$ For different frequencies, different subspaces are dominant, this accounts for the non-uniform divergence. The plateau appears when $F_{\text{min}}$ becomes dominant. $M_{\text{min}}$ is immune to displacement noise in this range of frequencies, i.e. it is an eigenspace of $M_{21} M_{21}^{\dagger}$ with an eigenvalue that is much smaller than shot noise, hence the plateau. This is discussed further in the supplemental.

Let us now consider thermal noise as well. The thermal noise is modeled as $\mathbf{\Delta x} \sim N \left(0,\frac{1}{2} \delta ^{2} \mathbb{1}  \right),$ 
where
$\delta^{2}(f)=2.7\cdot10^{-30}(1/f)^5 ~\text{meter}^{2}/\text{Hz}$
\cite{Saulson90}. Hence, the effect of thermal displacement noise is similar to the simplified model with a frequency-dependent $\delta$.

In the presence of both RPN and thermal displacement noise,
the optimal measurement quadratures are the quadratures of \cref{optimal_quad_rad_pres}, \new{ which are decoupled from RPN.} \new{Hence RPN is completely canceled and we are left only with the thermal noise.} The QFI thus takes the form of \cref{eq::simplified_QFI} with \new{a frequency-dependent} $\delta$. A plot of the corresponding sensitivity is presented in the inset of \cref{fig:sensitivity2}.

Measuring the phase quadratures, RPN is not canceled and the FI reads: $4\vplusph^{\dagger}\left(\mathbbm{1}+M_{21}M_{21}^{\dagger}+\delta^{2}A_{\text{ph}}A_{\text{ph}}^{\dagger}\right)^{-1}\vplusph$. The plot of the corresponding sensitivity profile  \new{(red solid line)} and a comparison with the QFI \new{(black dashed line)} is presented in the inset of \cref{fig:sensitivity2} \new{(b)}. Three different regimes can be observed in the plot, that correspond to three eigenspaces of the covariance matrix. For \new{low} enough frequencies, the DFS becomes dominant and the SD diverges as $f^{-2},$ instead of $f^{-5/2}.$ Before that, there is an intermediate regime where $M_{\text{min}}$ is dominant and a short plateau \new{exists}. The comparison \new{in \cref{fig:sensitivity2} (inset)} between the phase quadratures FI and the QFI shows that they coincide \new{at low} frequencies where the thermal noise is dominant but the QFI clearly outperforms \new{the phase quadratures FI at intermediate frequencies} where RPN is dominant.

\textit{\textbf{Effect of squeezing--- }}
We summarize the optimal schemes and sensitivities with squeezing. Given a squeezing factor of $e^{-r}$ the optimal QFI is: $4\vplusph^{\dagger}\left(e^{-2r}\mathbbm{1}+\delta^{2}A_{\text{ph}}A_{\text{ph}}^{\dagger}\right)^{-1}\vplusph,$ it can be achieved with squeezing of the phase quadratures and measuring the optimal quadratures of eq. \ref{optimal_quad_rad_pres}.
For phase quadratures measurement, the optimal FI is: $4\vplusph^{\dagger}\left(e^{-2r}\left(\mathbbm{1}+M_{21}M_{21}^{\dagger}\right)+\delta^{2}A_{\text{ph}}A_{\text{ph}}^{\dagger}\right)^{-1}\vplusph,$
achievable by squeezing the optimal quadratures. \new{These optimal squeezing quadratures and sensitivity bounds
are derived in the supplemental.}

The performance of the squeezed schemes, and comparison with the unsqueezed case, is shown in \cref{fig:squeezing}.
Observe that the gain from squeezing is not uniform and depends on the effectiveness of the DFI, i.e. on $\eta.$ We can define the gain from squeezing as ${\eta_{\text{gain}}=\frac{F_{\text{sq}}/F-1}{e^{2r}-1}},$ where $F_{\text{sq}} (F)$ is the FI with(out) squeezing. Clearly $0 \leq \eta_{\text{gain}} \leq 1,$ where $0$ corresponds to no gain and $1$ to maximal gain. We show in the supplemental that in the limit of large displacement noise $\eta_{\text{gain}}=\eta,$ hence $\eta_{\text{gain}}$ equals the fraction of information coming from the DFS. This is illustrated in the insets of \cref{fig:squeezing}. DFI is therefore necessary to gain from squeezing in the presence of large displacement noise. The improvement introduced by DFI is summarized in \cref{fig:squeezing} (c) where we compare the sensitivity with squeezed input given different readout combinations: a combination that maximizes the signal and the optimal combination that saturates QFI. The \new{sensitivity with optimal combination} considerably \new{outperforms the sensitivity with maximal-signal combination at low} frequencies due to two DFI properties: better scaling with $f$ ($f^{-2}$ compared to $f^{-2.5}$), and larger gain from squeezing.


\textit{\textbf{Extensions and conclusions---}} 
\new{The supplemental contains extensions of this triangular scheme to $n-$gons with $n$ mirrors. Such polygon schemes may lead to further sensitivity improvement.
The supplemental contains also an analysis of the Sagnac noise, i.e. a phase shift due to rotation. We 
show that the resulting sensitivity loss is small.}

To conclude, we developed new DFI schemes and derived general quantum precision limits, optimal measurements, and optimal squeezing quadratures.

\new{There are still several challenges and open questions. The main challenge is to incorporate suppression
of laser noise in this architecture.
The laser noise must be correlated between the different ports
and the challenge is to engineer such correlation.
Other challenges include further optimization over
the architecture and considering also detuning.}

\textit{\textbf{Acknowledgments---}} The authors are thankful to J. Preskill, 
Y. Drori, E. D. Hall, K. Kuns and L. McCuller for helpful discussions. T.G. acknowledges funding provided by the Institute for
Quantum Information and Matter and the Quantum Science and Technology Scholarship of the Israel Council for Higher Education. Y.C. acknowledges the support by the Simons Foundation (Award Number 568762). R.X.A is supported by NSF Grants No. PHY-1764464 and PHY-191267.

\bibliographystyle{unsrtnat}
\bibliography{bibliography}

\begin{widetext}

\newpage

\appendix

\setcounter{equation}{0}
\renewcommand{\theequation}{\thesection.\arabic{equation}}

\setcounter{figure}{0}
\renewcommand{\thefigure}{\arabic{figure}}

\renewcommand{\thesection}{\arabic{section}}

\begin{center}
{\bf {Supplementary Material}}
\end{center}
\section{Two photon formalism}
Assuming an initial phase of $0,$ an EM field can be written as \cite{Caves85}:
\begin{align*}
\hat{E}\left(t\right)=\left(A\left(t\right)+\hat{a}_{1}\left(t\right)\right)\cos\left(\omega_{0}t\right)+\hat{a}_{2}\left(t\right)\sin\left(\omega_{0}t\right)
\end{align*}
where $\hat{a}_{1,2}\left(t\right)$ are the Hermitian amplitude and phase quadrature operators respectively. They describe the amplitude and phase modulation of the field ($\omega_{0}$ is the carrier frequency). They satisfy $\langle \hat{a}_{1,2}\left(t\right)\rangle=0$ and their commutation relations are given by:
\begin{align*}
    \left[\hat{a}_{1}\left(t\right),\hat{a}_{2}\left(t'\right)\right]=-i\delta\left(t-t'\right)
\end{align*} 

We can further define $\hat{a}_{1,2}\left(\Omega\right)$ as the Fourier transform of $\hat{a}_{1,2}\left(t\right)$:
\begin{align*}
\hat{a}_{1,2}\left(\Omega\right)=\frac{1}{\sqrt{2\pi}}\int \hat{a}_{1,2}\left(t\right)e^{-i\Omega t}\;\text{dt}.
\end{align*}

Here, we observe that $\hat{a}_{1,2}\left(\Omega\right)$ are not Hermitian but they commute with the Hermitian conjugate of themselves. This means that they have an orthonormal eigenbasis but their eigenvalues are complex. Therefore, the commutation relations between them is given by:
\begin{align*}
\left[\left(\begin{array}{c}
\hat{a}_{1}\\
\hat{a}_{2}
\end{array}\right),\left(\hat{a}_{1}^{\dagger},\hat{a}_{2}^{\dagger}\right)\right]:=\left(\begin{array}{cc}
[\hat{a}_{1},\hat{a}_{1}^{\dagger}] & [\hat{a}_{1},\hat{a}_{2}^{\dagger}]\\{}
[\hat{a}_{2},\hat{a}_{1}^{\dagger}] & [\hat{a}_{2},\hat{a}_{2}^{\dagger}]
\end{array}\right)=i\left(\begin{array}{cc}
0 & 1\\
-1 & 0
\end{array}\right).
\end{align*}
where the $\Omega$ dependence is suppressed,
and we used the following notation for the commutation relations matrix: $\left[\mathbf{\hat{Q}},\mathbf{\hat{Q}}^{\dagger}\right]:=\left(\left[\hat{Q}_{i},\hat{Q}_{j}\right]\right)_{i,j}$. Hereafter we will use this notation. 

Therefore, we can interpret the above as two harmonic oscillators with:
\begin{align*}
&\hat{X}_{R}\left(\Omega\right)=\sqrt{2}\text{Re\ensuremath{\left(\hat{a}_{1}\right)},~~~\;\ensuremath{\hat{P}_{R}\left(\Omega\right)}=\ensuremath{\sqrt{2}\text{Re}\left(\hat{a}_{2}\right)}}\\
&\hat{X}_{I}\left(\Omega\right)~ =\sqrt{2}\text{Im\ensuremath{\left(\hat{a}_{1}\right)},~~~\;\ensuremath{\hat{P}_{I}\left(\Omega\right)}=\ensuremath{\sqrt{2}\text{Im}\left(\hat{a}_{2}\right)}}
\end{align*}

It follows that the commutation relations of two harmonic oscillators are given by:
\begin{align*}
\left[\left(\begin{array}{c}
\hat{X}_{R}\\
\hat{P}_{R}\\
\hat{X}_{I}\\
\hat{P}_{I}
\end{array}\right),\left(\begin{array}{cccc}
\hat{X}_{R},  \hat{P}_{R},  \hat{X}_{I},  \hat{P}_{I}\end{array}\right)\right]=\left(\begin{array}{cc}
\sigma_{y} & 0\\
0 & \sigma_{y}
\end{array}\right),
\end{align*}
where $\sigma_{x/y}$ are the Pauli $X/Y$ matrices and the $\Omega$ dependence is suppressed.

In a multichannel interferometer, we have a vector of input quadratures $\mathbf{\hat{Q}}=\left(\begin{array}{c} \mathbf{\hat{a}}_{1}, \mathbf{\hat{a}}_{2} \end{array}\right)^T,$ which defines a vector of Hermitian quadratures (position and momentum operators) given by $\mathbf{\hat{S}}=\left(\begin{array}{c} \mathbf{\hat{X}}_{R}, \mathbf{\hat{P}}_{R}, \mathbf{\hat{X}}_{I}, \mathbf{\hat{P}}_{I} \end{array}\right)^T.$

\section{Quantum Fisher information matrix}

 This section derives equation (3) in the main text and generalizes it to the case of multi-parameter estimation of $\mathbf{h}.$ This requires using the Quantum Fisher Information Matrix (QFIM) which lower bounds the covariance matrix of the estimators of $\mathbf{h}$: $\text{COV}\left(\mathbf{h}\right)\geq\mathcal{I}^{-1}.$  

In a multichannel interferometer, the output sideband fields are given by
\begin{align*}
\mathbf{\hat{Q}}_{\text{out}}=\left(\begin{array}{c} \mathbf{\hat{b}}_{1}\left(\Omega\right)\\ \mathbf{\hat{b}}_{2}\left(\Omega\right)
\end{array}\right)    
\end{align*}
with commutation relations:
\begin{align}
 \left[\left(\begin{array}{c}
\mathbf{\hat{b}}_{1}\\
\mathbf{\hat{b}}_{2}
\end{array}\right),\left(\mathbf{\hat{b}}_{1}^{\dagger},\mathbf{\hat{b}}_{2}^{\dagger}\right)\right]=i\left(\begin{array}{cc}
0 & \mathbbm{1}_{k}\\
-\mathbbm{1}_{k} & 0
\end{array}\right)
\label{eq:commutation_b}
\end{align}
where $k$ is the number of output fields and the $\Omega$ dependence is suppressed.

As above, the Hermitian quadratures are: 
\begin{align*}
\mathbf{\hat{S}}=\left(\begin{array}{c}
\mathbf{\hat{X}}_{R}\\
\mathbf{\hat{P}}_{R}\\
\mathbf{\hat{X}}_{I}\\
\mathbf{\hat{P}}_{I}
\end{array}\right)=\sqrt{2}\left(\begin{array}{c}
\text{Re}\left(\mathbf{\hat{b}}_{1}\right)\\
\text{Re}\left(\mathbf{\hat{b}}_{2}\right)\\
\text{Im}\left(\mathbf{\hat{b}}_{1}\right)\\
\text{Im}\left(\mathbf{\hat{b}}_{2}\right)
\end{array}\right),
\end{align*}
 which is the corresponding quadratures vector of $2k$ harmonic oscillators.

Given the input-output relations of $\mathbf{\hat{Q}}$ (eq. (1) in the main text), these relations for $\mathbf{\hat{S}}$ are given by:
\begin{align}
\mathbf{\hat{S}}_{\text{out}}=M'\mathbf{\hat{S}}_{\text{in}}+\mathcal{V}'\mathbf{h}'+A'\mathbf{\Delta x'},    
\end{align}
where for all complex vectors $u\;(=\mathbf{h},\mathbf{\Delta x})$ we have $u'=\sqrt{2}\left(\begin{array}{c}
\text{Re}\left(u\right)\\
\text{Im}\left(u\right)
\end{array}\right)$ and for all complex matrices $\mathcal{M}\;(=M,\mathcal{V},A)$ we have $\mathcal{M}'=\left(\begin{array}{cc}
\text{Re}\left(\mathcal{M}\right) & -\text{Im}\left(\mathcal{M}\right)\\
\text{Im}\left(\mathcal{M}\right) & \text{Re}\left(\mathcal{M}\right)
\end{array}\right).$
 $\mathbf{h},\mathbf{\Delta x},M,\mathcal{V} \text{ and }A$ are as defined in the main text. Note that we expanded the complex-valued vector of parameters, $\mathbf{h}$, to the real-valued vector $\mathbf{h'},$ which consists of $4$ parameters: $\text{Re}\left(h_{+}\right),\text{Im}\left(h_{+}\right),\text{Re}\left(h_{\times}\right),\text{Im}\left(h_{\times}\right).$

Since we consider an initial Gaussian state and the evolution is through a Gaussian channel, the final state of the output modes is also Gaussian and can be characterized by:
\begin{equation*}
\begin{aligned}
&\mathbf{d_{s}}\left(\mathbf{h'}\right)=\langle\mathbf{\hat{S}}\rangle\\
&\Sigma_{i,j}=\frac{1}{2}\langle \hat{S}_{i}\hat{S}_{j}+\hat{S}_{j}\hat{S}_{i}\rangle-\langle \hat{S}_{i}\rangle\langle \hat{S}_{j}\rangle
\end{aligned}
\end{equation*}
where all the information about $\mathbf{h'}$ is encoded in the first moment vector $\mathbf{d_{s}}(\mathbf{h'})$ (mean vector) and $\Sigma$ is the covariance matrix.
 
The Quantum Fisher Information Matrix (QFIM) about $\mathbf{h'}$ can be expressed using these first two moments \cite{Nichols18,Branford18}:
\begin{align}
    \mathcal{I}_{\mathbf{h'}}=2 (\partial_\mathbf{h'}\mathbf{d_{s}})^T\Sigma^{-1}(\partial_\mathbf{h'}\mathbf{d_{s}})=2\mathcal{V}'^{\dagger}\Sigma^{-1}\mathcal{V}',
\label{eq:QFI_1}    
\end{align}
where we used the fact that the state is Gaussian, that all the information is encoded in $\mathbf{d_{s}},$ and that $\partial_{\mathbf{h'}}\mathbf{d_{s}}=\mathcal{V}'.$

In the following subsection, we will show that the QFIM can be expressed in a more compact form involving only the mean values of $\mathbf{\hat{Q}}$ ($\mathbf{d_{q}}$), and the covariance matrix of $\mathbf{\hat{Q}}$ ($\Sigma_{q}$).

\subsection{Complex compact form of the QFIM }
\label{sec:complex_compact_form}
In this subsection we show that  eq. 3 in the main text is a compact form of supplemental eq. \ref{eq:QFI_1}.

Following ref.\cite{stoica05} let us first introduce the notion of circular symmetry of real matrices. A real symmetric matrix $A$ has a circular symmetry if it takes the form of:
\begin{align*}
A=\left(\begin{array}{cc}
A' & -\bar{A}\\
\bar{A} & A'
\end{array}\right).    
\end{align*}
If $A$ satisfies this symmetry then we can define its complex-compact form:
$A_{c}=A'+i\bar{A}.$
This mapping between complex Hermitian matrices and real symmetric matrices with this symmetry $M\leftrightarrow\left(\begin{array}{cc}
\text{Re}\left(M\right) & -\text{Im}\left(M\right)\\
\text{Im}\left(M\right) & \text{Re}\left(M\right)
\end{array}\right)$ is a homomorphism, i.e. it preserves multiplication:
\begin{align*} A_{c}=B_{c}C_{c}\Longleftrightarrow\left(\begin{array}{cc} A' & -\bar{A}\\ \bar{A} & A' \end{array}\right)=\left(\begin{array}{cc} B' & -\bar{B}\\ \bar{B} & B' \end{array}\right)\left(\begin{array}{cc} C' & -\bar{C}\\ \bar{C} & C' \end{array}\right).
\end{align*}
As a result, identity is mapped to identity:
\begin{align}
\mathbbm{1}_{k}\longleftrightarrow\left(\begin{array}{cc}
\mathbbm{1}_{k} & 0\\
0 & \mathbbm{1}_{k}
\end{array}\right)=\mathbbm{1}_{2k}
\end{align}
and $\left(A^{-1}\right)_{c}=A_{c}^{-1}.$

Let us now prove the following claim:\\
If the covariance matrix of the estimators of (real-valued) $\mathbf{h'}$ has a circular symmetry, i.e. it takes the form of: 
\begin{align*}
\text{COV}\left(\mathbf{h'}\right)=\left(\begin{array}{cc}
C' & -\bar{C}\\
\bar{C} & C'
\end{array}\right),    
\end{align*}
then the covariance matrix of the estimators of (complex-valued) $\mathbf{h}$ is the complex-compact form of $\text{COV}\left(\mathbf{h}'\right)$:
\begin{align*}
\text{COV}\left(\mathbf{h}\right)=C'+i\bar{C}.    
\end{align*}

{\it{Proof:}} To show this we need to show that for any $h_{\phi}=\cos\left(\phi\right)h_{+}+\sin\left(\phi\right)h_{\times}$:
\begin{align}
\frac{1}{2}\text{var}\left(\text{Re}\left(h_{\phi}\right)\right)+\frac{1}{2}\text{var}\left(\text{Im}\left(h_{\phi}\right)\right)=u_{\phi}^{T}\left(C'+i\bar{C}\right)u_{\phi},
\label{eq:circular_cov_proof}
\end{align}
with $u_{\phi}=\left(\begin{array}{cc}
\cos\left(\phi\right) & \sin\left(\phi\right)\end{array}\right)^{T}.$ Given the circular symmetry of $\text{COV}\left(\mathbf{h'}\right)$ we observe that
\begin{align*}
\text{var}\left(\text{Re}\left(h_{\phi}\right)\right)=\text{var}\left(\text{Im}\left(h_{\phi}\right)\right)=u_{\phi}^{T}C'u_{\phi}.    
\end{align*}
Since $\bar{C}$ is anti-symmetric $u_{\phi}^{T}\left(\bar{C}\right)u_{\phi}=0,$ and thus eq. \ref{eq:circular_cov_proof} is satisfied. Note that we could omit $\bar{C}$, but we keep it for brevity of notation afterwards.

This immediately implies that if $\mathcal{I}_{\mathbf{h'}}$ satisfies a circular symmetry:
\begin{align*}
\mathcal{I}_{\mathbf{h'}}=\left(\begin{array}{cc}
\mathcal{I}' & -\bar{\mathcal{I}}\\
\bar{\mathcal{I}} & \mathcal{I}'
\end{array}\right),    
\end{align*}
then the Cram\'{e}r-Rao bound for $\text{COV}\left(\mathbf{h}\right)$ is given by the complex-compact form of $\mathcal{I}_{\mathbf{h'}}$: 
$\text{COV}\left(\mathbf{h}\right)\geq\left(\mathcal{I}'+i\bar{\mathcal{I}}\right)^{-1},$ i.e. the QFIM about $\mathbf{h}$ is given by:
\begin{align*}
\mathcal{I}=\mathcal{I}'+i\bar{\mathcal{I}}.    
\end{align*}

Observe that by definition $\mathcal{V}'$ satisfies this circular symmetry. Hence if $\Sigma$ satisfies it:
\begin{align}
 \Sigma=\left(\begin{array}{cc} \Sigma' & -\bar{\Sigma}\\ \bar{\Sigma} & \Sigma' \end{array}\right),
 \label{eq:symmetry}
\end{align}
then $\mathcal{I}_{\mathbf{h'}}$ (eq. \ref{eq:QFI_1}) also satisfies it. Therefore given that $\Sigma$ has a circular symmetry the QFIM about $\mathbf{h}$ reads:    
\begin{align*}
\mathcal{I}=2\mathcal{V}^{\dagger} \Sigma_{c}^{-1} \mathcal{V},    
\end{align*}
where $\Sigma_{c}$ is the complex compact form of $\Sigma$:
\begin{align*}
\Sigma_{c}=\Sigma'+i\bar{\Sigma}.    
\end{align*}


The covariance matrix of $\mathbf{\hat{Q}}$ is defined as:
\begin{align*}
\left(\Sigma_{q}\right)_{i,j}=\frac{1}{2}\left\langle \left\{ \hat{Q}_{i},\hat{Q}_{j}^{\dagger}\right\} \right\rangle -\langle\hat{Q}_{i}\rangle\langle\hat{Q}_{j}^{\dagger}\rangle,
\end{align*}
with $\left\{ \hat{Q}_{i},\hat{Q}_{j}^{\dagger}\right\}:=\hat{Q}_{i}\hat{Q}_{j}^{\dagger}+\hat{Q}_{j}^{\dagger}\hat{Q}_{i}$  being the anti-commutator of $\hat{Q},\hat{Q}_{j}^{\dagger}.$

Given that $\Sigma$ satisfies the circular symmetry then $\text{COV}\left(\text{Re}\left(\hat{Q}_{i}\right),\text{Re}\left(\hat{Q}_{j}\right)\right)=\text{COV}\left(\text{Im}\left(\hat{Q}_{i}\right),\text{Im}\left(\hat{Q}_{j}\right)\right),$ $\text{COV}\left(\text{Re}\left(\hat{Q}_{i}\right),\text{Im}\left(\hat{Q}_{j}\right)\right)=-\text{COV}\left(\text{Im}\left(\hat{Q}_{i}\right),\text{Re}\left(\hat{Q}_{j}\right)\right).$

Hence:
\begin{align*}
\left( \Sigma_{q} \right)_{i,j}=\text{Cov\ensuremath{\left(\hat{Q}_{i}\hat{Q}_{j}^{\dagger}\right)}}=2\left[\text{COV\ensuremath{\left(\text{Re}\left(\hat{Q}_{i}\right)\text{Re}\left(\hat{Q}_{j}\right)\right)}}+i\text{COV\ensuremath{\left(\text{Im}\left(\hat{Q}_{i}\right)\text{Re}\left(\hat{Q}_{j}\right)\right)}}\right]
	=\left(\Sigma_{c}\right)_{i,j}
\end{align*}

We can thus write $\mathcal{I}$ as:
\begin{align}
\mathcal{I}=2\mathcal{V}^{\dagger} \Sigma_{q}^{-1} \mathcal{V}
\label{eq:qfi_general_form}
\end{align}.

This symmetry of the covariance matrix is satisfied in our problem given that the displacement noise process is stationary (see section \ref{sec:QFI_displacement} for details).
The initial state is a coherent state, hence $\Sigma_{i}=\frac{1}{2}\mathbbm{1}$, and the symmetry for this state is satisfied. In the interferometer, it undergoes a Gaussian channel which maps the covariance matrix to:
\begin{align*}
\Sigma=\mathcal{R}\Sigma_{i}\mathcal{R}^{\dagger}+\Lambda,
\end{align*}
where $\mathcal{R}=\left(\begin{array}{cc}
\text{Re}\left(M\right) & -\text{Im}\left(M\right)\\
\text{Im}\left(M\right) & \text{Re}\left(M\right)
\end{array}\right)$ with $M$ being the transfer matrix. $\Lambda$ is due to classical displacement noise (thermal, seismic, etc.). Given that the classical displacement noise is stationary i.i.d $\Lambda$ takes the form of (see section \ref{sec:QFI_displacement}):  $\Lambda=\frac{\delta^{2}}{2}\left(\begin{array}{cc}
\text{Re}\left(AA^{\dagger}\right) & -\text{Im}\left(AA^{\dagger}\right)\\
\text{Im}\left(AA^{\dagger}\right) & \text{Re}\left(AA^{\dagger}\right)
\end{array}\right),$ with A being the transfer matrix of the displacement noise. Since $\mathcal{R},\Sigma_{0},\Lambda$ satisfy this symmetry, $\Sigma$ also satisfies this symmetry and we can thus use:
\begin{align*}
\Sigma_{q}=\frac{1}{2} MM^{\dagger}+\frac{\delta^{2}}{2} AA^{\dagger}.
\end{align*}
Inserting this into eq. \ref{eq:qfi_general_form} yields:
\begin{align*}
\mathcal{I}=4\mathcal{V}^{\dagger}\left(M M^{\dagger}+\delta^{2}A A^{\dagger}\right)^{-1}\mathcal{V}.    
\end{align*}

We observe (from numerics) that the eigenvector of $\mathcal{I}$ with maximal eigenavalue corresponds to $h_{+}$, hence this is the polarization with maximal sensitivity.
Focusing on this maximal sensitivity polarization reduces the problem to a single complex parameter estimation of $h_{+}$, and thus the quantity of interest is the QFI about $h_{+}$. The single parameter QFI ($I$) is a special case of the multi-parameter QFIM and thus reads:
\begin{align*}
I=2\vplus^{\dagger}\Sigma_{q}^{-1}\vplus= 4\vplus^{\dagger}\left(M M^{\dagger}+\delta^{2}A A^{\dagger}\right)^{-1}\vplus.
\end{align*}
These are the expressions in eqs. (3) and (5) in the main text. 

Hereafter, we will focus mainly on the single parameter estimation of $h_{+},$ and will thus use this QFI expression.

\section{Fisher information with homodyne measurement}
For $2k$ output quadratures $\mathbf{\hat{Q}}_{\text{out}}=\left(\begin{array}{c}
\mathbf{\hat{b}_{1}}\\
\mathbf{\hat{b}_{2}}
\end{array}\right),$  let us consider a homodyne measurement of these $l\leq k$ {\it{commuting}} quadratures:
$T_{h}^{\dagger}\mathbf{\hat{Q}}_{\text{out}},$
where $T_{h}$ is a $2k \times l$ matrix.
The outcomes of this measurement have a $l$-dimensional complex Gaussian distribution with a mean vector $\sqrt{2} T_{h}^{\dagger}\mathcal{V}$ and a covariance matrix $\sigma_{h}=T_{h}^{\dagger}\Sigma_{q}T_{h}.$

The Fisher information (FI) about $h_{+}$ is therefore  \cite{Branford18}:
\begin{align}
F=2\vplus^{\dagger}T_{h}\left(T_{h}^{\dagger}\Sigma_{q}T_{h}\right)^{-1}T_{h}^{\dagger}\vplus.
\label{eq:FI_homodyne}
\end{align}

The space of qudrature operators is a $2k$-dimensional linear space. For convenience, we can represent these operators as $2k$-dimensional column vectors. A single quadrature $\mathbf{u}^{\dagger} \mathbf{\hat{Q}}_{\text{out}}$ is represented by the (column) vector $\mathbf{u},$ and our $l$ quadratures ,$T_{h}^{\dagger} \mathbf{\hat{Q}}_{\text{out}}$, are represented by the $l$ column vectors of the matrix $T_{h}.$ 
We then denote the projection operator onto the $l$ measured quadratures as $\Pi_{h},$ observe that $\Pi_{h}=T_{h}T_{h}^{\dagger}.$


We now show that this FI can be decomposed into the sum of 
the FI's of different subspaces of $\Pi_{h}$. Let us decompose $\Pi_{h}$ into orthogonal subspaces $\Pi_{h}=\underset{i}{\sum}\Pi_{h_{i}},$ and denote the FI given measurement of $\Pi_{h_{i}}$ quadratures as $F_{i}: $ $F_{i}=2\vplus^{\dagger}T_{h_{i}}\left(T_{h_{i}}^{\dagger}\sigma_{h}T_{h_{i}}\right)^{-1}T_{h_{i}}^{\dagger}\vplus.$ Given that $\sigma_{h}$ is block diagonal in this decomposition, then the measurements of $\Pi_{h_{i}}$ are statistically independent and thus $F=\underset{i}{\sum} F_{i}.$ Formally:
\begin{align}
F\overset{*}{=}2\vplus^{\dagger}\left(\underset{i}{\sum}\Pi_{h_{i}}\sigma_{h}\Pi_{h_{i}}\right)^{-1}\vplus\overset{\#}{=}\underset{i}{\sum}2\vplus^{\dagger}T_{h_{i}}\left(T_{h_{i}}^{\dagger}\sigma_{h}T_{h_{i}}\right)^{-1}T_{h_{i}}^{\dagger}\vplus
=\underset{i}{\sum} F_{i}
\end{align}
where $\left(*\right)$ is due to the fact that it is block diagonal and $\left( \# \right)$ is basically:
\begin{equation*} 
\left(\begin{array}{cccc}
\bm{\mathcal{V}}^{\dagger}_{1} & \bm{\mathcal{V}}^{\dagger}_{2} & \cdots & \bm{\mathcal{V}}^{\dagger}_{j}\end{array}\right)
\setlength{\fboxsep}{0pt}
\left(\begin{array}{@{}c@{}c@{}c@{\mkern-5mu}c@{\,}cc*{2}{@{\;}c}@{}}%
\noalign{\vskip 1.5ex}
  \fbox{\,$\begin{matrix}
 \sigma_{1}
  \end{matrix}$\,}
 \\[-0.4pt]
  & \hskip-0.4pt\fbox{\,$\begin{matrix}
 \sigma_{2}
  \end{matrix}$\,}\\[-0.5ex]
   & & & \;\ddots \\[-0.5ex]
   & & & & \fbox{\,$\begin{matrix}
  \sigma_{j}
  \end{matrix}$\,}\\
\end{array}\right)^{-1}
\left(\begin{array}{c}
\bm{\mathcal{V}}_{1}\\
\bm{\mathcal{V}}_{2}\\
\vdots\\
\bm{\mathcal{V}}_{j}
\end{array}\right)=\underset{i}{\sum}\bm{\mathcal{V}}_{i}^{\dagger}\sigma_{i}^{-1}\bm{\mathcal{V}}_{i}.
\end{equation*}

 We can use this fact to analyze DFI schemes.
For example, for phase quadratures measurement, the displacement free subspace (DFS) is an eigenspace of the covariance matrix, and thus the covariance matrix is block diagonal in the decomposition to the coupled subspace and the DFS. We thus have that $F=F_{\text{C}}+F_{\text{DFS}},$ where $F_{\text{C}}$ is the information from the coupled subspace and $F_{\text{DFS}}$ is the information from the DFS. The quantity: 
\begin{align}
\eta=\frac{F_{\text{DFS}}}{F_{\text{DFS}}+F_{\text{C}}}    
\end{align}
is the fraction of the information that comes from the DFS and thus quantifies the effectiveness of the DFI. 

\section{Optimal measurement basis}
\label{section: optimal_measurement}
We prove here that the optimal quadrature to be measured is $\Sigma_{q}^{-1} \vplus,$ i.e. the operator $\left(\Sigma_{q}^{-1}\vplus\right)\cdot\mathbf{\hat{Q}}_{\text{out}}.$
We then extend this to the multi-parameter case, proving that measuring the two quadratures $\Sigma_{q}^{-1} \mathcal{V}$ saturates the QFIM.



Consider the single parameter estimation of $h_{+}$. the mean vector is $\mathbf{d}_{q}= \mathbf{\mathcal{V}} h$  and the covariance matrix is $\Sigma_{q}$ (we use here the complex compact form). Measuring the quadrature $\mathbf{u}$ of this Gaussian state yields
the following FI about $h_{+}$ (special case of eq. \ref{eq:FI_homodyne}):
\begin{align}
 F=2 \frac{|\mathbf{u}\cdot\mathbf{\vplus}|^{2}}{\mathbf{u}^{\dagger}\Sigma_{q}\mathbf{u}}.  
\end{align}

From the Cauchy-Schwarz inequality,
\begin{align}
|(\sqrt{\Sigma_{q}}\mathbf{u})\cdot(\sqrt{\Sigma_{q}}^{-1}\mathbf{\vplus})|^{2}\leq\left(\mathbf{u}^{\dagger}\Sigma_{q}\mathbf{u}\right)\left(\mathbf{\vplus}^{\dagger}\Sigma_{q}^{-1}\mathbf{\vplus}\right)~\Rightarrow~2\frac{|\mathbf{u}\cdot\mathbf{\vplus}|^{2}}{\left(\mathbf{u}^{\dagger}\Sigma_{q}\mathbf{u}\right)}\leq2\left(\mathbf{\vplus}^{\dagger}\Sigma_{q}^{-1}\mathbf{\vplus}\right).
\end{align}
where the right-hand side of the inequality is the expression for QFI and equality is obtained if and only if, 
\begin{align}
\sqrt{\Sigma_{q}}\mathbf{u}\propto\sqrt{\Sigma_{q}}^{-1}\mathbf{\vplus}\Rightarrow\mathbf{u}\propto\Sigma_{q}^{-1}\mathbf{\vplus}.    
\end{align}

Hence, the QFI is saturated given that the distributed quadrature $\Sigma_{q}^{-1}\vplus$ is measured. $\square$

In general, the QFI is saturated by measuring a set of quadratures, if and only if $\Sigma_{q}^{-1}\vplus$ is contained in the subspace spanned by them.

Regarding the multi-parameter estimation of $h_{+},h_{\times}$ (or any other polarizations), 
we show that the QFIM is saturated by measuring the two quadratures $\Sigma_{q}^{-1}\mathcal{V},$ 
and this is therefore the optimal measurement. 

{\it{Proof:}}
Observe that for any projection operator $\Pi$:
\begin{equation}
\begin{aligned}
\mathcal{V}^{\dagger}\sqrt{\Sigma_{q}}^{-1}\left(\mathbbm{1}-\Pi\right)\sqrt{\Sigma_{q}}^{-1}\mathcal{V}\geq0\\
\Rightarrow\mathcal{V}^{\dagger}\sqrt{\Sigma_{q}}^{-1}\Pi\sqrt{\Sigma_{q}}^{-1}\mathcal{V}\leq\mathcal{V}^{\dagger}\Sigma_{q}^{-1}\mathcal{V}
\label{eq:general_cauchy_schearz}
\end{aligned}
\end{equation}
with equality if and only if $\Pi\sqrt{\Sigma_{q}}^{-1}\mathcal{V}=\sqrt{\Sigma_{q}}^{-1}\mathcal{V}.$
Taking the following projection operator: $\Pi=\sqrt{\Sigma_{q}}T_{h}\left(T_{h}^{\dagger}\Sigma_{q}T_{h}\right)^{-1}T_{h}^{\dagger}\sqrt{\Sigma_{q}}$,
and inserting it into eq. \ref{eq:general_cauchy_schearz} we obtain that:
\begin{align*}
2\mathcal{V}^{\dagger}T_{h}\left(T_{h}^{\dagger}\Sigma_{q}T_{h}\right)^{-1}T_{h}^{\dagger}\mathcal{V}\leq 2 \mathcal{V}^{\dagger}\Sigma_{q}^{-1}\mathcal{V}.
\end{align*}
The left term is exactly the homodyne FI.
Note that our $\Pi$ is a projection operator
onto the span of the column vectors of $\sqrt{\Sigma_{q}}T_{h},$
denoted as $\mathcal{C}\left(\sqrt{\Sigma_{q}}T_{h}\right).$
Hence equality is obtained iff $\mathcal{C}\left(\sqrt{\Sigma_{q}}^{-1}\mathcal{V}\right)\subseteq\mathcal{C}\left(\sqrt{\Sigma_{q}}T_{h}\right)$ and thus the minimal space of quadratures that saturate the inequality is: $ T_{h}=\Sigma_{q}^{-1}\mathcal{V}\Lambda,$
where $\Lambda$ is a normalization and orthogonalization matrix. $\square$ 

This was also proven in ref. \cite{Gessner20}.

Since the multi-parameter case requires commutativity of the quadratures given by the column vectors of $\Sigma_{q}^{-1} \mathcal{V},$ we prove a useful claim -
if the quadratures given by the column vectors of $\mathcal{V}$ commute and $\Sigma_{q}$ is a conjugate symplectic matrix then the quadratures given by $\Sigma_{q}^{-1} \mathcal{V}$ commute and thus the QFIM is achievable.

Proof: Since conjugate symplectic matrices form a group then $\Sigma_{q}$ conjugate symplectic $\rightarrow \Sigma_{q}^{-1}$ conjugate symplectic , i.e. denoting $\mathcal{W}=\left(\begin{array}{cc}
0 & \mathbbm{1}_{k}\\
-\mathbbm{1}_{k} & 0
\end{array}\right)    
$: 
\begin{align*}
\left( \Sigma_{q}^{-1} \right) ^{\dagger} \mathcal{W} \Sigma_{q}^{-1} =\mathcal{W}.   
\end{align*}
Hence: 
\begin{align*}
&\mathcal{V} \; \text{commute} \leftrightarrow \mathcal{V}^{\dagger}\mathcal{W}\mathcal{V}=0 \\
& \Rightarrow\left(\Sigma_{q}^{-1}\mathcal{V}\right)^{\dagger}\mathcal{W}\left(\Sigma_{q}^{-1}\mathcal{V}\right)=\mathcal{V}^{\dagger}\mathcal{W}\mathcal{V}=0.
\end{align*}
Therefore $\Sigma_{q}^{-1} \mathcal{V}$ commute.$\square$

In our problem, the column vectors of $\mathcal{V}$ are in the phase quadratures, hence they commute. Therefore in order to show commutativity of the optimal quadratures it suffices to show that $\Sigma_{q}$ is symplectic.

\section{QFI with Thermal displacement noise}
\label{sec:QFI_displacement}

Displacement of optical components 
leads to a noisy displacement of the quadratures, i.e. in Heisenberg picture: $\mathbf{\hat{Q}}\rightarrow\mathbf{\hat{Q}}+A\mathbf{\Delta x}$,
where $\mathbf{\Delta x}$ is a multivariate Gaussian random variable.

The new state under the action of this noise is a Gaussian mixture of states
\begin{align}
 \rho=\int p\left(\mathbf{\Delta x}\right)\rho\left(\mathbf{\Delta x}\right)\;\text{d}\mathbf{\Delta x}   
\end{align}
and therefore is also Gaussian.

 Note that while $\mathbf{\Delta x} \left( t \right)$ is a real vector, $\mathbf{\Delta x} \left( \Omega \right)$ is complex. The transformation of the Hermitian quadratures vector $\mathbf{\hat{S}}$ is therefore given by,
\begin{equation}
 \mathbf{\hat{S}}\rightarrow\mathbf{\hat{S}}+A'\mathbf{\Delta x'}   
\end{equation}
where
\begin{equation}
 A'=\left(\begin{array}{cc}
\text{Re}\left(A\right) & \text{-Im}\left(A\right)\\
\text{Im}\left(A\right) & \text{Re}\left(A\right)
\end{array}\right),\;\Delta x'=\sqrt{2}\left(\begin{array}{c}
\text{Re}\left(\Delta x\right)\\
\text{Im}\left(\Delta x\right)
\end{array}\right)   
\end{equation}

Since $\langle\mathbf{\Delta x}\rangle=0$ , the vector of the first moments $\mathbf{d_{s}}$ is unchanged. The covariance matrix however changes to:
\begin{equation}
\begin{aligned}
\Sigma=&\Sigma_{i}+\langle(A'\mathbf{\Delta x'})(A'\mathbf{\Delta x'})^{\dagger}\rangle\\
=&\Sigma_{i}+A'\Sigma_{\mathbf{\Delta x'}}A'^{\dagger},
\label{eq:cov_disp_noise}
\end{aligned}
\end{equation}
where $\Sigma_i$ is the covariance matrix of the states in the absence of displacement noise and $\Sigma_{\Delta x'}=\langle\mathbf{\Delta x'}(\mathbf{\Delta x'})^{\dagger}\rangle$ is the covariance matrix of $\mathbf{\Delta x'}.$

We assume that the displacement noise $\left\{ \mathbf{\Delta x}\left(t\right)\right\} _{t}$ is a Gaussian stationary process, where the different $\Delta x_{i}\left(t\right),\Delta x_{j}\left(t\right)$ are i.i.d. Therefore $\mathbf{\Delta x}'\left(\Omega\right)$ is also a Gaussian random variable with a covariance matrix of:
\begin{align*}
&\langle \left( \text{Re}\left(\Delta x\right) \right)^{2} \rangle=2\underset{0}{\overset{T}{\int}} \; \underset{0}{\overset{T}{\int}}\langle\Delta x\left(t_{1}\right)\Delta x\left(t_{2}\right)\rangle\cos\left(\Omega t_{1}\right)\cos\left(\Omega t_{2}\right)\;dt_{1}dt_{2}\approx T\underset{0}{\overset{T}{\int}}C\left(\tau\right)\cos\left(\Omega\tau\right)\;d\tau \\
&\langle \left( \text{Im}\left(\Delta x\right) \right) ^{2}\rangle=2\underset{0}{\overset{T}{\int}} \; \underset{0}{\overset{T}{\int}}\langle\Delta x\left(t_{1}\right)\Delta x\left(t_{2}\right)\rangle\sin\left(\Omega t_{1}\right)\sin\left(\Omega t_{2}\right)\;dt_{1}dt_{2}\approx T\underset{0}{\overset{T}{\int}}C\left(\tau\right)\cos\left(\Omega\tau\right)\;d\tau\\
&\langle\text{Re}\left(\Delta x\right)\text{Im}\left(\Delta x\right)\rangle=2\underset{0}{\overset{T}{\int}} \; \underset{0} {\overset{T}{\int}}\langle\Delta x\left(t_{1}\right)\Delta x\left(t_{2}\right)\rangle\cos\left(\Omega t_{1}\right)\sin\left(\Omega t_{2}\right)\;dt_{1}dt_{2}\approx0,
\end{align*}
and of course all the correlations of $\Delta x_{i}', \Delta x_{j}' (i\neq j)$ vanish. Therefore $\Sigma_{\Delta x'}=\frac{\delta^{2}}{2}\mathbbm{1},$ and thus $\mathbf{\Delta x'}\sim N\left(0,\frac{\delta^{2}}{2}\mathbbm{1}\right).$ 

Since $\Sigma_{i}=\frac{1}{2} M' M'^{\dagger}$ we get that:
\begin{align}
\Sigma=\frac{1}{2} \left(M'M'^{\dagger}+\delta^{2}A'A'^{\dagger} \right).  \label{eq:full_cov_matrix}  
\end{align}


The QFIM therefore reads,
\begin{equation}
\begin{aligned}
I=&2\partial_{\mathbf{h}}\mathbf{d_{s}}^{\dagger}\left(M' M'^{\dagger}+\delta^{2}A'A'^{\dagger}\right)^{-1} \partial_{\mathbf{h}}\mathbf{d_{s}}\\	
=&2\mathcal{V}'^{\dagger}\left(M' M'^{\dagger}+\delta^{2}A'A'^{\dagger}\right)^{-1}\mathcal{V}'
\end{aligned}
\end{equation}

Alternatively, the form of the covariance matrix can be also directly derived from the Wigner function. Due to the displacement noise, we have:
\begin{align}
W\left[\rho\right]=\int p\left(\mathbf{\Delta x}\right)W\left[\rho\left(\mathbf{\Delta x}\right)\right]\;\text{d}\mathbf{\Delta x}.
\label{eq:averaging_wigner}
\end{align}

For $k$ output quadratures, the Wigner function per realization of $\mathbf{\Delta x}$ is:
\begin{align*}
W(\mathbf{\Delta x})=\frac{1}{\left(2\pi\right)^{2k}\sqrt{\text{Det}\left(\sigma\right)}}\exp\left[-\frac{1}{2}\left(\mathbf{S}-d_{\mathbf{h}}-d_{\mathbf{\Delta x}}\right)^{\dagger}\Sigma_{i}^{-1}\left(\mathbf{S}-d_{\mathbf{h}}-d_{\mathbf{\Delta x}}\right)\right],    
\end{align*}


Since $p\left(\mathbf{\Delta x}\right)$ is Gaussian, then the averaging (eq. \ref{eq:averaging_wigner}) is basically a convolution of two Gaussian distributions, 
\begin{align*}
 N(\mathbf{d_{s}},\Sigma_i) \circledast N\left(0,\text{Cov\ensuremath{\left(d_{\mathbf{\Delta x}}\right)}}\right)=N(\mathbf{d_{s}},\Sigma_i) \circledast N \left(0, \delta^{2} A' A'^{\dagger} \right) = N(\mathbf{d_{s}},\Sigma_{i}+\delta^{2}A' A'^{\dagger})
\end{align*}

Therefore,
\begin{align*}
W=\frac{1}{\left(2\pi\right)^{2n}\sqrt{\text{Det}\left(\Sigma\right)}}\exp\left[-\frac{1}{2}\left(\mathbf{S}-d_{\mathbf{h}}\right)^{\dagger}\Sigma^{-1}\left(\mathbf{S}-d_{\mathbf{h}}\right)\right],    
\end{align*}
with $\Sigma=\Sigma_{i}+\delta^{2}A' A'^{\dagger}.$

We can therefore observe that the full covariance matrix, eq. \ref{eq:full_cov_matrix}, satisfies the symmetry of eq. \ref{eq:symmetry}. We have shown in sec. \ref{sec:complex_compact_form} that if the covariance matrix satisfies the circular symmetry then the QFI can be written in the complex compact form, this justifies our use of this form:
\begin{align}
I=4 \vplus^{\dagger}\left(MM^{\dagger}+\delta^{2}AA^{\dagger}\right)\vplus.    
\end{align}

On a brief note - a DFS is defined as the kernel of the (general) noise term $A_{\text{ph}} \Sigma_{\Delta x} A_{\text{ph}}^{\dagger}.$ Note that $\text{ker} \left(A_{\text{ph}}^{\dagger} \right)$ is contained in this subspace. furthermore if $\Sigma_{\Delta x}$ is a full rank matrix, the DFS is equal to $\text{ker} \left(A_{\text{ph}}^{\dagger} \right).$

\section{FI and QFI with radiation pressure}
\label{sec:rad_pressure}

\subsection{Derivation of transfer matrix}
\label{subsection:radiation_pressure}
Let us first write how Radiation Pressure Noise (RPN) enters into the equations \cite{Corbitt05}.
Resonance conditions are assumed.  
Given $\bf{\hat{a}},\bf{\hat{d}}$ fields that hit a mirror, $\bf{\hat{a}}=\left(\begin{array}{c}
\hat{a}_{1}\\
\hat{a}_{2}
\end{array}\right),$
$\bf{\hat{d}}=\left(\begin{array}{c}
\hat{d}_{1}\\
\hat{d}_{2}
\end{array}\right)$,
the reflected fields $\bf{\hat{b}},\bf{\hat{c}}$ satisfy
: 
\begin{align}
\left(\begin{array}{c}
\bm{\bm{\hat{b}}}\\
\bm{\hat{c}}
\end{array}\right)=M_{\text{mirror}}\left(\begin{array}{c}
\bm{\hat{a}}\\
\bm{\hat{d}}
\end{array}\right)-2\left(\frac{\omega_{0}}{c}\right)\frac{\sqrt{R}}{\sqrt{\hbar\omega_{0}}}\Delta \hat{x}\left(\begin{array}{c}
\boldsymbol{D_{a}^{*}}\\
\boldsymbol{D_{d}^{*}}
\end{array}\right),
\label{eq:general_disp}
\end{align}
with 
$M_{\text{mirror}}=\left(\begin{array}{cccc}
-\sqrt{R} &  & \sqrt{T}\\
 & -\sqrt{R} &  & \sqrt{T}\\
\sqrt{T} &  & \sqrt{R}\\
 & \sqrt{T} &  & \sqrt{R}
\end{array}\right)$
being the mirror transformation,
$\Delta \hat{x}$ is the displacement due to RPN. 
Assuming resonance ($\omega_{0} L/c=2 \pi n$, where $n$ is an integer):
$\boldsymbol{D_{j}}=\sqrt{2p_{j}}\left(\begin{array}{c}
1\\
0
\end{array}\right),$
$\boldsymbol{D_{j}^{*}}=\sqrt{2p_{j}}\left(\begin{array}{c}
0\\
-1
\end{array}\right)$,
where $p_j$ is the power of the $j$-th carrier field.
Eq. \ref{eq:general_disp} is the general way displacement noise is being propagated. In RPN $\Delta \hat{x}$ is an operator, and it is given by : 
\begin{equation*}
\begin{aligned}
\Delta \hat{x}=\frac{1}{m\Omega^{2}}\sqrt{\frac{\hbar\omega_{0}}{c^{2}}}\left[\left(\begin{array}{cc}
\bm{D_{a}^{t}} & \bm{-D_{d}^{t}}\end{array}\right)\left(\begin{array}{c}
\bm{\hat{a}}\\
\bm{\hat{d}}
\end{array}\right)+\left(\begin{array}{cc}
\boldsymbol{D_{b}^{t}} & -\boldsymbol{D_{c}^{t}}\end{array}\right)\left(\begin{array}{c}
\boldsymbol{\hat{b}}\\
\boldsymbol{\hat{c}}
\end{array}\right)\right]\\
=\frac{1}{m\Omega^{2}}\sqrt{\frac{2\hbar\omega_{0}}{c^{2}}}\left[\sqrt{p_{a}}\hat{a}_{1}-\sqrt{p_{d}}\hat{d}_{1}+\sqrt{p_{b}}\hat{b}_{1}-\sqrt{p_{c}}\hat{c}_{1}\right].
\end{aligned}
\end{equation*} 

Inserting this $\Delta \hat{x}$ into eq. \ref{eq:general_disp} we get two coupled sets of equations. For the amplitude quadratures:
\[
\left(\begin{array}{c}
\hat{b}_{1}\\
\hat{c}_{1}
\end{array}\right)=M_{\text{mirror}}\left(\begin{array}{c}
\hat{a}_{1}\\
\hat{d}_{1}
\end{array}\right),
\]

and for the phase quadratures: 
\begin{align}
\left(\begin{array}{c}
\hat{b}_{2}\\
\hat{c}_{2}
\end{array}\right)=M_{\text{mirror}}\left(\begin{array}{c}
\hat{a}_{2}\\
\hat{d}_{2}
\end{array}\right)-\frac{2\sqrt{2}\omega_{0}\sqrt{R}}{m\Omega^{2}c^{2}}\left[\sqrt{p_{a}}\hat{a}_{1}-\sqrt{p_{d}}\hat{d}_{1}+\sqrt{p_{b}}\hat{b}_{1}-\sqrt{p_{c}}\hat{c}_{1}\right]\left(\begin{array}{c}
\sqrt{2p_{a}}\\
\sqrt{2p_{d}}
\end{array}\right).
\label{eq:rad_pres_basic_dependence}
\end{align}

This implies a general structure for the multichannel case: the equations for the amplitude quadratures are closed and their solution is given by
\begin{align*}
\mathbf{\hat{b}_{1}}=M_{\text{int}}\mathbf{\hat{a}_{1}},
\end{align*}
where $M_{\text{int}}$ is the unitary transfer matrix of the interferometer which does not depend on the RPN.
The equations for the amplitude quadratures are coupled to the phase quadratures and their solution is given by:
\begin{align*}
\mathbf{\hat{b}_{2}}=M_{\text{int}}\mathbf{\hat{a}_{2}}+M_{21}\mathbf{\hat{a}_{1}}.
\end{align*}
Hence amplitude noise is being propagated into phase noise with a transfer matrix $M_{21}$.
The input-output relations therefore read:
\begin{align}
\left(\begin{array}{c}
\mathbf{\hat{b}_{1}}\\
\mathbf{\hat{b}_{2}}
\end{array}\right)=\left(\begin{array}{cc}
M_{\text{int}} & 0\\
M_{21} & M_{\text{int}}
\end{array}\right)\left(\begin{array}{c}
\mathbf{\hat{a}_{1}}\\
\mathbf{\hat{a}_{2}}
\end{array}\right),
\label{eq:M_rad_pressure}
 \end{align}
 
Two observations regarding the transfer matrix of eq. \ref{eq:M_rad_pressure} will be useful later:

1. $M_{21}$ can be expressed as a concatenation of two transfer matrices: $M_{21}=A  D_x ,$ where $A$ is the transfer matrix of mirror displacement vector $\mathbf{\Delta \hat{x}},$ and $D_x$ is the transfer matrix of the amplitude noise to the displacement vector: $\mathbf{\Delta \hat{x}}=D_x \mathbf{\hat{a}_{1}}.$ This fact will be used to 
in section \ref{sec:thermal+rad}.

2. Commutation relations have to be preserved (see eq. \ref{eq:commutation_b}). This implies that $M$ is a conjugate symplectic matrix and thus $M_{\text{int}}^{\dagger}M_{21}$ is Hermitian.

3. From eq. \ref{eq:rad_pres_basic_dependence} we observe that $M_{21} \propto 1/m \Omega^{2}.$
 



\subsection{QFI}
Given the general form of the transfer matrix $M$ (eq. \ref{eq:M_rad_pressure}), we can calculate the general form of the QFI and FI.
The QFI
is given by $4 \vplus^{\dagger}\left(MM^{\dagger}\right)^{-1}\vplus$ where
$MM^\dagger$ $(=2\Sigma_{q})$ equals to:
\begin{align}
MM^{\dagger}=\left(\begin{array}{cc}
\mathbbm{1} & M_{\text{int}}M_{21}^{\dagger}\\
M_{21}M_{\text{int}}^{\dagger} & \mathbbm{1}+M_{\text{21}}M_{21}^{\dagger}
\end{array}\right).
\label{eq:cov_rad_pressure}
\end{align}

Since $M_{\text{int}}$ is unitary we can observe that:
\begin{align*}
&\left(MM^{\dagger}\right)^{-1}=\left(\begin{array}{cc}
* & -M_{\text{int}}M_{21}^{\dagger}\\
* & \mathbbm{1}
\end{array}\right)\\
&\overset{\#}{\Rightarrow} I=4 \vplus^{\dagger} \vplus=4 \vplusph^{\dagger} \vplusph,
\end{align*}
where $(\#)$ is because: $\mathcal{V}=\left(\begin{array}{c}
0\\
\mathcal{V}_{\text{ph}}
\end{array}\right)$. The QFI thus obtains
the shot-noise limit. 
This is a generalization of the single channel optimal frequency-dependent readout scheme  \cite{Kimble01}: by measuring certain quadratures we can overcome the RPN.
Using the results of section \ref{section: optimal_measurement}, we know that the optimal quadratures to be measured are (up to normalization): $\Sigma_{q}^{-1} \vplus.$ The optimal quadrature is therefore:
\begin{align}
\mathbf{u} \propto \left(\begin{array}{c}
-M_{\text{int}}M_{21}^{\dagger} \vplusph\\
\vplusph
\end{array}\right),
\label{eq:rad_pres_comb}
\end{align}
i.e. measuring the operator $\mathbf{u}\cdot\mathbf{\hat{Q}}_{\text{out}}$ is optimal.
This $\mathbf{u}$ is a linear combination of the $k$ column vectors of the matrix $T_{\text{dec}}$:
\begin{align}
T_{\text{dec}}=\left(\begin{array}{c}
-M_{\text{int}}M_{21}^{\dagger}\\
\mathbbm{1}
\end{array}\right).
\label{eq:optimal_quadratures}
\end{align}
These $k$ column vectors correspond to $k$ quadratures decoupled from RPN.
To see explicitly that these quadratures are decoupled from RPN, note that the covariance matrix can be written as:
\begin{align*}
 \Sigma_{q} \propto \left(\begin{array}{cc}
\mathbbm{1} & M_{\text{int}}M_{21}^{\dagger}\\
M_{21}M_{\text{int}}^{\dagger} & \mathbbm{1}+M_{21}M_{21}^{\dagger}
\end{array}\right)=\left(\begin{array}{cc}
0 & 0\\
0 & \mathbbm{1}
\end{array}\right)
+\left(\begin{array}{c}
\mathbbm{1}\\
M_{21 }M_{\text{int}}^{\dagger}
\end{array}\right)\left(\begin{array}{cc}
\mathbbm{1} & M_{\text{int}}M_{21}^{\dagger}\end{array}\right),
\end{align*}
the space spanned by these quadratures is thus decoupled from RPN. Hence any combination of these quadratures is decoupled from this noise and thus yields an FI that does not diverge in the $f \rightarrow 0$ limit, the optimal combination of eq. \ref{eq:rad_pres_comb} obtains the shot noise limit.

Two remarks are now in order:
\begin{itemize}
\item We remark that the quadratures of eq. \ref{eq:optimal_quadratures} are not orthonormal, the orthonormalized quadratures are the $k$ column vectors of the following matrix:
\begin{align*}
\left(\begin{array}{c}
-M_{\text{int}}M_{21}^{\dagger}\\
\mathbbm{1}
\end{array}\right)\left(\mathbbm{1}+M_{21}M_{21}^{\dagger}\right)^{-1/2}    
\end{align*}
\item We note that these quadratures can be measured simultaneously only if they commute. Let show that they indeed commute:
\begin{align*}
\left(\begin{array}{cc}
-M_{21}M_{\text{int}}^{\dagger} & \mathbbm{1}\end{array}\right)\left(\begin{array}{cc}
0 & \mathbbm{1}\\
-\mathbbm{1} & 0
\end{array}\right)\left(\begin{array}{c}
-M_{\text{\text{int}}}M_{21}^{\dagger}\\
\mathbbm{1}
\end{array}\right)=-M_{\text{\text{int}}}M_{21}^{\dagger}+M_{21}M_{\text{\text{int}}}^{\dagger}=0,    
\end{align*}
where in the last equality we used the fact that $M_{21} M_{\text{int}}^{\dagger}$ is Hermitian.
\end{itemize}

We can thus either measure the quadratures of eq. \ref{eq:optimal_quadratures} or the optimal quadrature in eq. \ref{eq:rad_pres_comb}. The advantage in measuring the quadratures of eq. \ref{eq:rad_pres_comb} is that they are independent of $\vplus$. They are optimal for the estimation of any polarization, and thus optimal also for simultaneous estimation of $\vplus, \mathcal{V}_{\times}$ or any two other polarizations. 



\subsection{FI}
Let us now consider the case of measuring the phase quadratures. The correspondong covariance matrix is obtained by applying eq. \ref{eq:FI_homodyne}, i.e. keeping only the phase quadrature terms in the full covariance matrix (eq. \ref{eq:cov_rad_pressure}), which leaves us with $\sigma_{h}=1/2\left( \mathbbm{1}+M_{\text{21}}M_{21}^{\dagger} \right)$, and thus the FI is $F=4 \vplusph^{\dagger}\left(\mathbbm{1}+M_{\text{21}}M_{21}^{\dagger}\right)^{-1} \vplusph.$
$M_{\text{21}}M_{21}^{\dagger}$ is basically the displacement noise term. 

Since $M_{21}=A_{\text{ph}}  D_x ,$ we get:
\begin{align*}
M_{21}M_{21}^{\dagger}=A_{\text{ph}} D_x D_x^{\dagger} A_{\text{ph}}^{\dagger},    
\end{align*}
hence it takes the form of displacement noise (eq. \ref{eq:cov_disp_noise}) with $ D_x D_x^{\dagger}$ being the covariance matrix of the displacement vector $\mathbf{\Delta x}.$ 
We can immediately observe that the space decoupled from this noise is $\text{ker} \left( D_x^{\dagger} A_{\text{ph}}^{\dagger} \right).$
The DFS, $\text{ker}A_{\text{ph}}^{\dagger}$ , is thus contained in this subspace
and therefore decoupled from this noise.

In fig. \ref{fig:rad_pres_fig} we show the sensitivity profile given RPN for different homodyne measurements: optimal (eq. \ref{eq:rad_pres_comb}), phase quadratures and maximal signal combination (basically $\mathcal{V},$ which maximizes the signal and thus optimal in case there is only shot noise). For the optimal frequency-dependent combination (eq. \ref{eq:rad_pres_comb}), the SD coincides with the shot noise limit as expected.

\begin{figure}[h]
\begin{center}
\subfigure[]{\includegraphics[width=.65\textwidth]{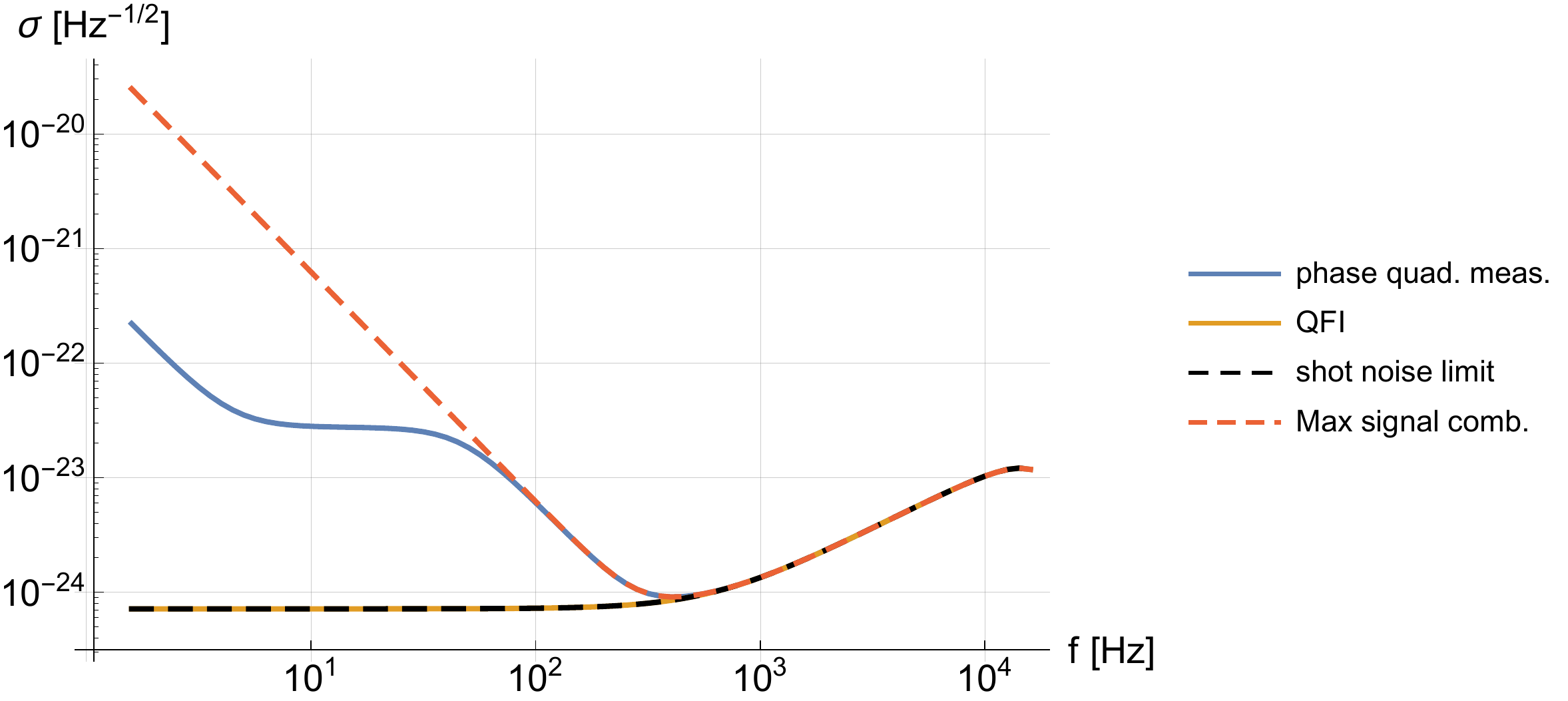}}
\end{center}
\caption{ Sensitivity profile with RPN for different measurement bases. The QFI coreesponds to optimal measurement (solid orange line) and saturates the shot noise limit (black dashed line). The solid blue line corresponds to phase quadratures measurement (and thus optimal combination of phase quadratures) and the dashed red line to the max-signal combination of phase quadratures, i.e. a combination that is optimal given only shot noise.    }
\label{fig:rad_pres_fig}
\end{figure}

It is interesting to compare the behavior of the FI with phase quadratures measurement and the behavior with the max-signal combination in fig. \ref{fig:rad_pres_fig}. Both diverge at low frequencies and clearly since the max-signal combination is not optimal its sensitivity is worse than the sensitivity of the phase quadratures measurement. While the max-signal combination diverges uniformly as $1/f^{2},$ the optimal phase quadratures combination has an intermediate range where the divergence stops and 
the sensitivity remains constant. This creates two orders of magnitude difference between the sensitivity with max-signal combination and with the optimal phase quadratures combination.

This plateau is due to a pseudo-DFS contained in the coupled subspace, $M_{\text{C}}.$ For any frequency, $M_{21}M_{21}^{\dagger}$ has $3$ eigenvalues: $0,t_{\text{min}},t_{\text{max}},$ where $t_{\text{min}} \ll t_{\text{max}}.$
The phase quadratures can be thus decomposed to the corresponding eigenspaces. The eigenspace of $0$ is the DFS, $M_{\text{DFS}},$ and the eigenspaces of $t_{\text{min}},t_{\text{max}}$ are denoted as $M_{\text{min}},M_{\text{max}}$ respectively. Note that $M_{\text{C}}=M_{\text{min}} \oplus M_{\text{max}}.$ Since the covariance matrix is diagonal w.r.t these subspaces the FI is a sum of the FI's of these subspaces:
\begin{align*}
F=F_{\text{DFS}}+F_{\text{C}}=F_{\text{DFS}}+F_{\text{max}}+F_{\text{min}},    
\end{align*}
where $F_{\text{max}}, F_{\text{min}}$ are the FI's achieved with $M_{\text{max}}, M_{\text{min}}$ respectively.
$\vplus$ is mostly in the subspace $M_{\text{max}},$ hence for frequencies higher than the plateau range $F \approx F_{\text{max}}.$ The SD that corresponds to $F_{\text{max}}$ goes as $1/f^{2}$ and coincides with the max-signal combination (see fig. \ref{fig:plateauexpl}). As $f$ gets smaller, $F_{\text{max}}$ drops as $f^4$ while $F_{\text{min}}$ remains the same (since $t_{min} \ll 1$). In this regime $M_{\text{min}}$ functions as a pseudo-DFS since the effect of displacement noise is much smaller than the shot noise. Therefore at some point $F_{\text{min}} > F_{\text{max}},$
and the FI coincides with the $F_{\text{min}}$ which remains the same. This is the plateau that can be observed in figs. \ref{fig:rad_pres_fig}, \ref{fig:plateauexpl}.
$t_{\text{min}}$ however also goes as $1/f^{4}$ and thus for low enough frequencies $t_{\text{min}} \gg 1$ and the SD continues to diverge as $1/f^{2}.$
This is shown in fig \ref{fig:plateauexpl} where the FI is shown along with the contribution of $F_{\text{max}}, F_{\text{min}},$ $F_{\text{DFS}}.$ We can see the crossing between $F_{\text{max}}$ and $F_{\text{min}}$ that takes place at the beginning of the plateau. 

\begin{figure}[h!]
\begin{center}
\subfigure[]{\includegraphics[width=.5\textwidth]{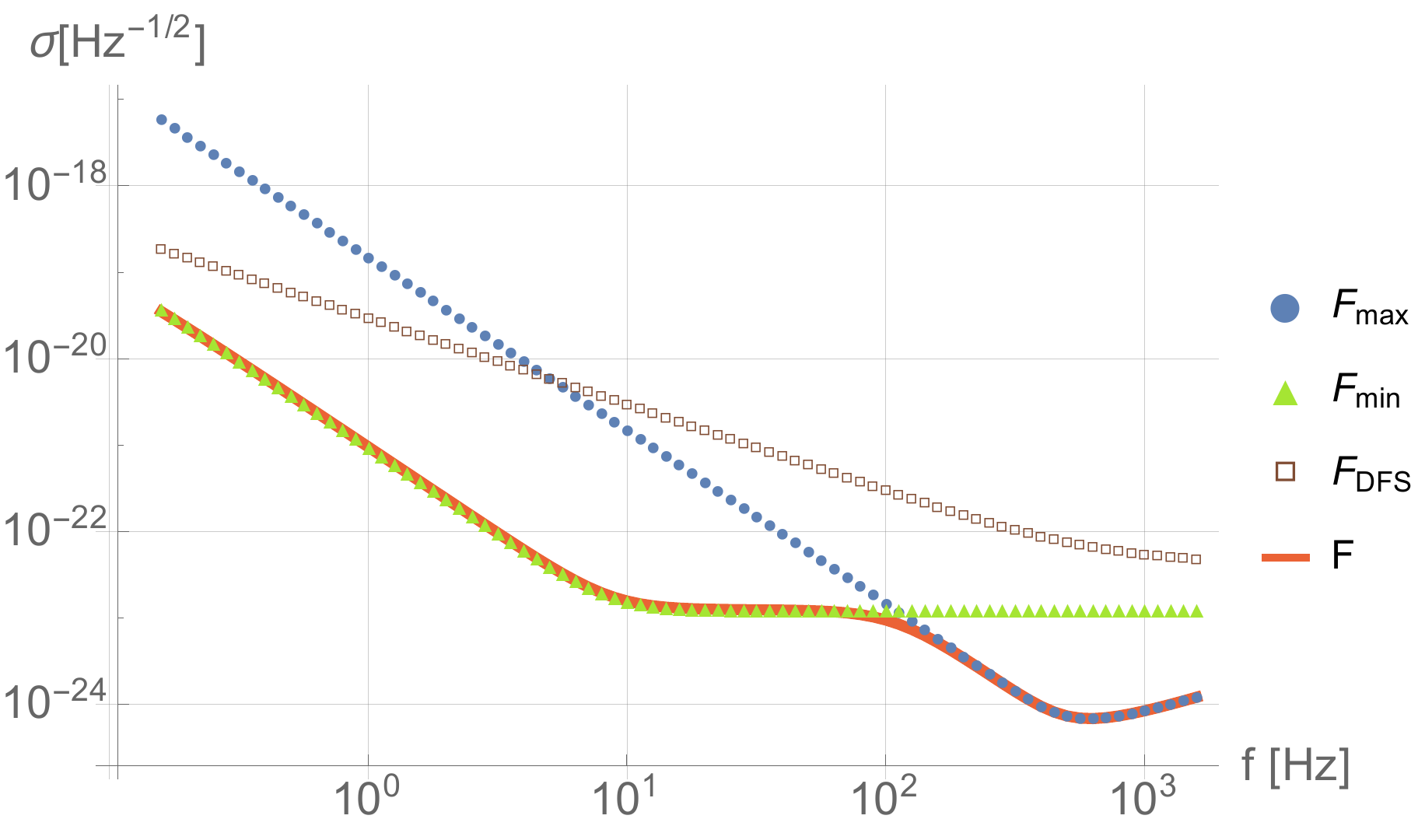}}
\end{center}
\caption{
Sensitivity profile that corresponds to phase quadratures measurement ($F$) along with $F_{\text{max}}$ (blue dots), $F_{\text{min}}$ (green triangles), and $F_{\text{DFS}}$ (brown squares), as defined in the text.  As shown in the text $F=F_{\text{max}}+F_{\text{min}}+F_{\text{DFS}}.$   }
\label{fig:plateauexpl}
\end{figure}

\section{FI and QFI with radiation pressure and thermal displacement noise}
\label{sec:thermal+rad}

\begin{figure}[b]
\begin{center}
\subfigure[]{\includegraphics[width=.49\textwidth]{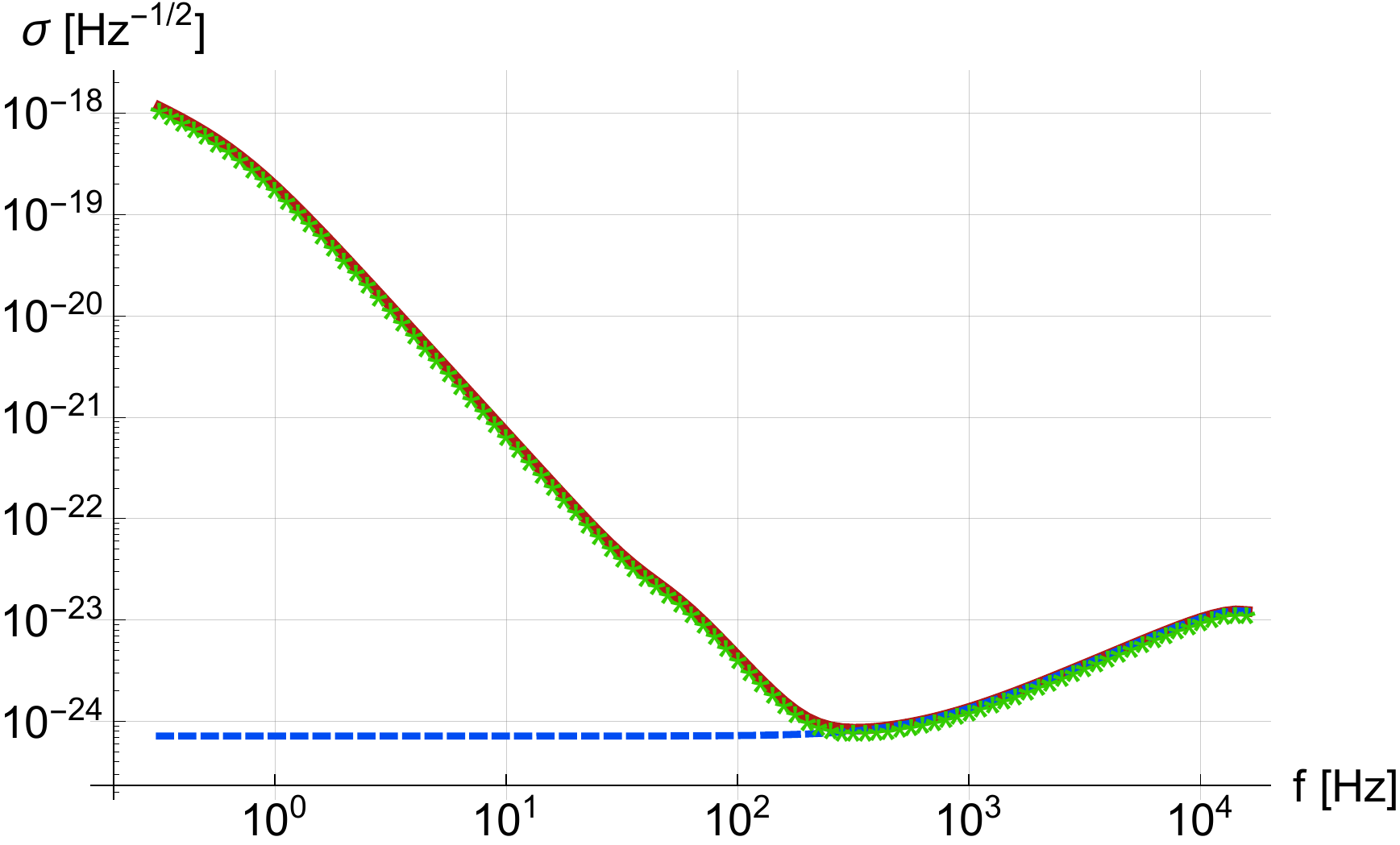}}
\subfigure[]{\includegraphics[width=.49\textwidth]{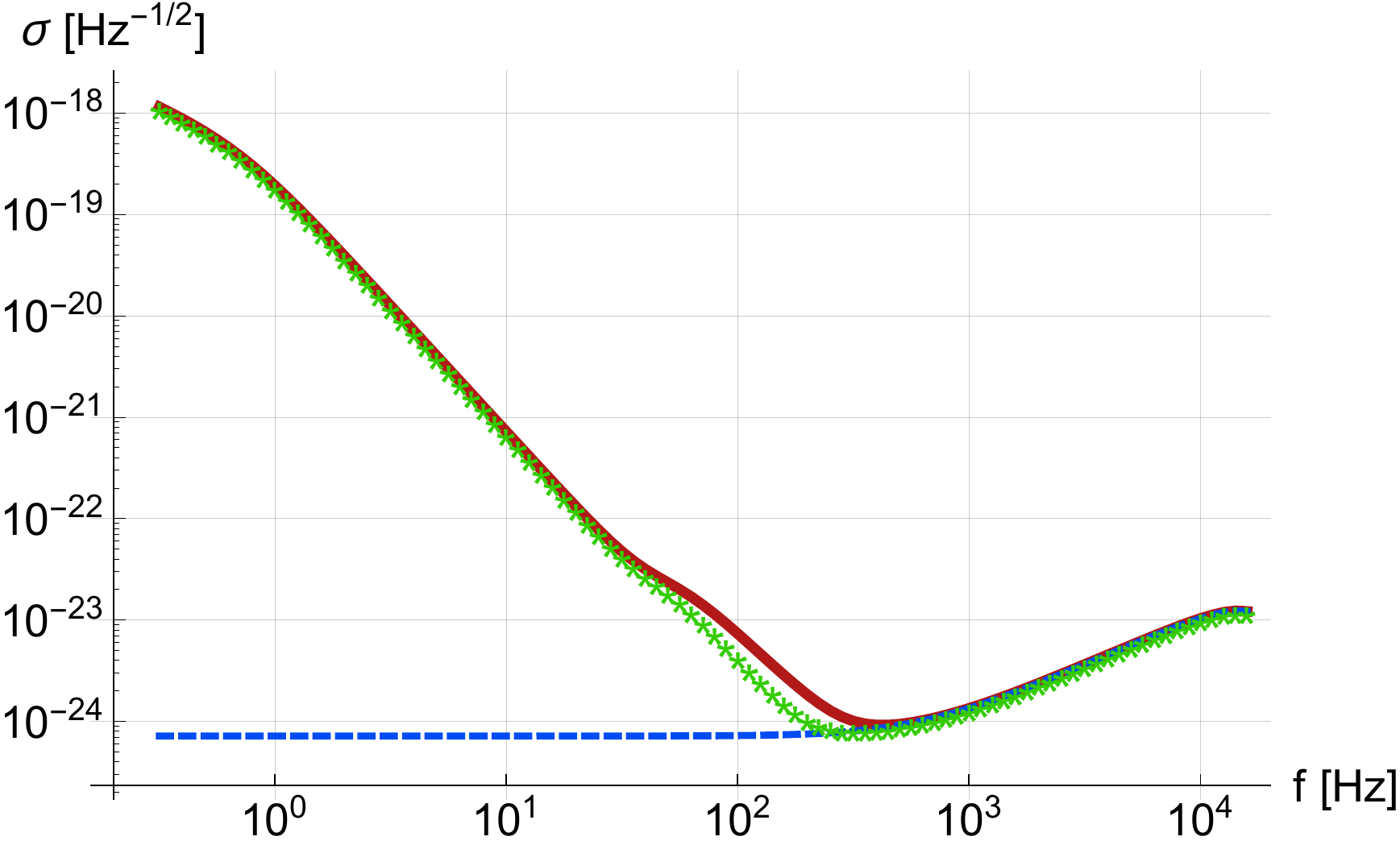}}
\end{center}
\caption{ Sensitivity profile given thermal noise, with and without radiation pressure.
(a) QFI: QFI with both thermal noise and RPN (red, solid line) coincides with QFI given only thermal noise (green diamonds). Blue dashed line corresponds to the shot noise limit.
(b) FI with phase quadrature measurement: red solid line corresponds to both thermal and RPN, green diamonds to only thermal noise and blue, dashed line to shot noise limit.  }
\label{fig:therm_rad_pres_fig}
\end{figure}

In realistic scenarios we have both RPN and thermal displacement noise, i.e. a covariance matrix of $\Sigma_{q}=1/2 \left(M M^{\dagger}+\delta^{2}A A^{\dagger} \right).$
The QFI in this case is the same as with only thermal displacement noise since the RPN can be removed using the same optimal frequency-dependent readout (eqs. \ref{eq:rad_pres_comb}, \ref{eq:optimal_quadratures}). Let us show this formally:
\begin{align*}
\Sigma_{q}=1/2\left(\begin{array}{cc}
\mathbbm{1} & M_{\text{int}}M_{21}^{\dagger}\\
M_{21}M_{\text{int}}^{\dagger} & \mathbbm{1}+M_{\text{21}}M_{21}^{\dagger}+\delta^{2} A_{\text{ph}} A_{\text{ph}}^{\dagger}
\end{array}\right).
\end{align*}

 With a similar calculation as before:
 \begin{align*}
&\Sigma_{q}^{-1}=2\left(\begin{array}{cc}
* & -M_{\text{int}}M_{21}^{\dagger}\left(\mathbbm{1}+\delta^{2}A_{\text{ph}} A_{\text{ph}}^{\dagger}\right)^{-1}\\
* & \left(\mathbbm{1}+\delta^{2}A_{\text{ph}}A_{\text{ph}}^{\dagger}\right)^{-1}
\end{array}\right)\\
&\Rightarrow I= 4  \vplusph^{\dagger}\left(\mathbbm{1}+\delta^{2}A_{\text{ph}}A_{\text{ph}}^{\dagger}\right)^{-1} \vplusph.
\end{align*}
Hence RPN is removed with this optimal readout, this is shown in fig. \ref{fig:therm_rad_pres_fig} (a). The optimal quadrature is 
\begin{align*}
u\propto\left(\begin{array}{c}
-M_{\text{int}}M_{21}^{\dagger}\tilde{\mathcal{V}}\\
\tilde{\mathcal{V}}
\end{array}\right),    
\end{align*}
where $\tilde{\mathcal{V}}=\left(\mathbbm{1}+\delta^{2}A_{\text{ph}} A_{\text{ph}} ^{\dagger}\right)^{-1}\vplusph.$
As in section \ref{sec:rad_pressure},  we can achieve this QFI also by measuring the quadratures decoupled from RPN of eq. \ref{eq:optimal_quadratures}. The only difference is that now the optimal linear combination of these quadratures is $\tilde{\mathcal{V}}$ instead of $\vplus.$ 

Measuring the phase quadratures the FI reads:
\begin{align*}
&F=4  \vplusph^{\dagger}\left(\mathbbm{1}+M_{21}M_{21}^{\dagger}+\delta^{2}A_{\text{ph}} A_{\text{ph}}^{\dagger}\right)^{-1}\vplusph \\
& \overset{*}{=}4\vplusph^{\dagger}\left(\mathbbm{1}+A_{\text{ph}}\left[D_{x}D_{x}^{\dagger}+\delta^{2} \mathbbm{1}\right]A_{\text{ph}}^{\dagger}\right)^{-1}\vplusph.
\end{align*}
There are two displacement noise terms: $M_{21}M_{21}^{\dagger}$ due to RPN and $\delta^{2} A_{\text{ph}} A_{\text{ph}}^{\dagger}$ due to thermal displacement noise. In $(*)$ we use the fact that $M_{21}=A_{\text{ph}} D_x ,$ therefore the two noise terms can be written as a single displacement noise $\mathbf{\Delta x},$ with a covariance matrix of $1/2 \left( \delta^{2}\mathbbm{1}+D_x D_x^{\dagger} \right).$ Since this covariance matrix is full rank the DFS is $\text{ker}\left( A_{\text{ph}}^{\dagger} \right)$  (in our triangular cavity example it is $3$ dimensional).

The behavior of the FI for phase quadratures measurement is shown in the red line of fig. \ref{fig:therm_rad_pres_fig} (b): there is an intermediate regime where the RPN is dominant and the FI does not coincide with the FI of only thermal noise. For low enough frequencies however the thermal noise becomes dominant and the FI coincides with the FI of only thermal noise.  

A more detailed analysis of the behavior of the FI is presented in fig. \ref{fig:plateauexpl2}. Similar to the behavior observed 
in section \ref{sec:rad_pressure}, the divergence is not uniform 
, but splits into three different regimes. Like in section \ref{sec:rad_pressure}, we can decompose the phase quadratures to:
$M=M_{\text{max}}\oplus M_{\text{min}}\oplus M_{\text{DFS}},$ and define the corresponding FI's: $F_{\text{max}},F_{\text{min}}, F_{\text{DFS}}.$ Since these subspaces are eigenspaces of the covariance matrix we have: $F=F_{\text{max}}+F_{\text{min}}+F_{\text{DFS}}.$ In fig. \ref{fig:plateauexpl2} we examine the contribution of these FI's. We can observe three regimes: for relatively high frequencies $F_{\text{max}}$ is dominant and coincides with $F.$ $M_{\text{max}}$ is a single dimensional subspace that has maximal noise (largest eigenvalue of the covariance matrix), but also yields maximal signal. Therefore in the limit of small noise $F$ coincides with it. Around $f\approx 10^{2}$ Hz, $F_{\text{min}}$ becomes larger than $F_{\text{max}},$ this transition can be observed in fig. \ref{fig:plateauexpl2}. After this 
crossing $F_{\text{min}}$ becomes the dominant contribution. Like in fig. \ref{fig:plateauexpl}, there is a small range of frequencies with plateau, where $v_{\text{min}}\approx 1.$
For lower frequencies ($f \approx 1$) the noise becomes larger and the dominant contribution is from the DFS: $F \approx F_{\text{DFS}}.$

\begin{figure}[h!]
\begin{center}
\subfigure[]{\includegraphics[width=.54\textwidth]{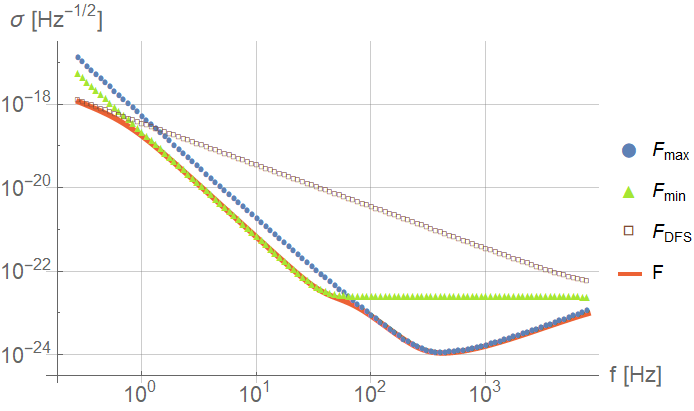}}
\end{center}
\caption{ Similar to fig. \ref{fig:plateauexpl} given thermal displacement noise and RPN with phase quadratures measurement. Three different regimes of the FI: the solid red line correspond to $F$, the blue dots to $F_{\text{max}},$ green triangles to $F_{\text{min}}$ and brown squares to $F_{\text{DFS}}.$ The different FI's coincide with total FI in three different regimes.    }
\label{fig:plateauexpl2}
\end{figure}

\section{Effect of squeezing \label{sec:squeezing}}
So far the initial state was taken to be a coherent state, i.e. $\Sigma_{i}=\frac{1}{2} \mathbbm{1}$. By squeezing some of the quadratures the sensitivity can be improved. The covariance matrix with squeezing is $\Sigma_{i}=\frac{1}{2} e^{2r}\Pi_{1}+ \frac{1}{2} e^{-2r}\left(\mathbbm{1}-\Pi_{1}\right),$ i.e. $k$ commuting quadratures are squeezed, where $\Pi_{1}$ is their projection operator. As a result the conjugate $k$ quadratures are antisqueezed. A trivial upper bound of the QFI corresponds to squeezing all the quantum noise by $e^{-2r}$, i.e.:
\begin{align}
I\leq2\vplus^{\dagger}\left(\Sigma_{\text{min}}\right)^{-1}\vplus,
\label{eq:sq_QFI_bound}
\end{align}
where $2\Sigma_{\text{min}}=e^{-2r}MM^{\dagger}+\delta^{2}AA^{\dagger},$ i.e. the quantum part ,$MM^{\dagger}$, is squeezed by $e^{-2r}$. This bound of the QFI implies a bound on the FI with any homodyne measurement. The bound for the FI given phase quadratures measurement is:
\begin{equation}
\begin{aligned}
&F\leq2\vplus^{\dagger}T_{\text{ph}}\left(T_{\text{ph}}^{\dagger}\Sigma_{\text{min}}T_{\text{ph}}\right)^{-1}T_{\text{ph}}^{\dagger}\vplus\\
&=4\vplusph^{\dagger}\left(e^{-2r}\left(\mathbbm{1}+M_{21}M_{21}^{\dagger}\right)+\delta^{2}A_{\text{ph}}A^{\dagger}_{\text{ph}}\right)^{-1}\vplusph,
\label{eq:sq_phase_quad_bound}
\end{aligned}
\end{equation}
i.e. squeezing the quantum part, $\mathbbm{1}+M_{21}M_{21}^{\dagger},$ by $e^{-2r}$.
We show a brief proof of these bounds. For brevity we omit the $\delta^{2}AA^{\dagger}$ term which does not change the proof:
\begin{align*}
&2\Sigma_{q}=M\left[e^{2r}\Pi_{1}+e^{-2r}\left(\mathbbm{1}-\Pi_{1}\right)\right]M^{\dagger}\geq e^{-2r}MM^{\dagger}=2\Sigma_{\text{min}}\\
& \overset{*}{\Rightarrow}\Sigma^{-1}\leq\Sigma_{\text{min}}^{-1}\\
&\overset{**}{\Rightarrow}\vplus^{\dagger}\left(\Sigma\right)^{-1}\vplus\leq\vplus^{\dagger}\left(\Sigma_{\text{min}}\right)^{-1}\vplus.
\end{align*}
$(*)$ is due to the fact that for any positive semi-definite matrices $B_{1},B_{2}$: $B_{1} \geq B_{2} \Rightarrow B_{1}^{-1} \leq B_{2}^{-1}.$ $(**)$ is immediate from definition. 
 Similarly:
 \begin{align*}
 \vplus^{\dagger}T_{\text{ph}}\left(T_{\text{ph}}^{\dagger}\Sigma T_{\text{ph}}\right)^{-1}T_{\text{ph}}^{\dagger}\vplus\leq\vplus^{\dagger}T_{\text{ph}}\left(T_{\text{ph}}^{\dagger}\Sigma_{\text{min}}T_{\text{ph}}\right)^{-1}T_{\text{ph}}^{\dagger}\vplus.
 \end{align*}
 
 Let us now show that both of these bounds are achievable. We first show that by an optimal choice of squeezed quadratures, $\Pi_1$, the phase quadratures bound of eq. \ref{eq:sq_phase_quad_bound} is achievable.
 
\subsection{Optimal squeezing for phase quadratures measurements}
To saturate eq. \ref{eq:sq_phase_quad_bound} we need to find a set of commuting quadratures such that squeezing them yields this bound. Let us first find suitable quadratures and then verify that they commute. We therefore need to find a projection operator $\Pi_{1}$ such that
\begin{align*}
M\left(e^{2r}\Pi_{1}+e^{-2r}\left(\mathbbm{1}-\Pi_{1}\right)\right)M^{\dagger}=\left(\begin{array}{cc}
* & *\\
* & e^{-2r}\left(\mathbbm{1}+M_{21}M_{21}^{\dagger}\right)
\end{array}\right)    
\end{align*}
This implies that $\Pi_{1}$ needs to satisfy:
\begin{align*}
M\Pi_{1}M^{\dagger}=\left(\begin{array}{cc}
* & *\\
* & 0
\end{array}\right),    
\end{align*}
i.e. no antisqueezing in the phase quadratures.
Following the derivation for single input single output \cite{Kimble01}, 
we use an ansatz:
\begin{align*}
\Pi_{1}=\left(\begin{array}{cc}
\cos\left(\mathcal{A}\right)^{2} & \cos\left(\mathcal{A}\right)\sin\left(\mathcal{A}\right)\\
\cos\left(\mathcal{A}\right)\sin\left(\mathcal{A}\right) & \sin\left(\mathcal{A}\right)^{2}
\end{array}\right),    
\end{align*}
where $\mathcal{A}$ is a Hermitian matrix that we seek to find. It can be checked that this defines a projection operator ($\Pi_{1}^2=\Pi_{1}$) onto the space spanned by $\left(\begin{array}{c}
\cos\left(\mathcal{A}\right)\\
\sin\left(\mathcal{A}\right)
\end{array}\right).$
We now observe:
\begin{align*}
& M\left(\begin{array}{cc}
\cos\left(\mathcal{A}\right)^{2} & \cos\left(\mathcal{A}\right)\sin\left(\mathcal{A}\right)\\
\cos\left(\mathcal{A}\right)\sin\left(\mathcal{A}\right) & \sin\left(\mathcal{A}\right)^{2}
\end{array}\right)M^{\dagger}=\left(\begin{array}{cc}
* & *\\
* & \left(M_{21}\cos\left(\mathcal{A}\right)+M_{\text{int}}\sin\left(\mathcal{A}\right)\right)\left(\cos\left(\mathcal{A}\right)M_{21}^{\dagger}+\sin\left(\mathcal{A}\right)M_{\text{int}}^{\dagger}\right)
\end{array}\right)  \\
&\Rightarrow M_{21}\cos\left(\mathcal{A}\right)+M_{\text{int}}\sin\left( \mathcal{A} \right)=0  \\
&\mathcal{A}=\arctan\left(-M_{\text{int}}^{\dagger}M_{21}\right).
\end{align*}
Note that $-M_{\text{int}}^{\dagger}M_{21}$ is Hermitian hence $\mathcal{A}$ is well defined. 
It is therefore optimal to squeeze the quadratures projected by $\mathbbm{1}-\Pi_{1}$:
\begin{align*}
&\left(\begin{array}{c}
-\sin\left(\mathcal{A}\right)\\
\cos\left(\mathcal{A}\right)
\end{array}\right)=\left(\begin{array}{c}
-\sin\left(-\arctan\left(M_{\text{int}}^{\dagger}M_{21}\right)\right)\\
\cos\left(-\arctan\left(M_{\text{int}}^{\dagger}M_{21}\right)\right)
\end{array}\right)\\
&=\left(\begin{array}{c}
M_{\text{int}}^{\dagger}M_{21}\\
\mathbbm{1}
\end{array}\right)
\left(\mathbbm{1}+\left(M_{\text{int}}^{\dagger}M_{21}\right)^{2}\right)^{-1/2}.
\end{align*}

It is straightforward to see that these quadratures commute: for any $\mathcal{A}$
\begin{align*}
\left(\begin{array}{cc}
-\sin\left(\mathcal{A}\right) & \cos\left(\mathcal{A}\right)\end{array}\right)\left(\begin{array}{cc}
0 & \mathbbm{1}_{k}\\
-\mathbbm{1}_{k} & 0
\end{array}\right)\left(\begin{array}{c}
-\sin\left(\mathcal{A}\right)\\
\cos\left(\mathcal{A}\right)
\end{array}\right)=0.    
\end{align*}

Numerical results of the SD without squeezing and with optimal squeezing are presented in fig. \ref{fig:squeezingfig}. 

\subsection{QFI with squeezing}
It is simple to show that the QFI satisfies the bound of eq. \ref{eq:sq_QFI_bound}. The calculation is identical to the calculations in sections \ref{sec:rad_pressure} and \ref{sec:QFI_displacement}. Given squeezing of phase quadratures:
\begin{align*}
&\Sigma=1/2\left(\begin{array}{cc}
M_{\text{int}} & 0\\
M_{21} & M_{\text{int}}
\end{array}\right)\left(\begin{array}{cc}
e^{2r} \mathbbm{1} & 0\\
0 & e^{-2r} \mathbbm{1}
\end{array}\right)\left(\begin{array}{cc}
M_{\text{int}}^{\dagger} & M_{21}^{\dagger}\\
0 & M_{\text{int}}^{\dagger}
\end{array}\right) +\frac{\delta^{2}}{2} AA^{\dagger} \\
&=1/2\left(\begin{array}{cc}
e^{2r} \mathbbm{1} & M_{\text{int}}M_{21}^{\dagger}e^{2r}\\
M_{21}M_{\text{int}}^{\dagger}e^{2r} & e^{-2r} \mathbbm{1}+M_{21}M_{21}^{\dagger}e^{2r}+\delta^{2} A_{\text{ph}}A_{\text{ph}}^{\dagger}
\end{array}\right)
\Rightarrow \Sigma^{-1}=2\left(\begin{array}{cc}
* & -M_{\text{int}}M_{21}^{\dagger}\left(e^{-2r} \mathbbm{1}+\delta^{2}A_{\text{ph}}A_{\text{ph}}^{\dagger}\right)^{-1}\\
* & \left(e^{-2r} \mathbbm{1}+\delta^{2}A_{\text{ph}}A_{\text{ph}}^{\dagger}\right)^{-1}
\end{array}\right)\\
&\Rightarrow I=4\vplusph^{\dagger}\left(e^{-2r} \mathbbm{1}+\delta^{2}A_{\text{ph}}A_{\text{ph}}^{\dagger}\right)^{-1}\vplusph.
\end{align*}
The quadrature that saturates this bound is:
\begin{align*}
\left(\begin{array}{c}
-M_{\text{int}}M_{21}^{\dagger}\\
\mathbbm{1}
\end{array}\right)\left(e^{-2r} \mathbbm{1}+\delta^{2}A_{\text{ph}}A_{\text{ph}}^{\dagger}\right)^{-1}\vplusph
\end{align*}

The QFI sensitivity with and without squeezing is presented in fig. \ref{fig:squeezingfig}.
It is interesting to compare this QFI to the phase quadratures FI. Without squeezing the QFI is significantly better than the FI in the regime of dominant RPN (in this illustration $50$ Hz-$500$ Hz). Optimal squeezing removes this difference almost completely since it squeezes the RPN while keeping the thermal displacement noise intact. We can also observe that in both cases the gain from squeezing is not uniform: in some frequency regimes this gain is much larger than in other regimes. This is explained in the next subsection.

\subsection{$\eta$ and squeezing}
As mentioned in the main text, for a given choice of homodyne measurement we can define the gain factor
\begin{align*}
\eta_{\text{gain}}=\frac{\frac{F_{\text{sq}}}{F}-1}{e^{2r}-1},    
\end{align*}
where $F_{\text{sq}}\, (F)$ is the FI with(out) squeezing. 

For phase quadratures measurement we defined $\eta$ in the main text 
as:
\begin{align*}
\eta=\frac{F_{\text{DFS}}}{F},    
\end{align*}
where $F_{\text{DFS}}$ is the FI coming from the DFS.
This is well defined for phase quadratures measurement, since in this case the quadratures space contains a DFS, $h_{\text{DFS}}\subseteq h,$ and the DFS is an eigenspace of the relevant covariance matrix $\sigma_{h}.$
Furthermore for phase quadratures measurement we can decompose: $h=h_{\text{DFS}}\oplus h_{\text{C}},$ where $h_{\text{C}}$ is the coupled subspace, and thus $F=F_{\text{DFS}}+F_{\text{C}},$
where:
\begin{align*}
&F_{\text{DFS}}=4\mathcal{V}_{\text{DFS}}^{\dagger}\mathcal{V}_{\text{DFS}},\\ 
&F_{\text{C}}=4\mathcal{V}_{\text{C}}^{\dagger}\left(\mathbbm{1}+M_{21}M_{21}^{\dagger}+\delta^{2}A_{\text{ph}}A_{\text{ph}}^{\dagger}\right)^{-1}\mathcal{V}_{\text{C}}
\end{align*}
where $\mathcal{V}_{\text{DFS}}=\Pi_{\text{DFS}} \mathcal{V}_{\text{ph}},$ $\mathcal{V}_{\text{C}}=\Pi_{\text{C}} \mathcal{V}_{\text{ph}}.$

Optimal squeezing changes the covariance matrix to $\frac{e^{-2r}}{2}
\left( \mathbbm{1+M_{21}M_{21}^{\dagger}}\right)+\frac{\delta^{2}}{2} A_{\text{ph}}A_{\text{ph}}^{\dagger},$ and thus after squeezing:
\begin{align*}
&F_{\text{DFS}}=4e^{2r}\mathcal{V}_{\text{DFS}}^{\dagger}\mathcal{V}_{\text{DFS}}\\
&F_{\text{C}}=4\mathcal{V}_{\text{C}}^{\dagger}\left(e^{-2r}\left(\mathbbm{1}+M_{21}M_{21}^{\dagger}\right)+\delta^{2}A_{\text{ph}}A_{\text{ph}}^{\dagger}\right)^{-1}\mathcal{V}_{\text{C}}. \end{align*}
Assuming large displacement noise, i.e. large enough $\delta,$ we can approximate:
\begin{align*}
F_{\text{C}} \approx 4\mathcal{V}_{\text{C}}^{\dagger}\left(\delta^{2}A_{\text{ph}}A_{\text{ph}}^{\dagger}\right)^{-1}\mathcal{V}_{\text{C}},    
\end{align*}
where $\left(\delta^{2}AA^{\dagger}\right)^{-1}$ refers to the pseudo-inverse of $\delta^{2} AA^{\dagger}$ and we can use it since $\mathcal{V}_{\text{C}}\perp\text{ker}\left(A_{\text{ph}}A_{\text{ph}}^{\dagger}\right)$ (by definition of the coupled subspace).  
Hence $F_{\text{DFS}}$ is increased by the squeezing factor $e^{2r},$ 
while in the limit of large displacement noise $F_{\text{C}}$ remains unchanged. Clearly only the DFS gains from squeezing and thus the gain from squeezing should go as the effectiveness of the DFI.
We can show this formally with $\eta_{\text{gain}}$:
\begin{align*}
& \frac{F_{\text{sq}}}{F}=\frac{e^{2r}F_{\text{DFS}}+F_{\text{C}}}{F_{\text{DFS}}+F_{\text{C}}}\\
&\eta_{\text{gain}}=\frac{\frac{F_{\text{sq}}}{F}-1}{e^{2r}-1}=\frac{\left(e^{2r}-1\right)F_{\text{DFS}}}{\left(e^{2r}-1\right)\left(F_{\text{DFS}}+F_{\text{C}}\right)}=\frac{F_{\text{DFS}}}{F}=\eta.
\end{align*}
Therefore in the limit of large displacement noise $\eta_{\text{gain}}=\eta,$ this is illustrated in fig. (3) in the main text.

\begin{figure}[h!]
\begin{center}
{\includegraphics[width=.6\textwidth]{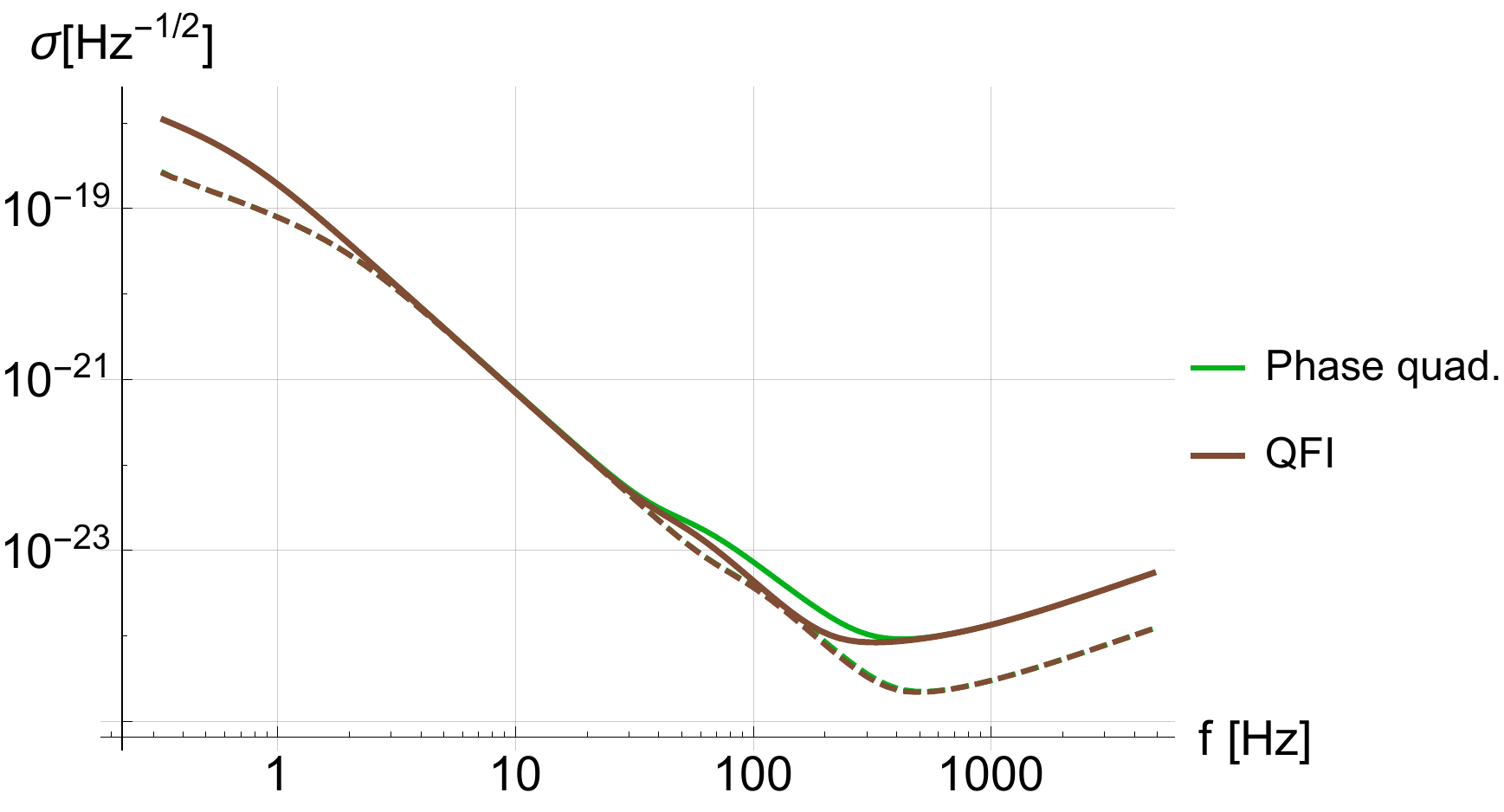}}
\end{center}
\caption{SD with optimal measurement (brown) and phase quadratures measurement (green). The solid lines correspond to the unsqueezed case and the dashed to optimal squeezing. 
In this illustration $r=1.5$. }
\label{fig:squeezingfig}
\end{figure}

\section{ Extensions}


\begin{figure}[h!]
\begin{center}
\subfigure[]{\includegraphics[width=5.5 cm]{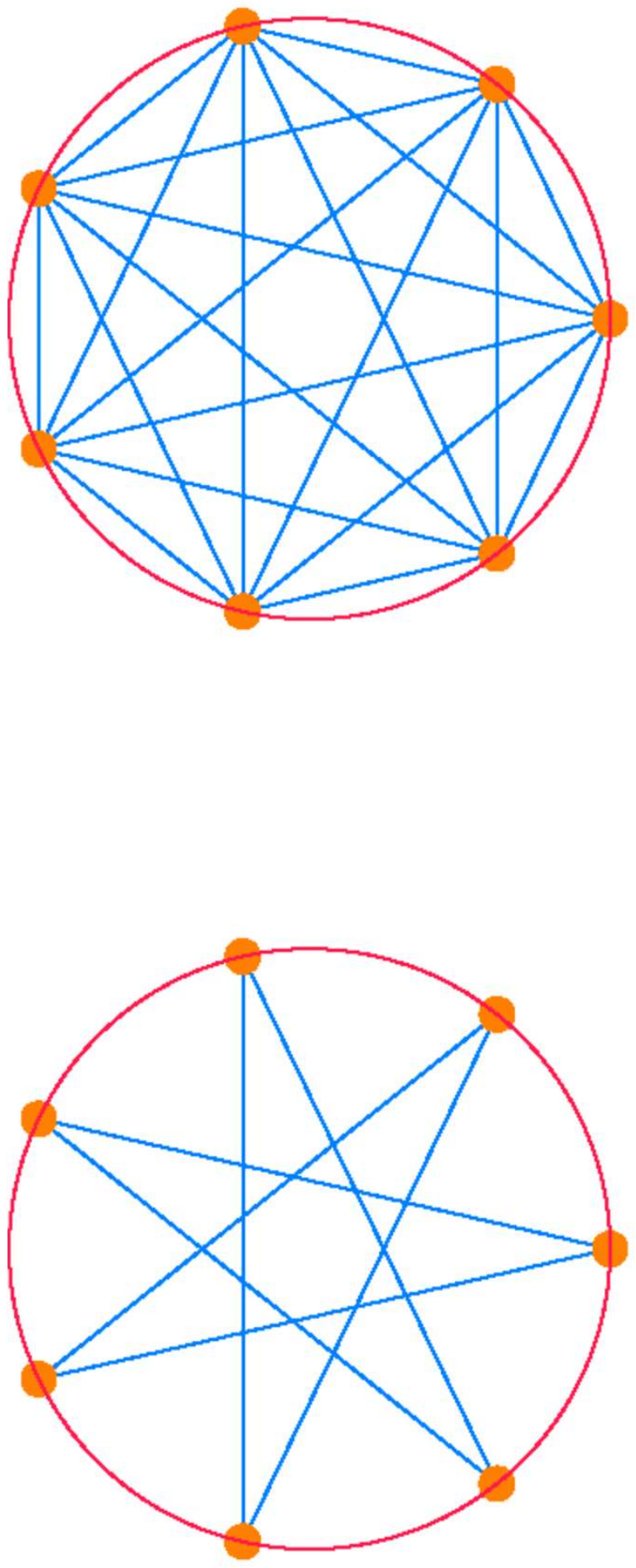}}
\subfigure[]{\includegraphics[width=.55\textwidth]{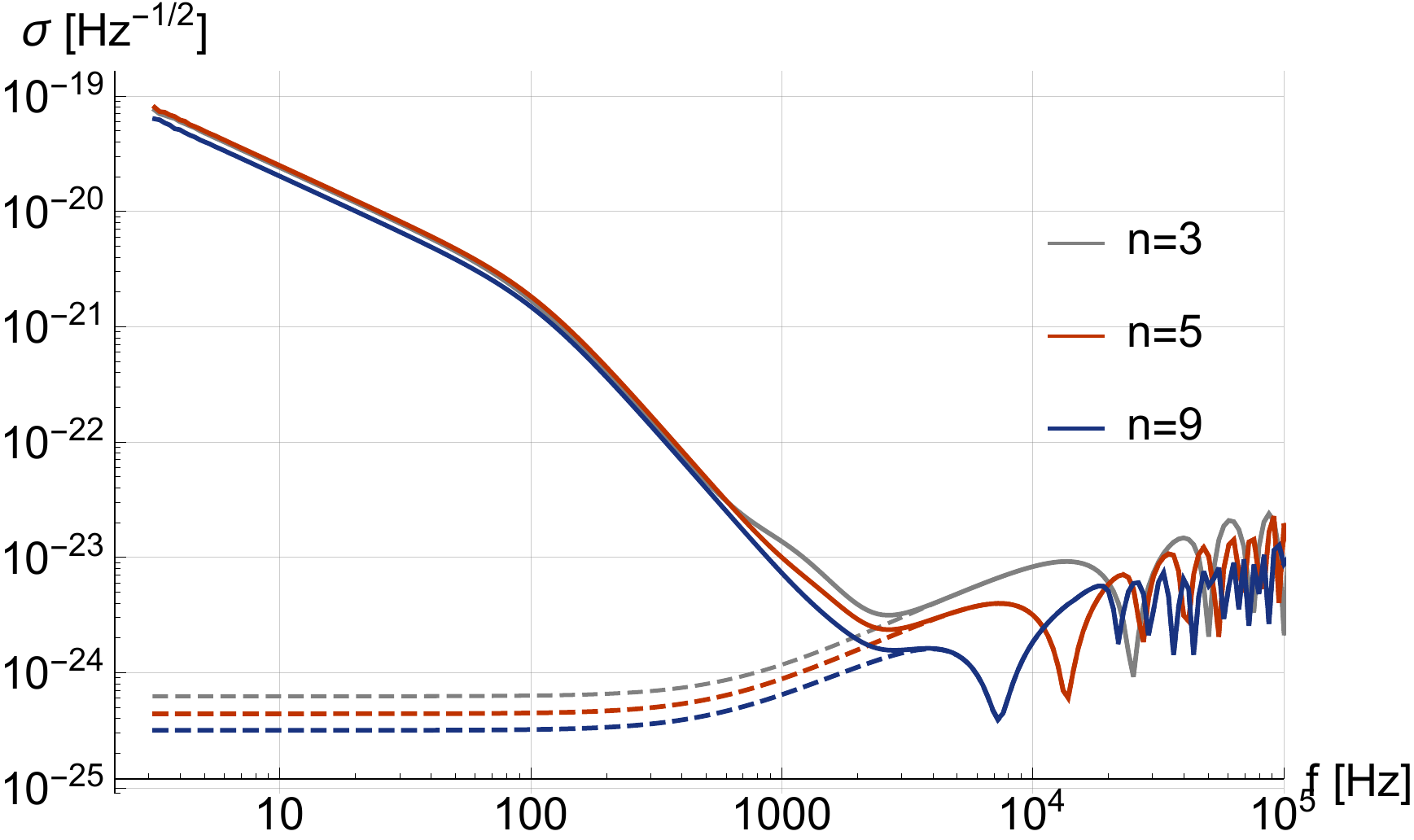}}
\end{center}
\caption{ (a) Extensions of the DFI triangular scheme to an arbitrary $n$-gon. The vertices represent the mirrors and the blue edges represent the light trajectories. 
(c) Comparison between the sensitivities of the different $n$-gon schemes: dashed lines correspond to shot-noise only and solid lines to thermal displacement noise as well. The blue, red and gray lines correspond to $n=3,5,9$ respectively. }
\label{fig:supp_extensions}
\end{figure}

The triangular scheme can be extended to any $n-$gon with $n$ mirrors, as illustrated in fig. \ref{fig:supp_extensions}. The $n$ mirrors are illustrated as $n$ vertices
of an $n-$gon embedded on a circle. For each polygon the
different light trajectories correspond to different billiard
trajectories hitting the vertices, hence there are $n-1$
possible trajectories that correspond to $n-1$ reflection
angles $\left\{ \frac{2\pi}{n}l\right\} _{l=1}^{n-1}.$
The two schemes illustrated in fig. \ref{fig:supp_extensions} correspond to all
$n-1$ trajectories (top), i.e. $n \left( n-1 \right)$ input and output fields, and two cyclic trajectories with maximal length (bottom), angles $\pi\frac{n-1}{n},\pi\frac{n+1}{n}$ for odd $n$.
We consider here only the schemes with two maximal length cyclic trajectories. Therefore the number of input/output fields is $2n$ and the phase quadratures DFS is $n$-dimensional. 
In fig. \ref{fig:supp_extensions} the triangular scheme ($n=3$) is compared to larger number of mirrors: $n=5,9,$ where thermal displacement noise is considered. In this comparison, all the $n$-gons have the same radius and the intracavity power is the same (taken to be $10^{5}\frac{T}{\left(1-\sqrt{R}\right)^{2}}\approx3.8\;\text{MW}$). 

It can be observed in fig. \ref{fig:supp_extensions} that assuming only shot noise the sensitivity in the DC (static signal) limit becomes better as $n$ is increased. More precisely, in DC the QFI goes as $I \propto n.$ The reason is that the length of the trajectory goes as $n$ which accounts for a factor of $n$ in the response vector $\mathcal{V},$ on the other hand there is more loss due to the increased number of mirros which accounts for a factor of $\frac{1}{\sqrt{n}}.$ Altogether $\mathcal{V}$ goes as $\sqrt{n}$ and thus the QFI as $n.$ However once we add thermal displacement noise the sensitivities of the schemes coincide and only a very minor improvement is observed for larger $n.$

Further extensions, such as using all the $n-1$ trajectories as showm in fig. \ref{fig:supp_extensions} hold the promise for improved sensitivities, as they have larger DFS. We leave these extensions for future work. 

\section{ Newtonian and Seismic Noise}
Apart from thermal and radiation pressure noise, interferometers suffer also from newtonian and seismic noise \cite{Saulson84,Hughes98}.

Like thermal noise, they can be modelled as random displacement vector $\mathbf{\Delta x}_{\text{NN}}, \mathbf{\Delta x}_{\text{S}}$ propagated to the output quadratures by a transfer matrix $A.$ Denoting the covariance matrices of $\mathbf{\Delta x}_{\text{NN}}, \mathbf{\Delta x}_{\text{S}}$ as  $\sigma_{\text{S}},$ $\sigma_{\text{NN}},$
we get that the QFI with these noises reads:
\begin{align*}
I=4\vplus^{\dagger}\left(MM^{\dagger}+A\left(\delta^{2}\mathbbm{1}+\sigma_{\text{NN}}+\sigma_{\text{S}}\right)A^{\dagger}\right)^{-1}\vplus.    
\end{align*}
Unlike thermal noise, the assumption of uncorrelated noise between mirrors is no longer correct for these noises. For example, two nearby mirrors will have correlated seismic and newtonian noises.

\section{Calculation outline}
\label{sec:calculation_outline}
The QFI and QFIM are expressed using the transfer matrices $\mathcal{V}~,~M,~A.$
In this section, we outline how these transfer matrices are calculated:

{\bf{Step 1: solve the equations for the carrier amplitudes }}

Let us denote the input carrier amplitudes entering the cavity as $\{ A_{\text{in}, j} \}_{j=1}^{6}$,
the carrier amplitudes hitting/ reflecting 
inside the cavity as $\{ A_{\text{hit},j} \}_{j=1}^{6}, \{ A_{\text{ref},j} \}_{j=1}^{6} $ respectively, and the outgoing amplitudes as $\{ A_{\text{out},j} \}_{j=1}^{6}.$  

Given the input carrier amplitudes $\{ A_{\text{in},j} \}_{j=1}^{6}$ we can solve for 
$\{ A_{\text{hit},j} \}_{j=1}^{6}, \{ A_{\text{ref},j} \}_{j=1}^{6}, \{ A_{\text{out},j} \}_{j=1}^{6} $ using the following set of input-output linear equations:

{\it{Mirror equations:}}
\begin{align*}
\left(\begin{array}{c}
A_{\text{out},j}\\
A_{\text{ref},j}
\end{array}\right)=\left(\begin{array}{cc}
-\sqrt{R} & \sqrt{T}\\
\sqrt{T} & \sqrt{R}
\end{array}\right)\left(\begin{array}{c}
A_{\text{in},j}\\
A_{\text{hit},j}
\end{array}\right)
\end{align*}

{\it{Propagation equations:}}
\begin{align*}
A_{\text{hit},j+1}=A_{\text{ref},j}    
\end{align*}

Where $R,T$ are the reflectivity and transmissivity of the mirrors and we assume symmetric configuration of identical transmissivities.
Given that all input amplitudes are also the same $(A_{\text{in},j}=E)$ the analytical solution for $A_{\text{hit},j}, A_{\text{ref},j}, A_{\text{out},j}$ is

\begin{align*}
&A_{\text{hit},j}=A_{\text{ref},j}=E\frac{\sqrt{T}}{1-\sqrt{R}}=E\frac{\sqrt{1+\sqrt{R}}}{\sqrt{1-\sqrt{R}}}\\
&A_{\text{out},j}=E.
\end{align*}

The solution to $A_{\text{hit},j}.
A_{\text{out},j}$ will be inserted in the equations of the sideband fields.

Regarding power amplification in our cavity: the ratio between the carrier power inside and outside the cavity is:
\begin{align*}
\frac{p_{\text{hit}}}{p_{\text{in}}}=\frac{p_{\text{ref}}}{p_{\text{in}}}=\frac{T}{\left(1-\sqrt{R}\right)^{2}}.    
\end{align*}
In our triangular scheme the total intracavity power is $2 p_{\text{hit}},$ the factor of $2$ is due to the clockwise and counter-clockwise trajectories. The total input power is $6 p_{\text{in}},$ because there are in total $6$ input fields. This yields the gain enhancement factor: $\frac{T}{3\left(1-\sqrt{R}\right)^{2}}.$

{\bf{Step 2: solve the equations of the sideband fields }}
The input and output sideband fields are $\left(\begin{array}{c}
\mathbf{\hat{a}_{1}}\\
\mathbf{\hat{a}_{2}}
\end{array}\right),$ $\left(\begin{array}{c}
\mathbf{\hat{b}_{1}}\\
\mathbf{\hat{b}_{2}}
\end{array}\right)$ 
respectively, as before. We introduce the intracavity sideband fields $\left(\begin{array}{c}
\mathbf{c}_{\text{hit},1}\\
\mathbf{c}_{\text{hit},2}
\end{array}\right),$ $\left(\begin{array}{c}
\mathbf{\hat{c}}_{\text{ref},1}\\
\mathbf{\hat{c}}_{\text{ref},2}
\end{array}\right)$ that hit and reflect from the mirrors respectively. The transfer matrices are obtained by solving the following set of linear equations:

{\it{Mirror equations :}}

{\it{No radiation pressure:}}

\begin{align}
\left(\begin{array}{c}
\hat{b}_{1,j}\\
\hat{b}_{2,j}\\
\hat{c}_{\text{hit},1,j}\\
\hat{c}_{\text{hit},2,j}
\end{array}\right)=\left(\begin{array}{cccc}
-\sqrt{R} &  & \sqrt{T}\\
 & -\sqrt{R} &  & \sqrt{T}\\
\sqrt{T} &  & \sqrt{R}\\
 & \sqrt{T} &  & \sqrt{R}
\end{array}\right)\left(\begin{array}{c}
\hat{a}_{1,j}\\
\hat{a}_{2,j}\\
\hat{c}_{\text{ref},1,j}\\
\hat{c}_{\text{ref},2,j}
\end{array}\right)-\Delta x\cos\left(\pi/6\right)\sqrt{R}\left(\frac{\omega_{0}}{c}\right)\left(\begin{array}{c}
0\\
A_{\text{in},j}\\
0\\
A_{\text{hit},j}
\end{array}\right)
\label{eq:mirror_eq}
\end{align}
or {\it{with radiation pressure}} (see also \ref{subsection:radiation_pressure}:  
)
\begin{align}
\begin{split}
&\left(\begin{array}{c}
\hat{b}_{1,j}\\
\hat{b}_{2,j}\\
\hat{c}_{\text{hit},1,j}\\
\hat{c}_{\text{hit},2,j}
\end{array}\right)=\left(\begin{array}{cccc}
-\sqrt{R} &  & \sqrt{T}\\
 & -\sqrt{R} &  & \sqrt{T}\\
\sqrt{T} &  & \sqrt{R}\\
 & \sqrt{T} &  & \sqrt{R}
\end{array}\right)\left(\begin{array}{c}
\hat{a}_{1,j}\\
\hat{a}_{2,j}\\
\hat{c}_{\text{ref},1,j}\\
\hat{c}_{\text{ref},2,j}
\end{array}\right)-\Delta x \cos \left( \pi/6 \right)\sqrt{R}\left(\frac{\omega_{0}}{c}\right)\left(\begin{array}{c}
0\\
A_{\text{in},j}\\
0\\
A_{\text{hit},j}
\end{array}\right)-\\
& 2\sqrt{2}\frac{\omega_{0}\sqrt{R}}{m\Omega^{2}c^{2}}\left[\sqrt{I_{\text{in}}}\hat{a}_{1,j}-\sqrt{I_{\text{hit}}}\hat{c}_{\text{hit},1,j}+\sqrt{I_{\text{out}}}\hat{b}_{1,j}-\sqrt{I_{\text{hit}}}\hat{c}_{\text{ref},1,j}\right]\left(\begin{array}{c}
0\\
-\sqrt{2I_{\text{hit}}}\\
0\\
-\sqrt{2I_{\text{hit}}}
\end{array}\right).
\label{eq:mirror_eq_rad_pressure}
\end{split}
\end{align}

{\it{Propagation equations:}}
\begin{align}
\left(\begin{array}{c}
\hat{c}_{\text{hit},1,j+1}\\
\hat{c}_{\text{hit},2,j+1}
\end{array}\right)=e^{-i\Omega\frac{L}{c}}\left(\begin{array}{c}
\hat{c}_{\text{ref},1,j}\\
\hat{c}_{\text{ref},2,j}
\end{array}\right)+A_{\text{ref},j}\left(\frac{\omega_{0}}{c}\right)\left(\begin{array}{c}
0\\
h_{j}\left(f\right)
\end{array}\right),
\label{eq:propagation_eq}
\end{align}
where 
$h_{j} \left( f \right)$ is the response to the GW in the $j$-th arm.
To write it explicitly let us introduce necessary notation:
the GW signal travels along an arbitrary direction given by the unit vector,
\begin{equation*}
\mathbf{k}=\left(\sin\left(\theta\right)\cos\left(\phi\right),\sin\left(\theta\right)\sin\left(\phi\right),\cos\left(\theta\right)\right),    
\end{equation*}
and the orthogonal unit vectors to $\bm{k}$ are
\begin{eqnarray*}
\begin{split}
&\mathbf{u}=\left( -\cos\left(\theta\right)\cos\left(\phi\right),-\cos\left(\theta\right)\sin\left(\phi\right),\sin\left(\theta\right) \right),\\
&\mathbf{v}=\left( -\cos\left(\phi\right),\sin\left(\phi\right),0 \right).
\end{split}
\end{eqnarray*}
$h_{j}\left(f\right)$ is then given by \cite{Kawamura04}:
\begin{equation}
h_{j}\left(f\right)=\zeta_{j}\left(f\right)\left(~h_{+}\langle n_{j}|e_{+}|n_{j}\rangle~+~h_{\times}\langle n_{j}|e_{\times}|n_{j}\rangle~\right),
\label{eq:h_j}
\end{equation}
where $e_{+}=\mathbf{u}\otimes \mathbf{u}-\mathbf{v}\otimes \mathbf{v},~ e_{\times}=\mathbf{u}\otimes \mathbf{v}+\mathbf{v}\otimes \mathbf{u}$ are the polarization tensors and $|n_{j}\rangle$ is the unit vector of the $j$-th arm. The factor $\zeta_{j}\left(f\right)$ contains the dependence on the frequency and GW direction and is given by
\begin{eqnarray}
\begin{split}
\zeta_{j}\left(f\right)=E \left(\frac{\omega_{0}}{c}\right)L~\text{sinc}\left(\epsilon(1-\mathbf{n}_{j}\cdot\mathbf{k})\right)\exp\left(i\frac{1}{L}\left(\mathbf{x}_{j,1}+\mathbf{x}_{j,2}\right)\cdot\mathbf{k}\epsilon-i\epsilon\right),
\end{split}
\label{eq:zeta_j}
\end{eqnarray}
with $\epsilon=2 \pi f L/2$ and $\mathbf{x}_{j,1},\mathbf{x}_{j,2}$ are the positions of the mirrors.

Equations \ref{eq:mirror_eq_rad_pressure}, \ref{eq:propagation_eq}
form a set of $36$ linear equations for $36$ variables $\mathbf{\hat{b}}_{1},\mathbf{\hat{b}}_{2},\mathbf{\hat{c}}_{\text{ref},1},\mathbf{\hat{c}}_{\text{ref},2},\mathbf{\hat{c}}_{\text{hit},1},\mathbf{\hat{c}}_{\text{hit},2}.$
We can recast these equations as:
\begin{align*}
&L\left(\begin{array}{c}
\mathbf{\hat{b}}_{1}\\
\mathbf{\hat{b}}_{2}\\
\mathbf{\hat{c}}_{\text{ref},1}\\
\mathbf{\hat{c}}_{\text{ref},2}\\
\mathbf{\hat{c}}_{\text{hit},1}\\
\mathbf{\hat{c}}_{\text{hit},2}
\end{array}\right)=O\left(\begin{array}{c}
\mathbf{\hat{a}}_{1}\\
\mathbf{\hat{a}}_{2}
\end{array}\right)+K_{1}\mathbf{\Delta x}+K_{2}\mathbf{h}\\
&\Rightarrow
\left(\begin{array}{c}
\mathbf{\hat{b}}_{1}\\
\mathbf{\hat{b}}_{2}\\
\mathbf{\hat{c}}_{\text{ref},1}\\
\mathbf{\hat{c}}_{\text{ref},2}\\
\mathbf{\hat{c}}_{\text{hit},1}\\
\mathbf{\hat{c}}_{\text{hit},2}
\end{array}\right)=L^{-1}O\left(\begin{array}{c}
\mathbf{\hat{a}}_{1}\\
\mathbf{\hat{a}}_{2}
\end{array}\right)+L^{-1}K_{1}\mathbf{\Delta x}+L^{-1}K_{2}\mathbf{h},
\end{align*}
where $L,O,K_{1},K_{2}$ are matrices of size $36\times36,36\times12,36\times3,36\times2$ respectively.
Hence the transfer matrices $M,A,\mathcal{V}$ are given by the first $12$ rows of $L^{-1}O,L^{-1}K_{1},L^{-1}K_{2}$ respectively.

\section{GW response: analytical expressions of transfer matrix $\mathcal{V}$\label{sec:GWresponse}}

We calculate the GW response, i.e. the vector of the first moments $\mathbf{d}=\mathcal{V} \mathbf{h}$, from which we can get the shot noise limit 4$\mathcal{V}_{\text{ph}}^\dagger \mathcal{V}_{\text{ph}}.$ Since the responce to the GW is in the phase quadratures, it suffices to calculate the $6$-dimensional $\mathcal{V}_{\text{ph}}$ and the corresponding $\mathbf{d}_{\text{ph}}=\mathcal{V}_{\text{ph}} \mathbf{h}$. We calculate it for the symmetric triangular scheme depicted in fig. (1) in the main text. Let us introduce the required notation: $L$-length of the arms, $A_{\text{in}}$-amplitude of the incoming fields (identical for all fields), $\omega_{0}$-frequency of the carrier fields, $c$-speed of light. All the mirrors have the same power reflectivity $R$ (and power transmissivity $T=1-R$).

As shown in section
\ref{sec:calculation_outline},
The GW response in the $j-$th arm, $h_{j} \left(f \right)$, is then given by eq. \ref{eq:h_j}, with $\zeta_{j} \left( f\right)$ given in eq. \ref{eq:zeta_j}.
We can observe that as $f \rightarrow 0$, $\zeta_{j}\left(f\right) \rightarrow L $ for all arms, and the response is simplified to
\begin{equation*}
  h_{j}\left(f\right)=A_{\text{in}} \left( \frac{\omega_{0}}{c}  \right)L\left(h_{+}\langle n_{j}|e_{+}|n_{j}\rangle+h_{\times}\langle n_{j}|e_{\times}|n_{j}\rangle\right)  
\end{equation*}

We consider a cyclic trajectory, hence each $h_{j}$ propagates to the $k$-th output mode in the following way:
\begin{eqnarray*}
\begin{split}
\left(d_{\text{ph}}\right)_{k} = \frac{T }{(1-\sqrt{R})(1- R'^{1.5})}\cdot (R' h_k+\sqrt{R'} h_{k+1}+h_{k+2}),
\end{split}
\end{eqnarray*}
with $R'=Re^{i 2 \Omega L}.$
Hence the full vector of output modes (clockwise and counter-clockwise) reads:
\begin{equation}
\mathbf{d}_{\text{ph}}=\frac{(1+\sqrt{R})}{(1-R'^{3/2})}\left(\begin{array}{c}
\left(\begin{array}{c}
R'h_{1}+\sqrt{R'}h_{2}+h_{3}\\
R'h_{2}+\sqrt{R'}h_{3}+h_{1}\\
R'h_{3}+\sqrt{R'}h_{1}+h_{2}
\end{array}\right)_{\text{L}}\\
\left(\begin{array}{c}
h_{1}+\sqrt{R'}h_{2}+R'h_{3}\\
h_{2}+\sqrt{R'}h_{3}+R'h_{1}\\
h_{3}+\sqrt{R'}h_{1}+R'h_{2}
\end{array}\right)_{\text{R}}
\end{array}\right)
\label{eq:GW_response_1}
\end{equation}

From here we can obtain the shot-noise limit. The QFIM about $h_{+},h_{\times}$ in the shot-noise limit is given by:
\begin{equation}
\mathbf{h}^{\dagger}\mathcal{I}\mathbf{h}=\mathbf{d}_{\text{ph}}^{\dagger}\mathbf{d}_{\text{ph}}, 
\end{equation}
with $\mathbf{h}=\left(h_{+},h_{\times}\right)^{T}.$

We thus calculate $\mathbf{d}_{\text{ph}}^{\dagger}\mathbf{d}_{\text{ph}}$ using eq. \ref{eq:GW_response_1}:
\begin{eqnarray}
\begin{split}
\mathbf{d}_{\text{ph}}^{\dagger}\mathbf{d}_{\text{ph}}=A_{\text{in}}^{2}\left(\frac{\omega_{0}}{c}\right)^{2}\frac{T}{\left(1-\sqrt{R}\right)^{2}}\frac{T}{\left|1-R'^{1.5}\right|^{2}}\left[l_{1}\left(R\right)\underset{j=1}{\overset{3}{\sum}}h_{j}^{2}+l_{2}\left(R'\right)\underset{j,m>j}{\sum}h_{j}h_{m}\right],
\end{split}
\label{eq::general_freq_fi}
\end{eqnarray}
where $l_{1}\left(R\right)=\frac{1-R^{3}}{1-R}$
and $l_{2}\left(R'\right)=2\text{Re}\left(\sqrt{R'}\frac{1-R'^{1.5}}{1-\sqrt{R'}}\right).$

Let us now focus on the DC limit ($f \ll c/L$ and thus $R=R'$ ). In this limit eq. \ref{eq::general_freq_fi} becomes:
\begin{eqnarray}
\begin{split}
\mathbf{d}_{\text{ph}}^{\dagger}\mathbf{d}_{\text{ph}}=2 A_{\text{in}}^{2}\left(\frac{\omega_{0}}{c}\right)^{2}\frac{\left( 1+\sqrt{R} \right)^{2}}{\left(1-R^{1.5}\right)^{2}} \cdot \left[\frac{1-R^{3}}{1-R}\underset{j}{\sum}h_{j}^{2}+2\sqrt{R}\frac{1-R^{1.5}}{1-\sqrt{R'}}\underset{m>j}{\sum}h_{j}h_{m}\right].
\end{split}
\label{eq:bilinear}
\end{eqnarray}

Note that in the relevant limit of $T \ll 1:$
\begin{align*}
\frac{1-R^{3}}{1-R},\sqrt{R}\frac{1-R^{1.5}}{1-\sqrt{R}}\approx3R+\mathcal{O}\left(T^{2}\right)
\end{align*}
hence this expression can be further simplified to:
\begin{equation}
\mathbf{d}_{\text{ph}}^{\dagger}\mathbf{d}_{\text{ph}} \approx 6A_{\text{in}}^{2}\left(\frac{\omega_{0}}{c}\right)^{2}\frac{\left(1+\sqrt{R}\right)^{2}}{\left(1-R^{1.5}\right)^{2}}R\left(\underset{j}{\sum}h_{j}\right)^{2}.
\label{eq:simplified_bilinear}
\end{equation}

The coefficients of the bilinear terms of $h_{+}, h_{\times}$ in eqs. \ref{eq:bilinear},\ref{eq:simplified_bilinear}
correspond to the FIM elements
(eq. \ref{eq::general_freq_fi}). It can be shown 
that the coefficient of $h_{+} h_{\times}$
vanishes , hence the FIM is diagonal in the $h_{+},h_{\times}$ basis.
Note that 
\begin{eqnarray}
\begin{split}
&\underset{j}{\sum}h_{j}=L h_{+}\underset{j}{\sum}\langle n_{j}|e_{+}|n_{j}\rangle+L h_{\times}\underset{j}{\sum}\langle n_{j}|e_{\times}|n_{j}\rangle=\\
&\frac{3}{2}L h_{+}\text{Tr}_{x-y}\left(e_{+}\right)+\frac{3}{2}L h_{\times}\text{Tr}_{x-y}\left(e_{\times}\right)=\\
& \frac{3}{2} L h_{+}\text{Tr}_{x-y}\left(e_{+}\right).    
\end{split}
\label{eq:simplification1}
\end{eqnarray}
In the second equality we use the fact that $\underset{j}{\sum}|n_{j}\rangle \langle n_{j}|$ is proportional to the projection operator on the $x-y$ plane: $\underset{j}{\sum}|n_{j}\rangle \langle n_{j}|=\frac{3}{2} \Pi_{x-y},$ , hence $\underset{j}{\sum}\langle n_{j}|\bullet|n_{j}\rangle=\frac{3}{2}\text{Tr}_{x-y}\left(\bullet\right).$
In the third equality we use the fact that $\text{Tr}_{x-y}\left(e_{\times}\right)=0,$ because:
\begin{equation*}
\text{Tr}_{x-y}\left(e_{\times}\right)=2\langle u|\Pi_{x-y}|v\rangle=0.    
\end{equation*}
Hence combining eqs. \ref{eq:simplified_bilinear}, \ref{eq:simplification1}, we get that the minimal variance is about $h_{+}$ (maximal eigenvalue of the FIM) and it reads:
\begin{align*}
&\sigma=\Delta h_{+}=\frac{1}{2\sqrt{ || V^{\dagger}\mathcal{V} || }}\\
&\Rightarrow \Delta h_{+}\approx\frac{1}{2 A_{\text{in}}\left(\frac{\omega_{0}}{c}\right)L|\text{Tr}_{x-y}\left(e_{+}\right)|\sqrt{13.5\frac{\left(1+\sqrt{R}\right)^{2}}{\left(1-R^{1.5}\right)^{2}}R}}\\
&=\frac{1}{2 A_{\text{in}}\left(\frac{\omega_{0}}{c}\right)L\sin\left(\theta\right)^{2}\sqrt{13.5\frac{\left(1+\sqrt{R}\right)^{2}}{\left(1-R^{1.5}\right)^{2}}R}}.
\end{align*}

\section{Displacement response: analytical expression of transfer matrix $A$ \label{sec:DFS}}

Each mirror displacement $\Delta x_{j}$ is coupled to a single output mode: $\underset{i}{\sum}\left( A_{\text{ph}} \right)_{i,j}\Delta x_{j}$, where $A_{\text{ph}}$ is the transfer matrix. Let us define the vector $d_{\Delta x}=A_{\text{ph}} \mathbf{\Delta x}.$ We find the propagation into the clockwise cyclic trajectory and from symmetry we get the propagation into the counter clockwise trajectory: It can be observed that for every $j,k$ the coefficient of $\Delta x_{j}$ in $(d_{\Delta x})_{j+k,\text{L}}$ is identical to its coefficient in $(d_{\Delta x})_{j-k,\text{R}}$, where by by $j\pm k$ we mean mod $3$. Observe that for every $j\neq0$ the coefficient of $\Delta x_{j}$ in $(d_{\Delta x})_{j+k,\text{L}}$ is equal to
\begin{equation*}
-A_{\text{in}}\left(\frac{\omega_{0}}{c}\right)\cos\left(\pi/6\right)\frac{\left(1+\sqrt{R} \right)\left(R'\right)^{k/2}}{\left(1-R'^{3/2}\right)},    
\end{equation*}
 the coefficient of $\Delta x_{j}$ in $b_{j} \left(k=0\right)$ is equal to
\begin{eqnarray*}
\begin{split}
&-A_{\text{in}}\left(\frac{\omega_{0}}{c}\right)\cos\left(\pi/6\right)\left(\sqrt{R}+\frac{  \left( 1+\sqrt{R} \right)\left(R'\right)^{3/2}}{\left(1-R'^{3/2}\right)}\right)\\
&=-A_{\text{in}}\left(\frac{\omega_{0}}{c}\right)\cos\left(\pi/6\right)\left(\frac{\sqrt{R}+R'^{3/2}}{1-R'^{3/2}}\right).
\end{split}
\end{eqnarray*}
Hence in the DC limit the transfer matrix of the displacement vector,$A_{\text{ph}}$, reads:
\begin{eqnarray}
\begin{split}
&A_{\text{ph}}=-A_{\text{in}}\left(\frac{\omega_{0}}{c}\right)\cos\left(\pi/6\right)\frac{\sqrt{R}}{\left(1-R^{3/2}\right)} \cdot \\
&\left(\begin{array}{c}
\left(\begin{array}{c}
1+R\\
1+\sqrt{R}\\
\sqrt{R}+R
\end{array}\right)_{\text{L}}\\
\left(\begin{array}{c}
1+R\\
\sqrt{R}+R\\
1+\sqrt{R}
\end{array}\right)_{\text{R}}
\end{array}\begin{array}{c}
\left(\begin{array}{c}
\sqrt{R}+R\\
1+R\\
1+\sqrt{R}
\end{array}\right)_{\text{L}}\\
\left(\begin{array}{c}
1+\sqrt{R}\\
1+R\\
\sqrt{R}+R
\end{array}\right)_{\text{R}}
\end{array}\begin{array}{c}
\left(\begin{array}{c}
1+\sqrt{R}\\
\sqrt{R}+R\\
1+R
\end{array}\right)_{\text{L}}\\
\left(\begin{array}{c}
\sqrt{R}+R\\
1+\sqrt{R}\\
1+R
\end{array}\right)_{\text{R}}
\end{array}\right)
\end{split}
\end{eqnarray}
Comparing with the DC limit of eq. \ref{eq:GW_response_1}, we can observe that the GW response is a linear combination of the vector of $A_{\text{ph}}$, hence the DC divergence of the SD.

\section{Sagnac effect}

\begin{figure}[b]
\begin{center}
{\includegraphics[width=.48\textwidth]{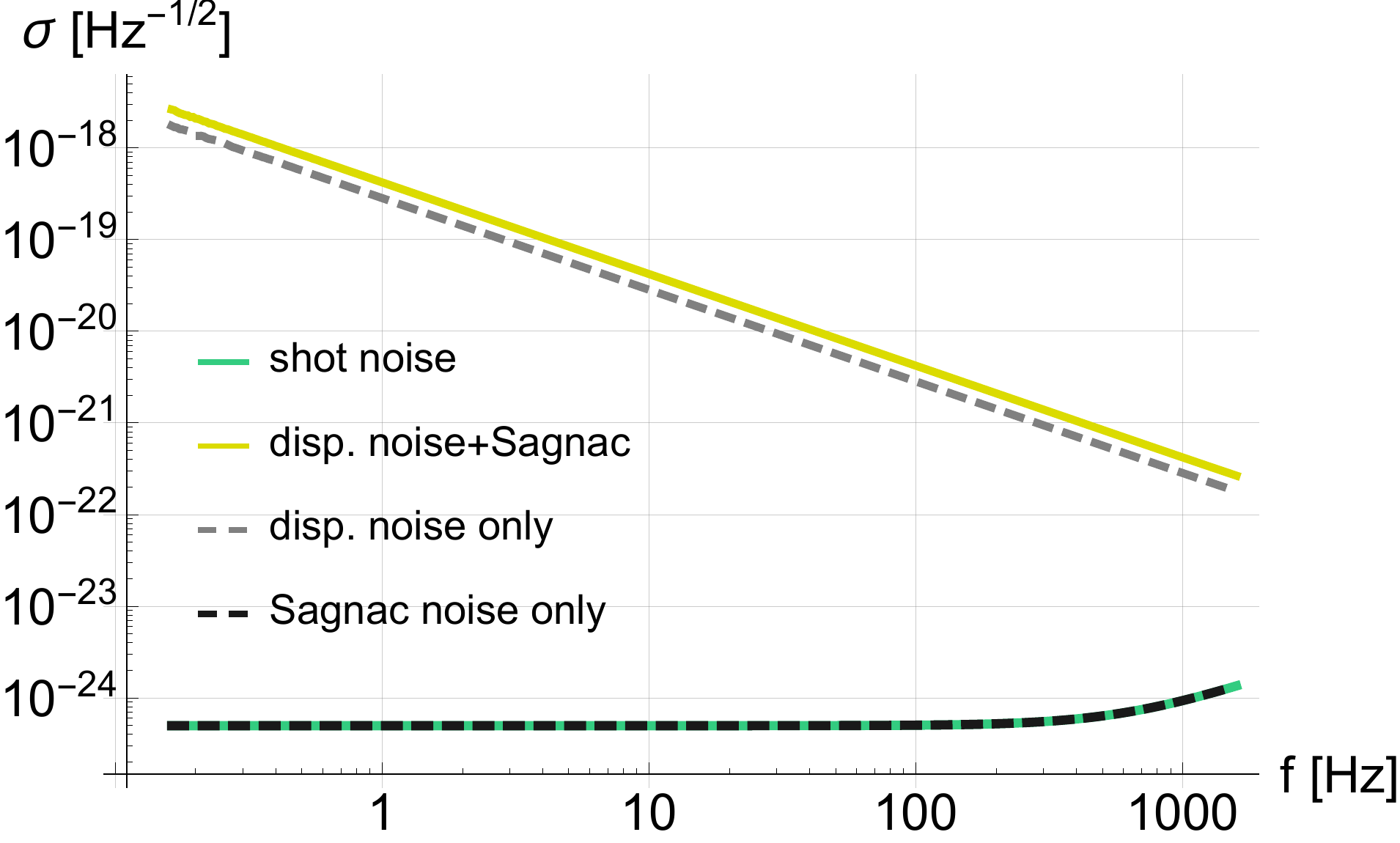}}
{\includegraphics[width=.48\textwidth]{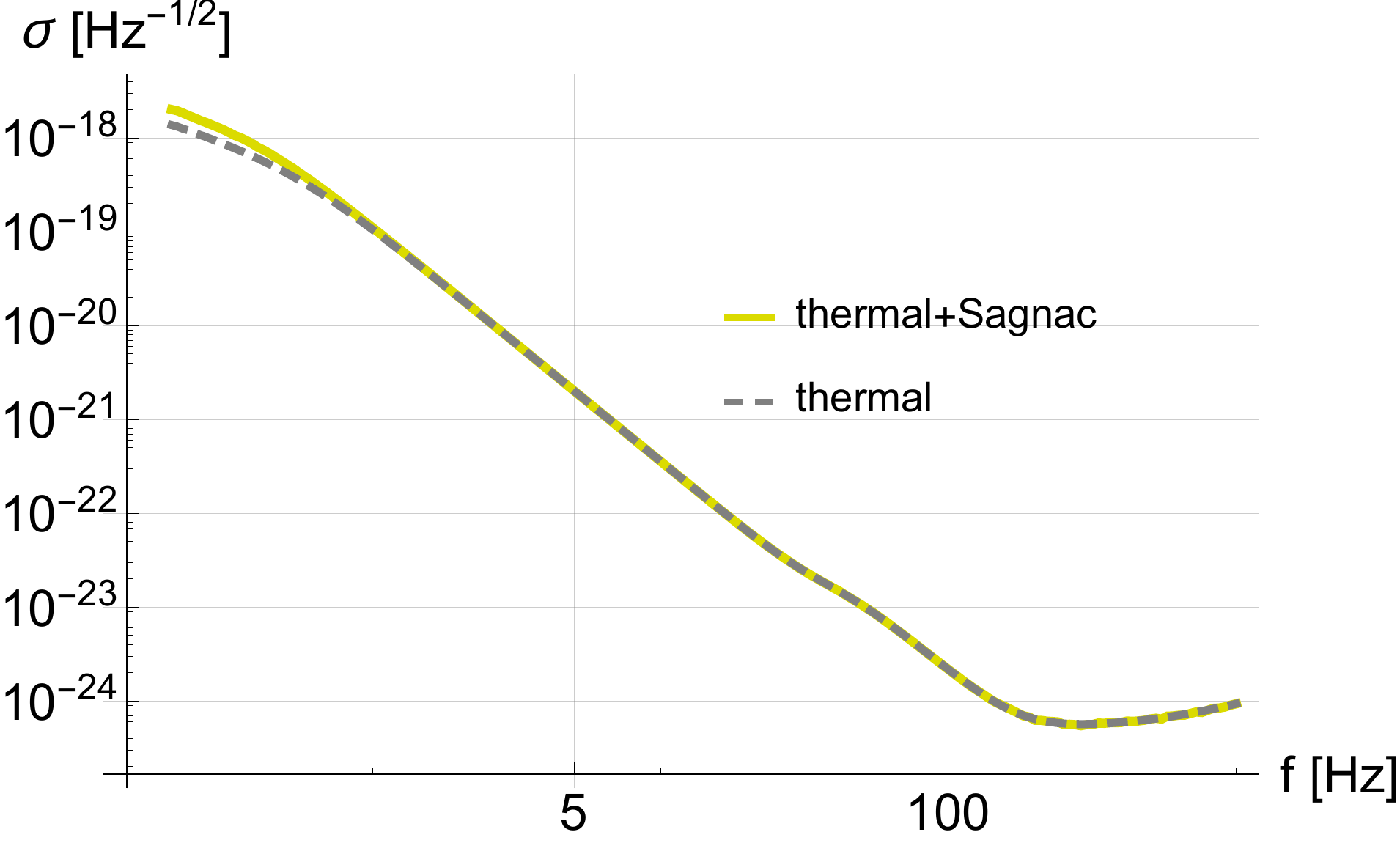}}
\end{center}
\caption{ left: sensitivity profiles with {\it{white infinite}} displacement and Sagnac noises. Bottom green (solid) line corresponds to shot noise limit. Top yellow (solid) line corresponds to both infinite displacement and Sagnac noises. The dashed gray (top) and black (bottom) lines correspond to displacement and Sagnac noise only respectively. 
Right: Comparison between thermal noise only (dashed,grey line) and thermal noise with infinite Sagnac noise (yellow, solid curve). The Sagnac noise almost does not change the sensitivity profile, its effect becomes non-negligible only in the limit of very poor sensitivity.}
\label{fig:sagnac}
\end{figure}

Since our proposed interferometers have a non-zero area, a rotation, such as the rotation of the earth, would induce a phase shift. This unwanted phase shift is an additional noise source that should be accounted for.

In this section we calculate this effect and find the contribution to the noise budget. Denoting the rotation axis as $z$ axis and the rotation frequency $\omega_{r}$, by moving to the rotating frame the first order correction (in $\omega_{r} L/c$) to the flat metric reads:
\begin{align}
h=\left(\begin{array}{cccc}
0 & \frac{\omega_{r}}{c}y & -\frac{\omega_{r}}{c}x & 0\\
\frac{\omega_{r}}{c}y & 0 & 0 & 0\\
-\frac{\omega_{r}}{c}x & 0 & 0 & 0\\
0 & 0 & 0 & 0
\end{array}\right).
\label{eq:h_sagnac}
\end{align}
Given a light that propagates in direction $\mathbf{n}$ and a carrier frequency $\omega_{0},$ the resulting phase shift reads:
\begin{align*}
&\partial_{\mathbf{n}}\phi_{s}=-\frac{\omega_{0}}{2c}h^{\mu\nu} n_{\mu} n_{\nu}\\
&\Rightarrow\phi_{s}=-\frac{\omega_{0}}{2c}\underset{0}{\overset{L}{\int}}n_{i}n_{j}h_{i,j}\left(ct_{0}+\xi,\boldsymbol{x_{0}}+\xi\boldsymbol{n}\right)\;\text{d}\xi.
\end{align*}
Inserting the correction to the metric due to rotation, eq. \ref{eq:h_sagnac}, yields:
\begin{align*}
&n_{i}n_{j}h_{i,j}\left(ct,\mathbf{r}\right)=\left(\begin{array}{cccc}
-1 & n_{x} & n_{y} & n_{z}\end{array}\right)\frac{\omega_{r}}{c}\left(\begin{array}{cccc}
0 & y & -x & 0\\
y & 0 & 0 & 0\\
-x & 0 & 0 & 0\\
0 & 0 & 0 & 0
\end{array}\right)\left(\begin{array}{c}
-1\\
n_{x}\\
n_{y}\\
n_{z}
\end{array}\right)\\
&=2\omega_{r}/c\left(\mathbf{n}_{x-y}\times\mathbf{r}_{x-y}\right),
\end{align*}
where $\mathbf{n}_{x-y},$ $\mathbf{r}_{x-y}$ are the projections of $\mathbf{n}$ and $\mathbf{r}$ onto the $x-y$ plane. From here we can calculate the phase shift generated in each arm (given a cyclic trajectory).
This expression can be further simplified for straight arms where the trajectory can be parameterized as $\mathbf{r}\left(\xi\right)=\mathbf{r}_{j}+\xi\hat{n_{j}},$ and thus $\hat{n}_{j}\times\mathbf{r}_{j}=\mathbf{r}_{j+1}\times\mathbf{r}_{j}.$
Summing over all the phases accumulated in a cyclic trajectory yields the well-known Sagnac phase shift $\underset{j}{\sum}\phi_{j}=-\frac{\omega_{r}\omega_{0}}{c^{2}}\underset{j}{\sum}\mathbf{r}_{j+1}\times\mathbf{r}_{j}=-2\frac{\omega_{r}\omega_{0}}{c^{2}}A.$
In our case, since all the ports are open, the shift in each output port will be different. Once we have the different phase shifts $\left\{ \phi_{j}\right\} _{j=1}^{3},$ they are being propagated 
to the output ports in the same manner as in eq. \ref{eq:GW_response_1}, yielding:
\begin{align}
\mathbf{d}_{\text{Sagnac}}=\frac{(1+\sqrt{R})}{(1-R'^{3/2})}\left(\begin{array}{c}
\left(\begin{array}{c}
R'\phi_{1}+\sqrt{R'}\phi_{2}+\phi_{3}\\
R'\phi_{2}+\sqrt{R'}\phi_{3}+\phi_{1}\\
R'\phi_{3}+\sqrt{R'}\phi_{1}+\phi_{2}
\end{array}\right)_{\text{L}}\\
-\left(\begin{array}{c}
\phi_{1}+\sqrt{R'}\phi_{2}+R'\phi_{3}\\
\phi_{2}+\sqrt{R'}\phi_{3}+R'\phi_{1}\\
\phi_{3}+\sqrt{R'}\phi_{1}+R'\phi_{2}
\end{array}\right)_{\text{R}}
\end{array}\right)
\label{eq:sagnac_response}
\end{align}

Given a fixed rotation frequency ($\Omega_{r}$) and axis, such as the rotation of earth, this effect results in a systematic dc noise. In that case it should not affect the sensitivity since: 1. as a systematic noise it can be deducted. 2. it appears only on dc, in which the SD diverges anyway.      

Let us address the case of random AC rotations where this noise cannot be accounted as systematic.
In this case, $\mathbf{d}_{\text{Sagnac}}$
is a gaussian noise and we include it in the QFIM expression along with the displacement noise: $I=4\mathcal{V}^{\dagger}\left(MM^{\dagger}+\delta^{2}AA^{\dagger}+\mathbf{d}_{\text{Sagnac}}\mathbf{d}_{\text{Sagnac}}^{\dagger}\right)^{-1}\mathcal{V}.$ 
Let us now examine the effect of this noise and in particular whether it has the same effect as displacement noise. 
The results are shown in fig. \ref{fig:sagnac}. The effect of Sagnac noise is quite negligible compared to the displacement noise. The curve of Sagnac noise alone, even infinite Sagnac noise, coincides with the shot noise limit.    
The reason for this is that in the dc limit the overlap between the Sagnac noise (eq. \ref{eq:sagnac_response}) and the GW signal (eq. \ref{eq:GW_response_1}) is Very small, they are almost orthogonal.
In this limit the GW signal is close to symmetric while the Sagnac noise is close to anti symmetric. The effect of the Sagnac noise becomes more pronounced in the limit of large displacement noise where the contribution of the small overlap is non-negligible.

\section{Comparison with standard Sagnac interferometers}

In this section we compare the performance of our triangular symmetric DFI scheme (fig. \ref{fig:comp_sagnac}(a) top) with the performance of the standard triangular Sagnac interferometer (fig. \ref{fig:comp_sagnac}(a) bottom). 
The standard triangular Sagnac interferometer has the same geometry as our DFI scheme: $3$ mirrors that form a triangle, but unlike the DFI scheme only one port is open and $2$ other ports are close. Hence the standard Sagnac scheme has two input/ output fields, while our DFI scheme has six input/output.

Since the Sagnac interferometer has $2$ outputs and $3$ mirrors it does not necessarily have a DFS. In practice, due to symmetry, the Sagnac interferometer has a DFS in dc, but does not have a DFS at higher frequencies (in dc $\text{rank} \left( A \right)=1$). We thus expect our DFI scheme to outperform the standard Sagnac scheme in the presence of strong displacement noise.

This expectation is confirmed in the numerical results shown in fig. \ref{fig:comp_sagnac}(b),
where we compare the performance of the interferometers given phase quadratures measurement with realistic noise profile (thermal noise and radiation pressure).
In this comparison the intracavity power is the same.
It can be observed that the symmetric DFI outperforms the Sagnac interferometer in almost the entire frequency range. At high frequencies, the symmetric DFI has a better shot noise limit, which accounts for a better sensitivity in this regime. At lower frequencies, the RPN and displacement noise are dominant. The noise resilient subspaces of the symmetric DFI, i.e. DFS and pseudo-DFS, enable better sensitivity in this regime. Interestingly, despite the fact that the Sagnac interferometer has a DFS in dc, its dc divergence is much worse than that of the symmetric DFI.

\begin{figure}[b]
\begin{center}
\subfigure[]{\includegraphics[width=.25\textwidth]{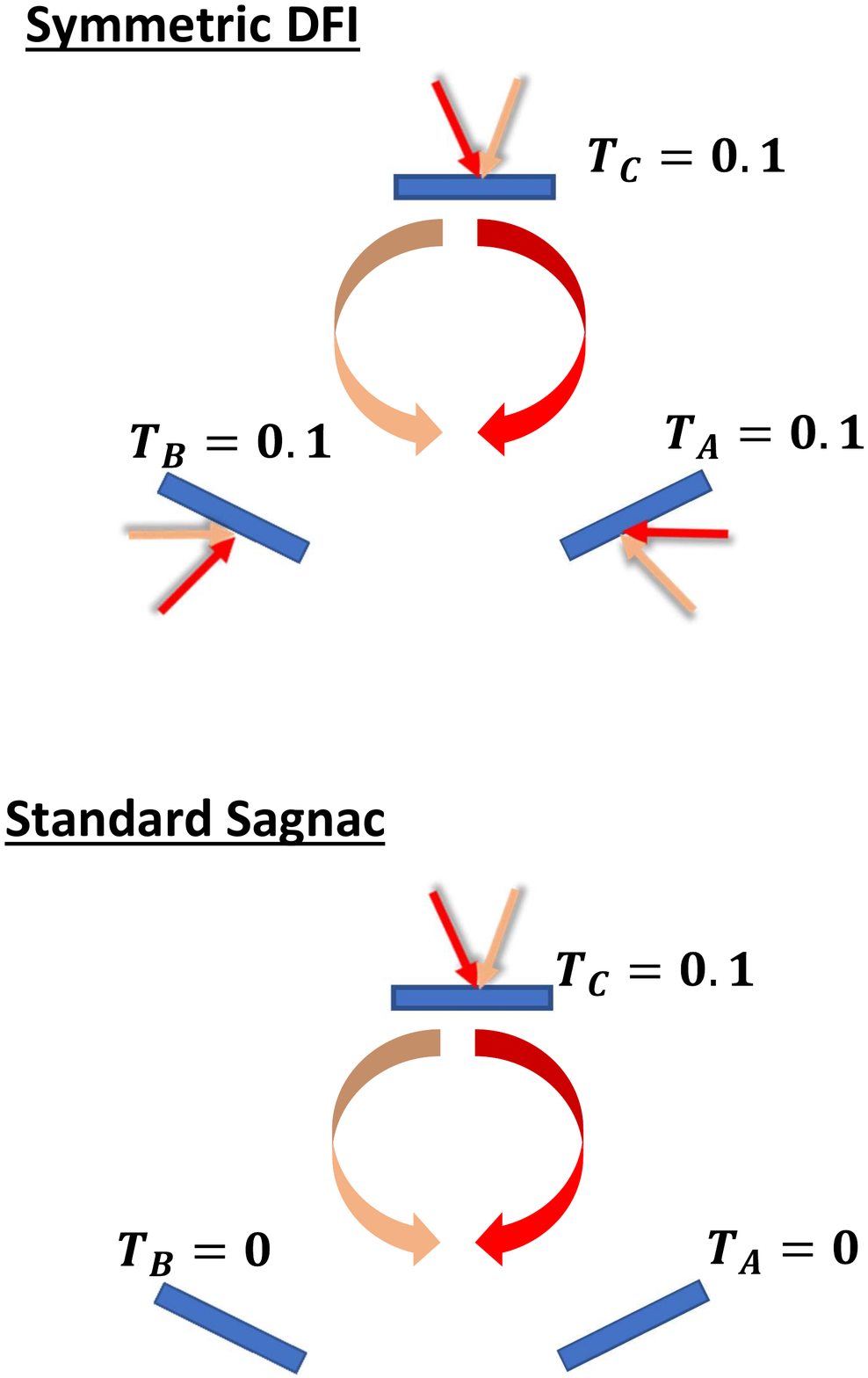}}
\subfigure[]{\includegraphics[width=.35\textwidth]{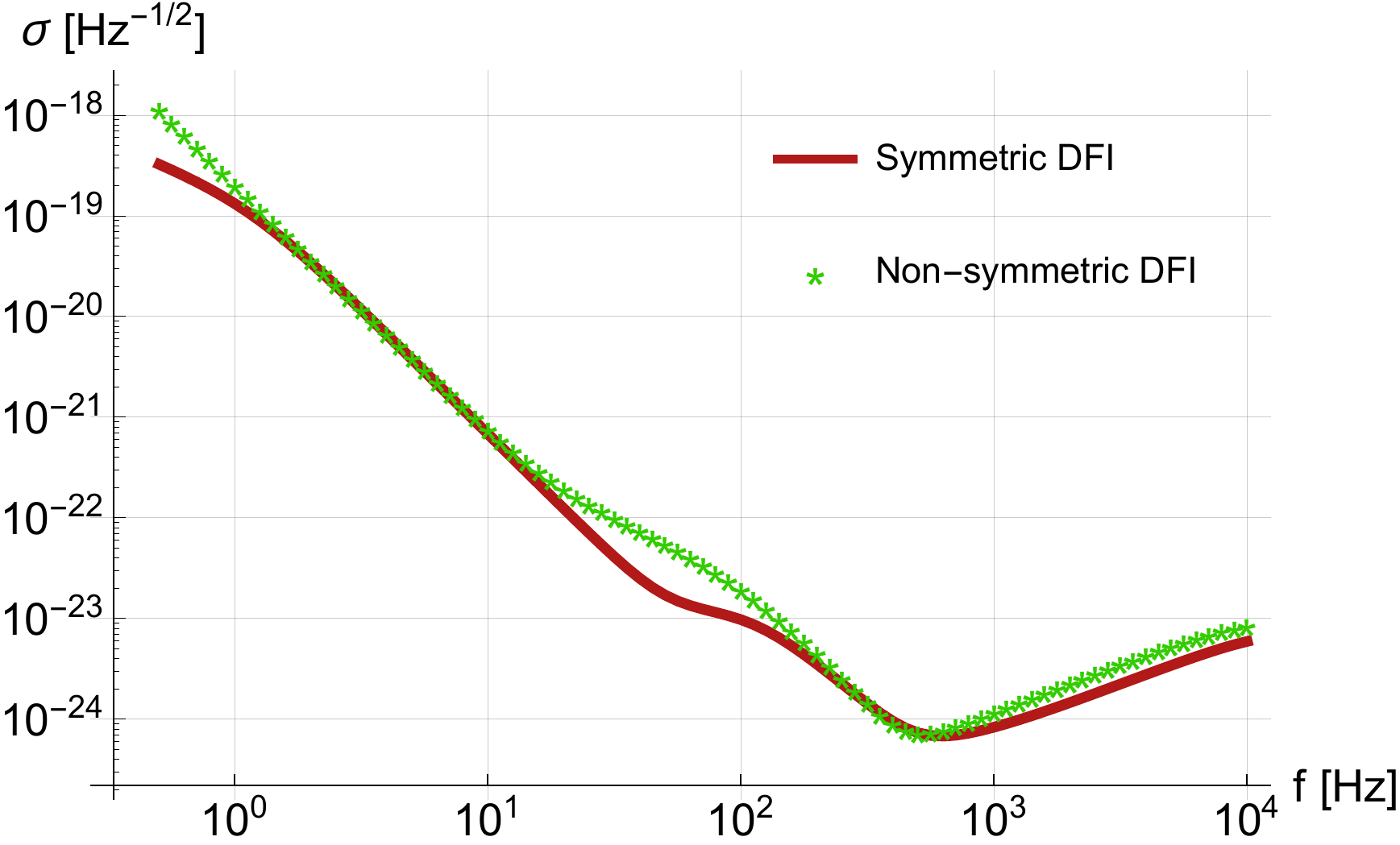}}
\subfigure[]{\includegraphics[width=.35\textwidth]{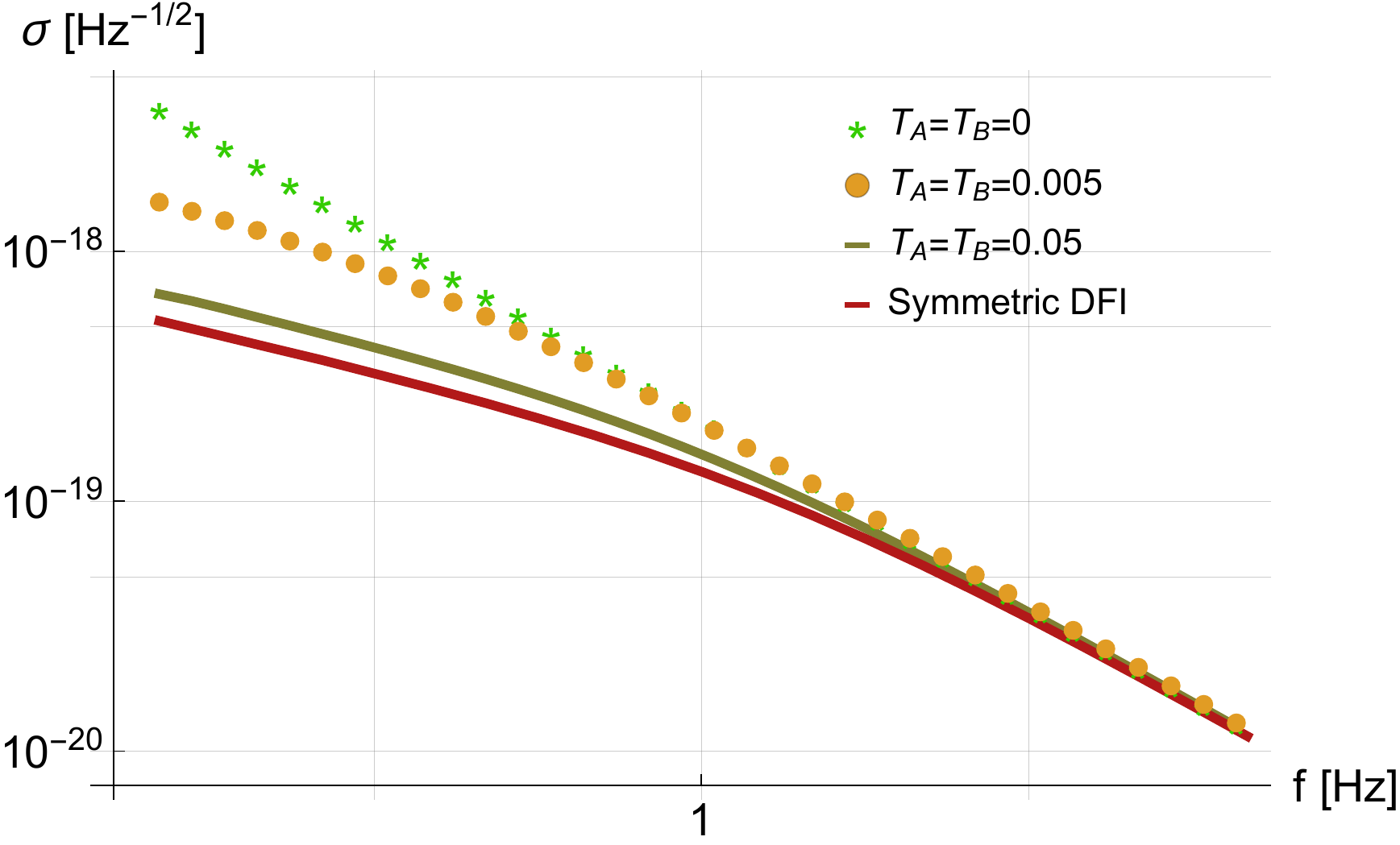}}
\end{center}
\caption{
(a) Top: our symmetric DFI interferometer, bottom: standard triangular Sagnac interferometer, where only one port is open. (b) Sensitivity comparison between the standard triangular Sagnac and the symmetric DFI given phase quadratures measurement and realistic noise profile (RPN and thermal noise). (c) sensitivity comparison, at low frequencies, between different transmissivity configurations, illustrating the optimality of the symmetric DFI.  }  
\label{fig:comp_sagnac}
\end{figure}

We remark that the standard Sagnac and symmetric DFI differ only in the mirrors transmissivities (see fig. \ref{fig:comp_sagnac}(a)).
This raises the question of what are the optimal transmissivities of the mirrors? 
We performed numerical optimization of the sensitivity as a function of the mirrors' transmissivities, with the constraint of $0\leq T_{A},T_{B},T_{C}\leq0.1$ and fixed intracavity power. We observed that the optimal values correspond to our symmetric DFI configuration: $T_{A},T_{B},T_{C}=0.1.$
These results are illustrated in fig. \ref{fig:comp_sagnac}(c) and they establish the optimality of the symmetric DFI scheme analyzed in the paper.


\end{widetext}


\end{document}